\documentclass[sigconf]{acmart}

\setcopyright{none}
\renewcommand\footnotetextcopyrightpermission[1]{}

\usepackage{enumitem}
\usepackage{multirow}
\usepackage{soul} 
\usepackage{subcaption}

\usepackage{graphicx}
\usepackage{array}
\usepackage{microtype}
\usepackage{booktabs}
\usepackage{caption}
\usepackage{tikz}
\usepackage{float}
\usepackage{stfloats}
\usepackage{balance}
\usetikzlibrary{matrix,positioning,calc}

\usepackage{fancyhdr}
\fancypagestyle{plain}{%
   \fancyhf{} %
   \fancyfoot[L]{ACM SIGCOMM Computer Communication Review }%
   \fancyfoot[R]{Volume 56 Issue 2, July 2026}%
}
\fancypagestyle{firstpagestyle}{%
   \fancyhf{} %
   \fancyfoot[L]{ACM SIGCOMM Computer Communication Review}%
   \fancyfoot[R]{Volume 56 Issue 2, July 2026}%
}

\begin{document}

\title{A DualPI2 Module for Mahimahi: Behavioral Characterization and Cross-Platform Analysis}

\author{Nawel Alioua}
\affiliation{
	\institution{UC Santa Barbara}
}
\email{nawel@ucsb.edu}

\author{Linghe Zhang}
\affiliation{
	\institution{UC Santa Barbara}
}
\email{linghe@ucsb.edu}

\author{Aneesh Garg}
\affiliation{
	\institution{UC Santa Barbara}
}
\email{aneeshgarg@ucsb.edu}

\author{Francis Y. Yan}
\affiliation{
	\institution{University of Illinois Urbana-Champaign}
}
\email{fyy@illinois.edu}

\author{Elizabeth Belding}
\affiliation{
	\institution{UC Santa Barbara}
}
\email{ebelding@ucsb.edu}

\begin{abstract}

Low Latency, Low Loss, and Scalable Throughput (L4S) is an emerging paradigm for latency control based on DualPI2 active queue management and scalable congestion control. While a Linux kernel implementation of DualPI2 is available, controlled and reproducible experimentation on L4S mechanisms can be facilitated by a modular, user-space alternative. In this paper, we present a DualPI2 module for the Mahimahi network emulator, designed to support extensible, component-level experimentation without kernel modification. We conduct a statistical behavioral characterization of the Mahimahi implementation by examining key metrics across diverse traffic patterns and network conditions, using the Linux kernel implementation as a reference baseline. Our analysis shows that behavioral alignment across execution environments is not automatic: identical DualPI2 parameterization does not guarantee identical dynamics. Instead, key control parameters exhibit environment-dependent sensitivity, leading to regime-dependent discrepancies across bandwidth-delay product (BDP) conditions. Through targeted parameter exploration, we identify configurations that improve cross-platform alignment in low BDP regimes, while revealing structural differences that persist under higher load. This work provides both a practical tool for experimental L4S research and empirical insight into cross-platform behavioral differences, highlighting the importance of systematic characterization and environment-aware parameter selection in emulation-based AQM studies.

\end{abstract}

\ccsdesc[500]{Networks~Network experimentation}
\ccsdesc[500]{Networks~Network performance analysis}
\ccsdesc[300]{Networks~Network architectures}
\ccsdesc[300]{Networks~Transport protocols}

\keywords{L4S, DualPI2, low-latency, network emulation, active queue management, ECN, statistical analysis}

\maketitle

\section{Introduction}\label{sec:intro}
The Low Latency Low Loss Scalable Throughput (L4S) architecture is a modern approach to network-wide congestion control that aims to reduce queuing delays while maintaining high throughput. L4S challenges the  long-held belief that these two objectives form a zero-sum tradeoff~\cite{rfc9330}.
This architectural shift is increasingly gaining traction amid two seemingly contradictory trends in modern networking: (i)~the growing availability of high-bandwidth access technologies (e.g., fiber, 5G), which continually push the frontier of achievable 
throughput~\cite{fcc-throughput}; and (ii)~the growing recognition among ISPs and network operators that increasing bandwidth alone does not solve the core problem of user-perceived performance, namely insufficient interactive responsiveness and high latency~\cite{ookla_webinar}. As a result, the focus of innovation has shifted from raw throughput to end-to-end latency and, by extension, to application-level quality of experience (QoE). This shift is especially important given the rise of latency-sensitive services like AR/VR, cloud gaming, and teleconferencing.

The persistent latency problem emerges as a direct consequence of bufferbloat, characterized by ``overly large, unmanaged, and uncoordinated buffers''~\cite{bufferbloat}. While early fixes focused on in-network active queue management (AQM) and shallow queue designs~\cite{bufferbloat-early-survey}, L4S builds on these principles and introduces a new paradigm: latency control through sender-side congestion responsiveness enabled by explicit and fine-grained congestion signaling (ECN), and supported by specialized AQMs like DualPI2~\cite{olga-dualpi2} at bottleneck routers.

Ten years after its early conceptualization, L4S is now being actively standardized at the IETF, with working group discussions, multiple experimental RFCs~\cite{rfc9330, rfc9331, rfc9332}, and early deployments from ISPs. While industry and standards activity have accelerated in recent years, academic exploration of L4S is only beginning to gain momentum.

A key enabler of broader academic engagement will be the availability of accessible experimentation tools. Most L4S studies rely on the Linux kernel implementation of DualPI2, which, while functional, requires elevated system privileges for empirical evaluation, and non-trivial kernel reconfiguration and recompilation for design exploration of alternative AQM components. This limits exploratory testing, slows iteration, and complicates reproducibility. With L4S still evolving, the broader networking community is well positioned not only to assess its performance, but also to contribute to its refinement. Supporting this collaborative development will require tooling that is modular and accessible to researchers and developers across platforms.

To address this need, we present a modular implementation of the DualPI2 AQM~\cite{olga-dualpi2} integrated into Mahimahi~\cite{mahimahi-atc, mahimahi-ccr}, a lightweight, user-space emulator already widely used for mobile and transport-level network evaluation. Unlike the monolithic nature of the kernel implementation—dicta\-ted by the internal structure and integration constraints of the Linux networking stack—our version separates the components of DualPI2 in alignment with the RFC architecture. This modularization encourages design experimentation, such as swapping in alternative schedulers or tuning marking strategies. We show that DualPI2 exhibits environment-dependent parameter sensitivity, such that default configurations do not ensure cross-platform behavioral consistency. Through targeted tuning, we identify configurations that improve cross-platform alignment under specific BDP regimes.

\vspace*{0.1in}
Our key contributions are as follows:
\begin{itemize}
  \item We present the first L4S-capable DualPI2 AQM implementation integrated into the Mahimahi emulator.

  \item We release the implementation of the DualPI2 Mahimahi module\footnote{\url{https://github.com/5G-VCA-CC/mahimahi}} and experimental artifacts\footnote{\url{https://github.com/5G-VCA-CC/Experimentation-Data-Scripts/tree/ccr26-submission}} to enable repeatable L4S experimentation without kernel modification.

  \item We statistically quantify the achieved throughput, queueing dynamics, ECN marking rates, and drop behavior of our implementation across controlled scenarios, highlighting similarities and differences relative to the Linux DualPI2 implementation.

  \item We analyze the sensitivity of DualPI2 behavior to its configuration across execution environments, identifying key parameters that govern alignment between Mahimahi DualPI2 and the Linux implementation. We show that targeted tuning improves behavioral consistency under specific network conditions and provide practical parameter recommendations for reproducible experimentation.

  \item We release the statistical testing framework as an independent project to enable similar validation and characterization work using any metric.\footnote{\url{https://github.com/5G-VCA-CC/mahimahi_validation.git}} 
  
\end{itemize}

\section{Background and Motivation}\label{sec:background_motiv}
In this section, we first outline the core design of L4S to ground our work in its technical context. We then examine recent industry deployment efforts, which signal growing momentum but also highlight the need for broader collaboration. Finally, we articulate the current limitations preventing open, reproducible experimentation with L4S, particularly within the academic community, and how our work seeks to address them.

\subsection{Overview of the L4S Architecture}
 
The L4S architecture, as standardized collectively across RFCs 9330, 9331 and 9332~\cite{rfc9330, rfc9331, rfc9332}, is comprised of two main components: a compliant congestion controller on the end hosts that responds to the explicit congestion signals, and in-network AQM support at the bottleneck for managing the coexistance of classic traffic and L4S traffic, in the form of  a Dual Queue Coupled AQM. 

The end-host congestion controller plays a central role in the L4S architecture, as it is responsible for translating explicit ECN-based congestion signals into rate adaptations that jointly achieve low latency, low loss, and scalable throughput. Unlike classic congestion control, which primarily relies on rare packet loss events and deep queues for stability, L4S operates with frequent, low-amplitude ECN marking and deliberately shallow queues at the bottleneck. These design choices fundamentally change the nature of the congestion signal exposed to the sender, requiring continuous and fine-grained rate adaptation in the absence of buffering-induced smoothing. Traditionally, ECN signals in ``classic ECN'' were treated as equivalent to packet drops, triggering the same abrupt congestion responses—such as halving the congestion window in New Reno. In contrast, L4S redefines the semantics of the ECN Congestion Experienced (CE) mark to indicate early, low-level congestion, intended to elicit smooth and responsive rate control rather than drastic backoff. This shift enables the coexistence of low latency and high throughput, but also demands that both endpoints and AQMs interpret ECN signaling in fundamentally new ways.

While the concept of scalable congestion control predates L4S~\cite{scalable-tcp}, the signaling and queueing model introduced by L4S makes scalable congestion control a necessity rather than an optimization: congestion controllers that depend on infrequent loss events or queue buildup for stability are ill-suited to achieve L4S' goals. In addition, loss-based controllers typically incur long convergence times following a loss event, particularly at high sending rates, which is incompatible with the L4S objective of maintaining both low latency and high utilization.

A congestion controller is~\emph{scalable} if it fulfills the following conditions regardless of the scale of the network bandwidth~\cite{koen-dctth, scalable-tcp}: 

\begin{enumerate}[label=\roman*)]
  \item the time it takes to double the transmission rate must be independent of the current rate. This ensures the scalable throughput aspect of L4S, even in high-bandwidth settings.
  
  \item the number of congestion signals per RTT remains constant, allowing the protocol to remain sensitive to these signals.
\end{enumerate}

\noindent Together, these properties enable scalable congestion controllers to respond proportionally to frequent ECN marks while maintaining stable operation at high rates and low queueing delay. In contrast, classic congestion control algorithms such as Reno or Cubic are optimized for loss-driven operation and implicitly rely on queue buildup and infrequent congestion events, leading to slower convergence after loss and larger delay oscillations.

On the deployability front, for a scalable congestion control protocol to be usable with L4S over the public Internet, it must also comply with the Prague L4S Requirements~\cite{prague-req, rfc9331}. By satisfying these requirements, a transport can safely use L4S signals on the Internet; its congestion control can exploit frequent and fine-grained feedback provided through AccECN~\cite{rfc9768}; and it is able to coexist with classic flows fairly at shared bottlenecks. In addition, it can fallback to classic (Reno-friendly) behavior on packet loss. Examples of L4S-compatible congestion controllers include TCP Prague~\cite{prague-req}, UDP Prague~\cite{udp-prague-github} and SCReAM~\cite{rfc8298, scream-paper}. 

\begin{figure}[ht]
\centering

\includegraphics[width=0.47\textwidth]{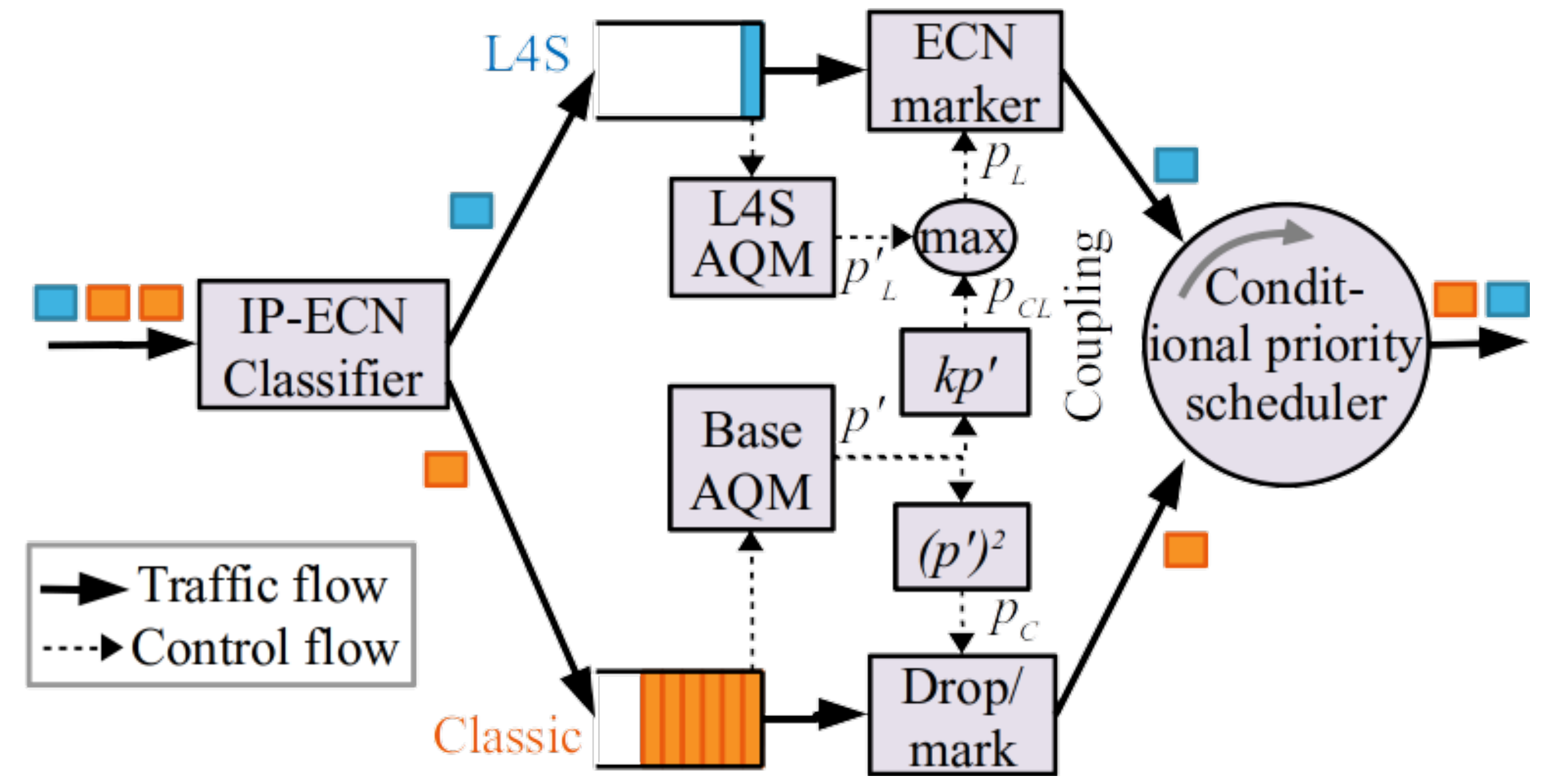}
\caption{Dual Queue Coupled AQM~\cite{koen-dualq, koen-dctth, olga-dualpi2}}
\label{fig:dualq-coupled-aqm}
\vspace*{-0.06in}
\end{figure}

The design of the Dual Queue Coupled (DQC) AQM, specified in RFC 9332~\cite{rfc9332} and depicted in Fig.~\ref{fig:dualq-coupled-aqm}, isolates traffic by way of two queues: a classic (C) queue serving legacy congestion-controlled flows, and an L4S (L) queue for flows employing scalable congestion control. Each queue is managed by its own AQM, which independently regulates queue delay around a target operating point by computing a base control probability, namely $P'$ for the C queue, and $P'_L$ for the L queue. Then, these probabilities respectively determine packet drops ($P_C$) for classic traffic, and ECN marking ($P_L$) for L4S traffic.

The L4S queue is given latency priority to minimize queuing delay for scalable flows; however, throughput fairness between classic and L4S traffic is enforced through a coupling mechanism between the two AQMs; this is realized by setting the final L4S marking probability as the maximum of its native marking probability $P'_L$ and a scaled version of the classic base drop probability $P_C=k*P'$.

The coupling ensures that L4S traffic remains responsive to congestion driven by classic flows, while still allowing the L4S AQM to independently maintain low queueing delay. Finally, a conditional priority scheduler arbitrates service between the C and L queues, providing latency priority to the L queue while using preset queue weights to control long-term throughput sharing with the C queue. 

DualPI2 AQM~\cite{olga-dualpi2, rfc9332} is the de facto Linux kernel reference implementation of the DQC AQM. It is derived from the PI$^2$ AQM~\cite{koen-pi2}, which itself is an evolution of the PIE AQM~\cite{pi-aqm}. The DualPI2 traffic classifier directs packets based on the value of the ECN field in the IP header as shown in Table~\ref{tab:ecn_codepoints}. Packets marked with ECT(1) or CE are directed to the L queue, while packets marked Not-ECT or ECT(0) are sent to the C queue.

\vspace*{-0.05in}
\begin{table}[ht]
  \centering
  \caption{ECN codepoints for L4S~\cite{rfc9331}}
  \vspace*{-0.1in}
  \label{tab:ecn_codepoints}
  \includegraphics[width=0.47\textwidth]{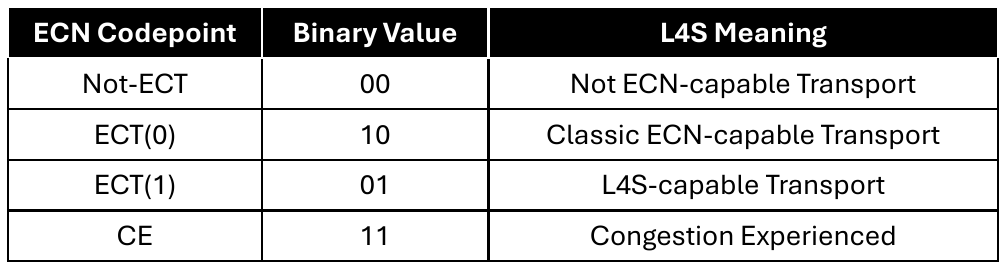}
  \vspace*{-0.05in}
\end{table}

The L queue in DualPI2 employs a low-latency AQM, referred to as the~\emph{Step AQM} in the Linux implementation, with a very shallow ECN marking target (default \texttt{step\_thresh} =  1 ms). In contrast, the C queue is managed by the PI$^2$ AQM and operates at a significantly higher target queuing delay (default \texttt{pi2\_target} = 15 ms), which reduces the drop probability and allows loss-based congestion controls to achieve high link utilization.

PI$^2$~\cite{koen-pi2} is a proportional–integral controller parameterized by an integral gain ($\alpha$) and a proportional gain ($\beta$). The integral component gradually eliminates persistent steady-state deviations from the target delay, while the proportional component reacts more rapidly to transient changes in excess load. Together, the proportional and integral components provide both fast response and convergent long-term behavior.

\subsection{Industry Deployment Efforts}
\label{sec:industry-efforts}

Historically, industry considerations of network performance have predominantly focused on increasing network capacity, also known as bandwidth. L4S emerged from an alternative strategy for improving quality of service, concentrating instead on the reduction of network latency, particularly loaded latency~\cite{ookla-loaded-latency}. Nokia, through its Bell Labs research arm, has played a leading role in L4S research and standardization, which was initiated in the mid-2010s~\cite{nokiaL4S}, and first introduced to the IETF community in 2016~\cite{l4s-first-id}. Research activity gradually expanded, notably with CableLabs integrating L4S mechanisms into their Data Over Cable Service Interface Specification (DOCSIS)~\cite{docsis}.  This, in turn,  lead to  the publication of the first Low Latency DOCSIS Internet Draft~\cite{lldocsis-internet-draft} in 2019. The following years saw the technology mature into practical deployment. Early 2025, Comcast became the first ISP to roll out low latency DOCSIS to its customers~\cite{comcast-xfinity, comcast-l4s-2025}, recently reaching the milestone of 10~million L4S-enabled homes, representing 360 million devices~\cite{livingood-10-million}.

Beyond fixed access networks, mobile network operators have also embraced L4S as a pathway to consistent low-latency services. In particular, Vodafone, in collaboration with Nokia Bell Labs, conducted one of the first end-to-end trials of L4S over passive optical broadband~\cite{vodafone}, yielding near-zero queuing delays under load. Another example of such partnerships was between Ericsson, Qualcomm and SoftBank, which demonstrated L4S-enabled ultra-low latency on commercial 5G Standalone networks, and integrated L4S with 5G scheduling to secure consistent latency for mission-critical services~\cite{softbank}. In the US, T-Mobile has begun deploying L4S capabilities across their 5G networks~\cite{tmobile-L4S}. T-Mobile frames L4S as a flexible mechanism for supporting application-aware latency behavior rather than a single, uniform service tier. By combining L4S with a 5G standalone core and network slicing, T-Mobile is deploying L4S across multiple slices, where each slice may be tuned differently depending on the latency tolerance and traffic characteristics of the applications it supports~\cite{tmobile-L4S}.

Application-layer adoption is also gaining momentum. Apple has added L4S support to FaceTime, showing improved video quality and reduced stalls under congestion~\cite{apple-l4s}. With native integration in iOS 17 and macOS Sonoma, apps using QUIC, HTTP/3, or TCP can benefit with little or no modification. Developers also have the option to implement scalable congestion control and ECN validation for fine-grained optimization. Meanwhile, Netflix is developing a receiver-driven rate adaptation algorithm designed to leverage signals like those provided by L4S~\cite{netflix}. Though still experimental, these efforts reflect growing alignment between modern applications and emerging low-latency network capabilities.

\vspace*{-0.2in}
\subsection{Obstacles to Open and Rapid L4S Experimentation}

This section describes the challenges associated with reproducible, design-oriented experimentation within the L4S ecosystem, and discusses how an emulation-based DualPI2 implementation can complement system-level deployments by enabling broader access to controlled experimentation. In particular, it highlights how such an approach opens new opportunities for both network- and application-level evaluation across diverse access environments.

\vspace{0.1in}
\noindent
{\bf System-Level L4S Implementations}.
Multiple implementations of L4S are now available in Linux~\cite{rfc9332}, FreeBSD~\cite{free-bsd-l4s}, 5G programmable RAN such as OpenAirInterface~\cite{5g-tc-ran} and srsRAN~\cite{l4span}. Among these, the Linux kernel DualPI2 queue discipline (qdisc), referenced in RFC 9332~\cite{rfc9332}, remains the most widely used implementation in the research community. While tailored for production deployments, the kernel implementation is not optimized for rapid experimentation or prototyping. In fact, the monolithic nature of kernel qdisc modules, particularly DualPI2, complicates the isolation of specific components of the architecture for experimentation, such as the scheduler, since it is tightly embedded within the dequeue logic. A more modular design would facilitate mapping between the code and conceptual components defined in the RFC, streamlining targeted architectural exploration and comparative studies. As a result, relying solely on the kernel module can raise the barrier to entry for new researchers or practitioners seeking to test modifications or evaluate design alternatives.

In contrast, a user-space DualPI2 implementation would support configuration via command-line arguments, enabling clean state resets between runs and simplifying test automation. Parameters can be easily modified and swept over large ranges without kernel interaction, supporting systematic exploration of algorithmic behaviors across operating conditions. Instrumentation is straightforward, and logging can be implemented directly in user space without specialized kernel tracing tools. Moreover, an implementation on Mahimahi would offer a version-stable experimental baseline. The kernel DualPI2 codebase remains under active development, and changes in the upstream Linux tree may affect module behavior or interface assumptions over time. The Mahimahi module serves as a frozen snapshot of the design at specific points, enabling consistent replication of experiments over long timescales. Together, these properties make the Mahimahi DualPI2 module a valuable complement to the kernel implementation, supporting a faster, more flexible design exploration, reproducible experimentation, promoting broader engagement with L4S development.

\vspace*{0.1in}
\noindent
{\bf Limited Testing Opportunities on Mobile Networks.}
Testing congestion control mechanisms over real mobile infrastructure presents numerous challenges. Cellular networks exhibit highly variable channel conditions due to factors such as user mobility, environmental interference, and scheduling dynamics at the radio access layer. These fluctuations introduce significant noise into experimental results, requiring repeated trials under diverse conditions to reach statistically meaningful conclusions. Moreover, access to mobile infrastructure with sufficient control and visibility into the network stack is often limited, and conducting tests on commercial networks can be both time-consuming and costly.

As a result, mobile networking research frequently turns to emulation. Tools like Mahimahi (\S\ref{sec:mahimahi-related-work}) are widely adopted for this purpose due to their ability to replay recorded mobile traces in controlled, repeatable environments. Mahimahi supports uplink and downlink bandwidth traces collected from real cellular networks, enabling researchers to evaluate protocols and applications under representative mobile conditions without the unpredictability of live deployments.

However, despite its widespread use in mobile networking research, Mahimahi currently lacks support for L4S mechanisms. As discussed in \S\ref{sec:industry-efforts}, L4S deployments in commercial mobile networks remain limited, making direct experimentation on physical infrastructure impractical. This absence of emulator support represents not only a missed opportunity, but a significant gap in the current L4S research landscape, particularly for mobile scenarios. Without accessible, trace-driven emulation capabilities, researchers in both academia and industry are constrained in their ability to rapidly evaluate, refine, and innovate on L4S in mobile contexts, ultimately slowing progress toward optimization and broader adoption.

\vspace*{0.1in}
\noindent
{\bf Challenges for Application-Level L4S Research.}
Of the prior L4S application research noted in this paper (\S\ref{sec:l4s-prior}), the majority is conducted on testbeds~\cite{l4sxr-steininger, split-render-xr, l4s-cloud-gaming-graff, l4s-volumetric, l4s-private-5g-video, rtc-role-l4s} or using simulators or emulators~\cite{real-time-stream}, some of which are proprietary~\cite{brunello-ieee, brunello-master}, which limits access and reproducibility.

In contrast, Mahimahi has been instrumental to numerous application QoE-focused research efforts, such as VR~\cite{salient-vr}, web browsing~\cite{vesper}, telemetry on cellular networks~\cite{ng-scope}, mobile application QoE~\cite{floo}, video conferencing~\cite{ace, mowgli, tambur}, and most notably, video streaming~\cite{pensieve, cellfusion, vantage, robust-dash, in-situ, salsify, neural-video, prior, low-latency-http, compact}.
However, Mahimahi currently lacks an L4S-capable AQM, preventing controlled evaluation of application-level L4S mechanisms within this widely adopted framework. If Mahimahi had DualPI2, it would enable rigorous, accessible, and mobile-aware L4S experimentation across these diverse application domains.

\vspace*{0.1in}
\noindent
{\bf Contribution.} This work extends Mahimahi with a DualPI2 implementation, allowing L4S evaluations to be conducted without requiring specialized kernel deployments. The DualPI2 AQM-in-emulator thus provides the missing link for application-level L4S research, leveraging Mahimahi’s widespread adoption to bridge the gap between transport-layer L4S development and application-layer evaluation.

Concretely, with a Mahimahi DualPI2 module, researchers can evaluate L4S-enabled applications against a DQC bottleneck without relying on dedicated testbeds, which may be constrained by availability, root privileges, and operational complexity. Further, our implementation enables evaluation without the need to modify or rebuild kernels, deploy custom operating systems, or configure ECN on physical network interfaces.
Finally, the modular design of our DualPI2 implementation (\S\ref{sec:design}) enables experimentation with structural variations (e.g., alternative schedulers or queue coupling logics), allowing researchers to systematically and directly study their impact on application QoE and to share reproducible configurations alongside their findings.

\section{Design and Implementation}\label{sec:design}

\begin{figure}[ht]
  \centering
  \includegraphics[width=0.47\textwidth]{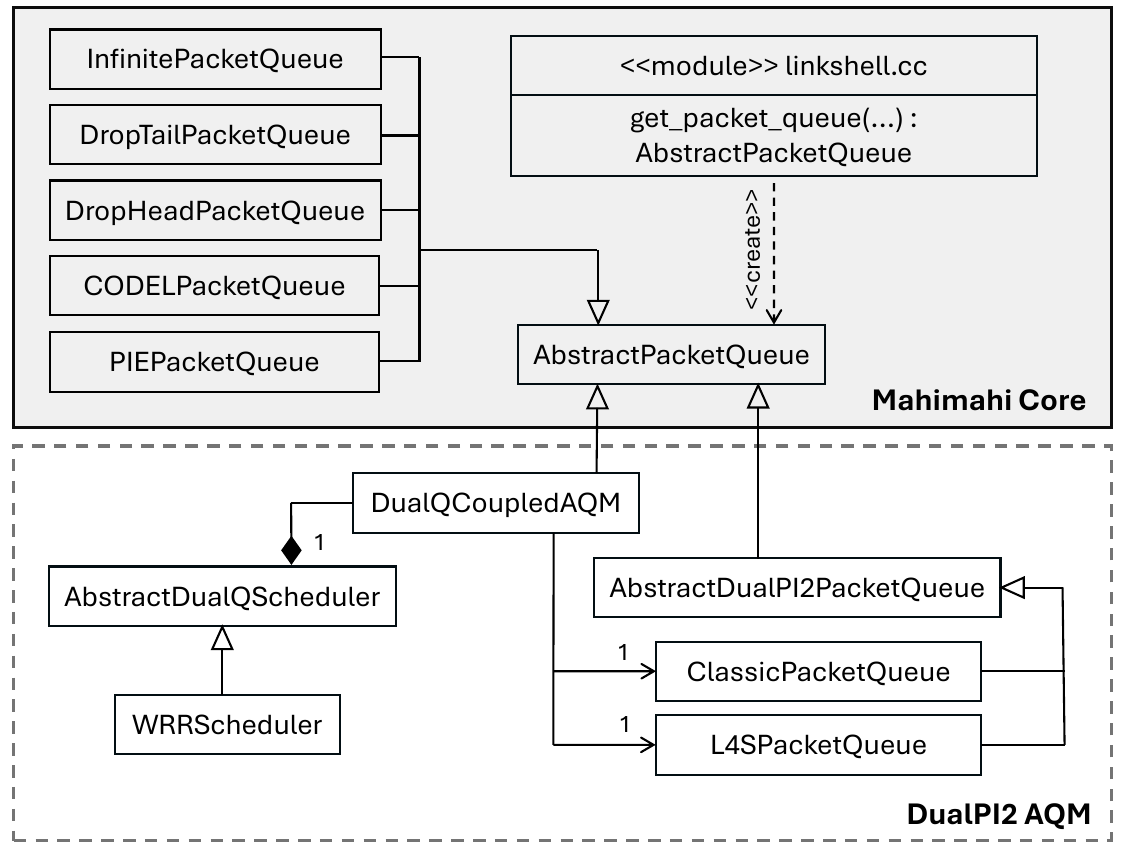}
  \caption{Simplified UML class diagram of the DualPI2 module in Mahimahi}
  \label{fig:uml-diagram}
\end{figure}

The DualPI2 AQM module in Mahimahi~\footnote{\url{https://github.com/5G-VCA-CC/mahimahi}}, depicted in Fig.~\ref{fig:uml-diagram}, consists of six additional classes. The main DualPI2 logic is contained in the \texttt{DualQCoupledAQM} class, which inherits from \texttt{AbstractPacketQue\-ue}. This class is used in the \texttt{LinkShell} module to instantiate and return the AQM choice specified by the user. 

The \texttt{DualQCoupledAQM} keeps two queues, the C queue and the L queue, which are instances of the \texttt{ClassicPacketQueue} and \texttt{L4SPacketQueue}, respectively. Both classes inherit from \texttt{Abstract\-DualPI2PacketQueue}. Although the initial intuition was to derive these two classes from \texttt{DroppingPacketQueue}, we decided to create a new class deriving directly from \texttt{AbstractPacketQueue}, because \texttt{DroppingPa\-cketQueue} requires setting the maximum queue length at the queue level. However, in the DualPI2 conceptual model, the maximum queue length is intended to be set and maintained at the AQM level rather than at the individual queue level, since the coupling mechanism of the AQM, in tandem with the scheduler, determines the effective share of bandwidth across both queues as if they formed a single pool~\cite{rfc9330}. This prompted the need for a new base class for the two queues, \texttt{AbstractDualPI2PacketQueue}, which leaves queue capacity management to the AQM class.

The general structure of the DualPI2 AQM was inspired by the pseudocode available in RFC 9332~\cite{rfc9332}. Any logic not explicitly specified in the RFC was carefully ported from the Linux kernel implementation\footnote{\url{https://github.com/L4STeam/linux/blob/testing/net/sched/sch_dualpi2.c}}, ensuring behavioral consistency while adapting it to Mahimahi’s architecture. The weighted round robin (WRR) scheduler is an example of such a module adapted from the kernel code. In our implementation, we first created the \texttt{AbstractDualQScheduler} class to provide an interface for individual scheduler classes, of which the \texttt{WRRScheduler} inherits.

Overall, we deliberately decompose the AQM into well-defined abstractions at both the scheduler and queue levels. Importantly, this modular design facilitates both the addition of new schedulers and the introduction of alternative logics for the C and L queues through their respective base classes, moving away from the monolithic design of the kernel code and explicitly promoting flexible experimentation with new algorithms.

\begin{figure}[ht]
  \centering
  \includegraphics[width=0.47\textwidth]{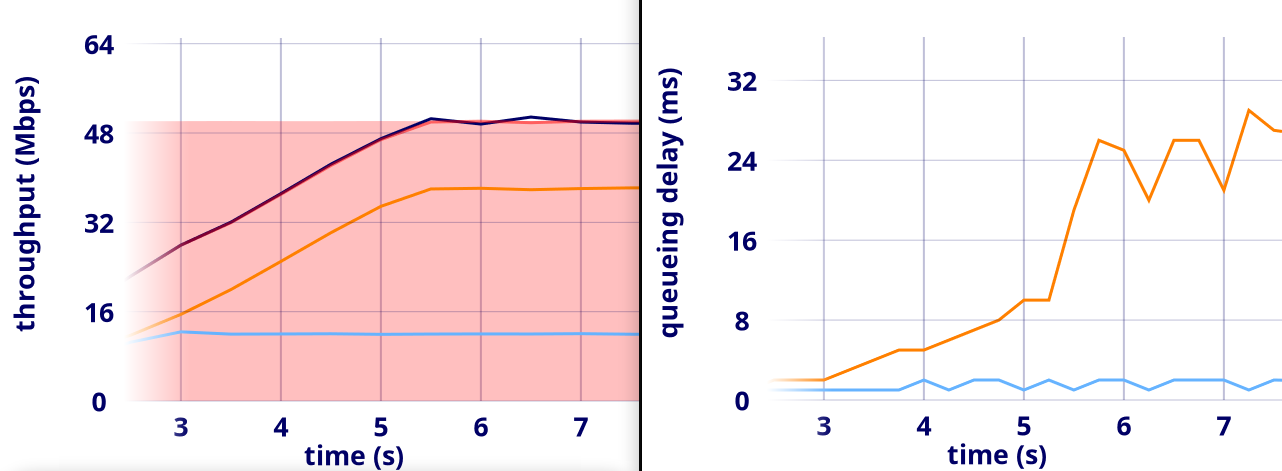}
  \caption{Distinguishing L4S traffic (blue) from classic traffic (orange) in Mahimahi live graphs; the dark blue line is the cumulative throughput.}
  \label{fig:mahimahi-dual-viz}
\end{figure}

\vspace*{0.1in}
On the visualization side, Mahimahi provides live graphs displaying available bandwidth, achieved throughput, and queue delay. To clearly distinguish between L4S and classic traffic, we extend these visualizations with additional lines that separately represent L4S traffic (in blue) and classic traffic (in orange). Fig.~\ref{fig:mahimahi-dual-viz} shows live graphs of an example dual-flow with separate lines for the L4S and classic flows.

\section{Behavioral Characterization and Comparative Analysis}\label{sec:validation}

To assess the fidelity our Mahimahi DualPI2 module with respect to the Linux kernel reference implementation, we employ a nonparametric permutation test to quantify similarities and differences in terms of four metrics and across multiple traffic and network scenarios. 
The following sections describe the metrics of interest and the distance measure, the statistical validation framework, and finally the traffic scenarios and network conditions considered.

\subsection{Metrics and Distance Measure}
\label{sec:metrics-dtw}

The metrics of interest span the throughput achieved in each traffic pattern; the internal state of the AQM, captured by the queue occupancy trace; and the AQM's outward signaling behavior, represented by ECN marks and packet drops.

\vspace*{0.1in}
\noindent
{\bf Throughput.} We use the \texttt{iperf3} utility~\cite{iperf3} as a traffic generator in our experiments and collect the average throughput value for each run directly from the final \texttt{iperf3} report. In dual-flow traffic scenarios, we use  \emph{cumulative throughput}, which is the sum of the throughput values achieved by the L4S flow and the classic flow. Since we extract a single average throughput value per run, the distance measure between two runs is simply the difference between their respective average throughput values.  

\vspace*{0.1in}
\noindent
{\bf State Similarity.} 
We define \emph{state similarity} as the degree to which queue occupancy traces are comparable between the two implementations at any given time. Inspired by the work of Ware et al.~\cite{ccanalyzer}, which demonstrates that a queue occupancy time series can be used to classify congestion control protocols based on their buffer-filling behavior (under a fixed AQM), we adopt the converse hypothesis: given identical traffic, differences in queue occupancy time series can be attributed to differences in AQM behavior and/or system-specific deviations. Therefore, if we can reject the hypothesis that the distance between queue occupancy traces produced by the Mahimahi and Linux DualPI2 implementations is beyond a certain threshold, we can conclude that cross-system deviations can reasonably be attributed to internal stochastic noise rather than structural divergence of the AQM logic. The statistical test is described in more detail in \S\ref{sec:stat-test}.

\vspace*{0.1in}
\noindent
{\bf Signal Similarity.}
Relying solely on state similarity for a comprehensive behavioral comparison is insufficient, as queue occupancy is an aggregate outcome influenced by traffic characteristics (e.g. rate and burstiness) as well as internal AQM behaviors that may obscure finer differences between implementations. Although we control for traffic characteristics, ensuring that they are identical across both the kernel and Mahimahi experiments, any differences in AQM signaling behavior need to be further investigated. As an active queue management technique, DualPI2 influences congestion response by choosing whether to drop packets or mark them with ECN CE. These actions constitute the \emph{signals} that the AQM sends through the network, influencing receiver feedback and, ultimately, sender adaptation. To evaluate whether the Mahimahi DualPI2 module produces equivalent signaling to the reference implementation, we collect and compare time series traces of ECN markings and packet drops, where each data point is the number of ECN marks or drops that occured during the last time interval.
Our goal is not to test for exact superposition or pointwise equality of traces from Mahimahi and the kernel reference, which is unrealistic in systems subject to unavoidable timing noise. Rather, we seek to assess practical differences between the two implementations. 

\vspace*{0.1in}
\noindent
{\bf Dynamic Time Warping as the Distance Measure for Time Series Data.}
The most fundamental building block of our evaluation methodology is the distance measure that compares two time series traces.
Similar to~\cite{ccanalyzer}, our approach for computing individual pairwise trace differences is based on Dynamic Time Warping (DTW)~\cite{dtw-seminal}, which takes two time series as input and returns a scalar distance that quantifies how dissimilar the two traces are. DTW is used to account for temporal dilation and contraction between traces and has been traditionally used in pattern-matching domains such as automatic speech recognition~\cite{dtw-speech}, data mining~\cite{dtw-scale, dtw-trillion}, financial time series analysis~\cite{dtw-inflation}, gesture recognition~\cite{dtw-gesture}, and bioinformatics~\cite{dtw-genome}.

\begin{figure}[b!]
  \centering
  \begin{subfigure}[t]{0.49\linewidth}
    \centering
    \includegraphics[width=\linewidth]{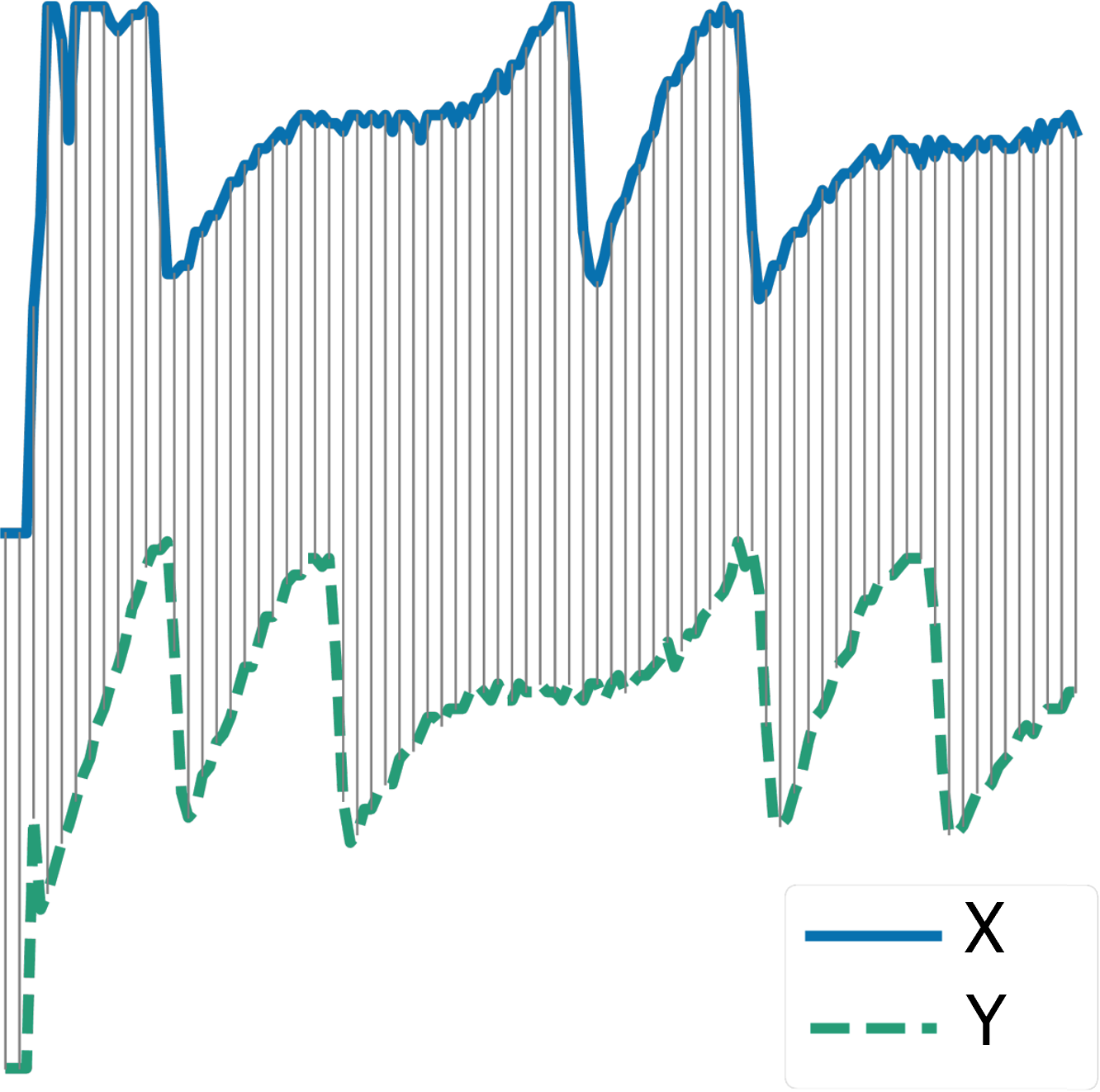}
    \caption{Euclidean distance = 2.47}
    \label{fig:eucli}
  \end{subfigure}
  \hfill
  \begin{subfigure}[t]{0.49\linewidth}
    \centering
    \includegraphics[width=\linewidth]{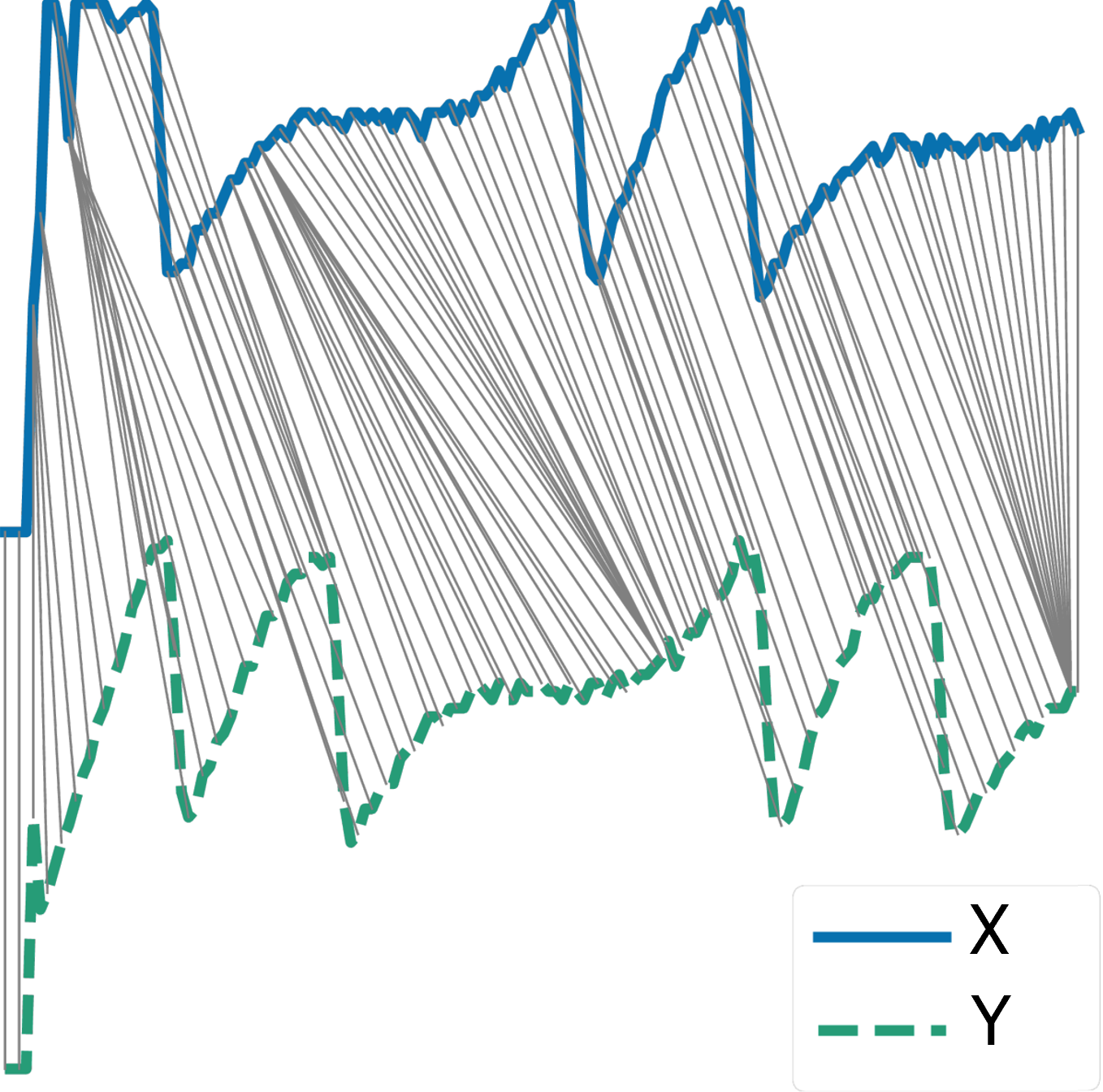}
    \caption{DTW distance = 0.76}
    \label{fig:dtw}
  \end{subfigure}

  \caption{Comparing two time series using the Euclidean distance vs. DTW distance. Adapted from~\cite{ccanalyzer}.}
  \label{fig:dtw-eucli-compare}
\end{figure}

To build intuition for DTW, we first consider a baseline approach using Euclidean Distance (ED). Let
$X=(x_1, \dots,x_n)$ and $Y=(y_1, \dots,y_m)$ be two time series of length $n$ and $m$, where $x_i$ and $y_i$ represent the value of the considered metric at time step $i$ in traces $X$ and $Y$, respectively. ED is computed as the sum of the squared differences between corresponding elements in the two traces. However, this one-to-one comparison often fails for network traces. Fig.~\ref{fig:dtw-eucli-compare} illustrates such a case: two queue occupancy traces exhibit similar patterns but with temporal misalignments. 

Fig.~\ref{fig:eucli} shows the inflated ED due to these misalignments, while Fig.~\ref{fig:dtw} shows the DTW distance, which captures structural similarity even when traces are temporally misaligned, yielding a smaller and more meaningful distance than the ED. Network traces may experience temporal dilation or contraction due to dynamic conditions like congestion, rate adaptation and system limitations. In our case, distortions may be caused by differences in scheduling behavior and timing granularity between Mahimahi and the Linux kernel. These effects cause otherwise similar traces to appear stretched or compressed in time. 

Unlike ED, which requires a strict point-to-point comparison, DTW allows one-to-many alignments between trace points. At a high level, DTW computes an optimal alignment (or \emph{warping path}) between the two sequences that minimizes a total cost. This cost is defined as the sum of absolute differences between aligned sample values along the computed alignment path.

Furthermore, trace collection intervals are not always exactly aligned to millisecond boundaries, which means that different flow traces may contain different numbers of sampled points. Without correction, DTW distances would be artificially inflated for longer traces simply because more points must be matched, and therefore more pointwise errors are accumulated along the warping path.

To account for this effect, we normalize the raw DTW distance by the length of the
optimal warping path $\pi$ (with $|\pi|$ denoting the number of alignment
steps). The normalized DTW distance, defined in \eqref{eq:dtw-norm}, yields a length-invariant measure of per-alignment-point deviation, ensuring that DTW distances are comparable across traces of differing durations and sampling densities.

\begin{equation}
\label{eq:dtw-norm}
\mathrm{DTW}_{\mathrm{norm}}(X, Y)
=
\frac{1}{|\pi|}
\sum_{(i,j)\in \pi}
\left| x_i - y_j \right|
\end{equation}

\subsection{Statistical Equivalence Testing}
\label{sec:stat-test}
To evaluate whether Mahimahi behaves similarly to the Linux kernel in terms of our four metrics (throughput, queue occupancy, ECN marks and packet drops), we employ a \emph{nonparametric exceedance-based validation test}. This test assesses whether cross-system differences are extreme relative to natural within-system variability. Specifically, we evaluate how frequently kernel–Mahimahi differences exceed a predefined tolerance threshold. The approach makes no assumptions regarding equality of variances or the underlying distributional shape of the populations. The only assumption is independence between experimental runs. We enforce this by fully resetting the experimental environment between runs, including termination of \texttt{iperf} processes, teardown of each environment, and a fixed quiescence period prior to the next trial to prevent residual state carryover. The sections below outline the key components of our testing methodology, which is publicly available.\footnote{\url{https://github.com/5G-VCA-CC/mahimahi_validation.git}}

\vspace*{0.1in}
\noindent \textbf{Defining the Distance Measure.} Let $\mathcal{M} = \{M_1, \dots, M_{100}\}$ and 
$\mathcal{K} = \{K_1, \dots, K_{100}\}$ denote the sets of traces 
collected from Mahimahi and the Linux kernel, respectively.
\newpage

\noindent $\bullet\ $ \textbf{Time-series metrics (Queue occupancy, ECN marks, packet drops).}
For temporal observations, we compute the normalized DTW distance between two traces $X$ and $Y$:

\vspace*{-0.01in}
\begin{equation}
\label{eq:distance-timeseries}
d(X, Y) = \mathrm{DTW}_{\mathrm{norm}}(X, Y)
\end{equation}
\vspace*{0.005in}

\noindent Equation (\ref{eq:dtw-norm}) defines the normalized DTW distance between the two traces $X$ and $Y$. We compute all inter-system pairwise distances for each group, as well as all cross-system distances between traces in $\mathcal{M}$ and $\mathcal{K}$. These collections form the empirical distance distributions used in our validation procedure.

\vspace*{0.08in}
\noindent $\bullet\ $ \textbf{Scalar metrics (average throughput).} For non–time-series metrics, the distance measure we use is the 
absolute difference between two scalar measurements $X$ and $Y$:

\vspace*{-0.01in}
\begin{equation}
\label{eq:distance-scalar}
d(X, Y) =  |X - Y|
\end{equation}
\vspace*{0.005in}

\noindent \textbf{Defining the Test Statistic and Exceedance-Based Validation.} Having constructed the inter-system and cross-system distance distributions, we assess whether Mahimahi behaves similarly to the Linux kernel by comparing cross-system discrepancies against natural inter-system variability. Let $\varepsilon_{\max}$ denote 
a tolerance threshold defined as the maximum of the 95th percentiles of the two inter-system distance distributions. 

\vspace{0.2em}
\begin{equation}
\label{eq:eps-max}
\varepsilon_{\max}
=
\max\!\left(
\mathrm{Quantile}_{0.95}(D_{\text{within}}^{\mathcal{M}}),
\;
\mathrm{Quantile}_{0.95}(D_{\text{within}}^{\mathcal{K}})
\right)
\end{equation}
\vspace{0.2em}

\noindent Where $D^{M}_{\text{within}}=\{ d(X_i^M, X_j^M) : i < j \}$ denotes the set of all pairwise distances within the Mahimahi group, and $D^{K}_{\text{within}}=\{ d(Y_i^K, Y_j^K): i < j \}$ denotes the set of all pairwise distances within the kernel group. $d(X, Y)$ is the distance measure defined in Equation (\ref{eq:distance-timeseries}) for time series inputs, and in Equation (\ref{eq:distance-scalar}) for scalar inputs.

Consequently, a cross-system difference is considered excessive only if it exceeds the greatest natural fluctuation exhibited by either system. This avoids unfairly evaluating cross-system discrepancies against the tighter variability of the less noisy implementation and ensures that only genuinely extreme deviations are flagged. 

We then estimate our test statistic, which is the true cross-system exceedance probability, denoted as $\widehat{p}_{\max}$, and defined as the probability that a randomly selected cross-system 
distance exceeds $\varepsilon_{\max}$.

\vspace{0.2em}
\begin{equation}
\label{eq:p-hat-max}
\widehat{p}_{\max}
=
\frac{1}{|D_{\text{cross}}|}
\sum_{d \in D_{\text{cross}}}
\mathbf{1}\{ d > \varepsilon_{\max} \}
\end{equation}
\vspace{0.2em}

\noindent $D_{\text{cross}}$ denotes the set of all cross-system distances and is defined as $D_{\text{cross}} = \{ d(X, Y) \}$, where $d(X, Y)$ is the distance measure defined in Equation (\ref{eq:distance-timeseries}) for time series inputs, and in Equation (\ref{eq:distance-scalar}) for scalar inputs. The $\widehat{p}_{\max}$ quantity is estimated 
empirically as the proportion of observed cross-system distances that satisfy $d > \varepsilon_{\max}$. The resulting estimator quantifies how often Mahimahi--kernel discrepancies are more extreme than variability observed internally within either system.

\vspace*{0.1in}
\noindent \textbf{Assessing statistical equivalence.} We interpret $\widehat{p}_{\max}$ under a practical behavioral 
similarity framework. Specifically, we define the null hypothesis $H_0$ as follows: 
\[
\begin{aligned}
H_0: & \text{ Mahimahi does not behave similarly to the Linux} \\[-0.7ex]
     & \text{ kernel under a 5\% exceedance criterion}
\end{aligned}
\]
\vspace*{-0.05in}
\[
\begin{aligned}
H_1: & \text{ Mahimahi behaves similarly to the Linux kernel} \\[-0.7ex]
     & \text{ under a 5\% exceedance criterion}
\end{aligned}
\]

\noindent Operationally, this corresponds to:\[
H_0:\ p_{\max} \ge 0.05
\quad\text{and}\quad
H_1:\ p_{\max} < 0.05
\]

\noindent Here, $p_{\max}$ denotes the true (population-level) cross-system
exceedance probability, whereas $\widehat{p}_{\max}$ is its empirical
estimate computed from the observed data.

\vspace*{0.05in}
If $\widehat{p}_{\max} \ge 0.05$, cross-system differences exceed natural inter-system variability beyond the acceptable threshold of 5\%, the cross-system distribution is significantly misaligned with the inter-system distribution, and we fail to reject $H_0$. In this case, the two implementations cannot be deemed behaviorally similar under this criterion. Conversely, if cross-system differences are within the acceptable threshold of variability of 5\%, the kernel DualPI2 AQM dynamics are preserved in the Mahimahi implementation, which rejects $H_0$ and concludes practical 
behavioral similarity when $\widehat{p}_{\max} < 0.05$. 

\vspace*{0.05in}
The statistical significance threshold is somewhat arbitrary and depends largely on the field of application. The value of 5\% was originally proposed by Fisher in his textbook ``Statistical Methods for Research Workers''~\cite{fisher1925}. Although some fields such as genomics and high-energy physics use far stricter thresholds, a value of 5\% is still considered a conventional significance threshold and is used in many experimental disciplines~\cite{redefine_signif}.

\subsection{Traffic Scenarios and Network Conditions}
\label{sec:traffic-netcond}

We collected data from identical experiments on both Mahimahi and Linux DualPI2 using the \texttt{iperf3}~\cite{iperf3} bandwidth probing tool, following three different traffic patterns:

\begin{itemize}
  \item[--] \textbf{Single L4S flow}, using TCP Prague congestion controller to isolate the behavior of the L queue.
  \item[--] \textbf{Single classic flow}, using a TCP Cubic flow to isolate the behavior of the C queue.
  \item[--] \textbf{Dual flow}, one TCP Prague and one TCP Cubic, to validate the coupling mechanism between the L queue and the C queue.
\end{itemize}

We explore all combinations of the above three flow types with three representative Bandwidth–Delay Product (BDP) classes summarized in Table~\ref{tab:exp-configs}, resulting in nine distinct experimental scenarios.

\begin{table}[ht]
\centering
\caption{Representative BDP classes used in our evaluation.}
\renewcommand{\arraystretch}{1.15}
\setlength{\tabcolsep}{4pt}
\begin{tabular}{|l|c|c|p{3.2cm}|}

\hline
\textbf{Regime} & \textbf{RTT} & \textbf{Bandwidth} & \textbf{Example Networks} \\
\hline\hline
Low-BDP  & 20 ms       & 12 Mbps  & Legacy DSL \\
\hline
Medium-BDP  & 40 ms   & 50 Mbps  & 4G LTE/5G; Starlink \\
\hline
High-BDP & 100 ms & 200 Mbps & 
Inter-continental fiber\\
\hline
\end{tabular}

\label{tab:exp-configs}
\end{table}

\noindent We collect 100 \texttt{iperf3} flow traces from Mahimahi DualPI2 and 100 from the Linux kernel DualPI2 for each of the BDP regimes, for each traffic pattern. We provide a justification for the sample size in Appendix~\ref{app:sample-size}. Each \texttt{iperf3} run is 30 seconds long. For every test scenario, we collect average per-run throughput and extract time series samples of queue occupancy, ECN marks, and packet drops every \texttt{tupdate} (default 16 ms~\cite{rfc9332}) for both implementations. We chose \texttt{tupdate} as our sampling rate because it is the frequency at which the key probabilities of the DualPI2 AQM are updated, effectively representing the ``internal clock'' of DualPI2. Our data collection scripts are publicly available.\footnote{\url{https://github.com/5G-VCA-CC/Experimentation-Data-Scripts/tree/ccr26-submission}}

\section{Validation Setup}
\label{sec:validation-setup}
This section outlines the measures taken to ensure that the validation of the Mahimahi AQM against its kernel counterpart is conducted fairly, and describes the experimental settings used to enable controlled comparisons of the AQM dynamics. 

We installed a precompiled kernel\footnote{kernel version \texttt{5.15.72-48b3db6b4-prague-111} available at \url{https://github.com/L4STeam/linux?tab=readme-ov-file}. More recent kernel versions may have been released since the completion of this work.} that contains the DualPI2 AQM and  TCP Prague modules. TCP Prague, with AccECN~\cite{rfc9768} enabled, is used for both the Mahimahi and the kernel experiments as the congestion control protocol of the L4S-enabled \texttt{iperf3} flow. The DualPI2 kernel module is the reference against which we validate our Mahimahi DualPI2 AQM. We set the same, default values for the DualPI2 parameters~\cite{rfc9332} for both implementations, namely $target=$ 15 ms,  $step\_thresh=$ 1 ms, $limit= max\_link\_rate \:\ast $ 250 ms, $tupdate=$ 16 ms, $alpha=$ 0.16, and $beta=$ 3.20. 

\vspace*{0.1in}
\noindent
{\bf Mahimahi–kernel Matching Setup.}
To realize the various configurations discussed in \S\ref{sec:traffic-netcond} on Mahimahi, we use two of Mahimahi's shells~\cite{mahimahi-ccr}: LinkShell and DelayShell. LinkShell emulates link dynamics using constant-rate packet delivery traces, while DelayShell enables the introduction of a fixed per-packet delay. For the Linux kernel runs, we use two network namespaces, sender and receiver, connected by a virtual link consisting of a veth pair (one veth interface in each namespace). At the sender egress we layer a Hierarchical Token Bucket (HTB) over the DualPI2 AQM to enforce the rate limit. To control end-to-end RTT in the kernel experiments, we emulate propagation delay using the Linux traffic control (\texttt{tc}) \texttt{NetEm}~\cite{netem} queuing discipline by applying a one-way delay on each of the sender and receiver namespaces. 

For rigorous validation, we augment the kernel DualPI2 implementation with additional counters exposed via the struct named \texttt{tc\_dualpi2\_xstats}, enabling direct observation of marking and drop behavior using \texttt{tc} tooling, which was also modified for this purpose. These changes reside in our Linux fork\footnote{https://github.com/gargAneesh/linux/tree/add-counters} (which is derived from the DualPI2 testing branch). The parsing changes to \texttt{tc} are in our \texttt{iproute2} fork.\footnote{https://github.com/gargAneesh/iproute2/tree/add\_counters} This setup provides a reproducible, Mahimahi-aligned bottleneck with a strict rate cap enforced by HTB, queuing/marking governed by DualPI2, and a configurable RTT introduced by \texttt{NetEm}. We also provide scripts to automate kernel module recompilation and loading, as well as namespace reinstallation.\footnote{https://github.com/gargAneesh/l4s\_scripts}

\section{Evaluation Results \& Discussion}
\label{sec:eval}
This section presents a comparative behavioral evaluation of Mahi\-mahi DualPI2 and the Linux kernel implementation using our statistical testing framework. We begin with dual-flow experiments under the default DualPI2 configuration to establish a baseline comparison. Guided by the observed discrepancies, we then investigate the sensitivity of key parameters in single-flow experiments within Mahimahi. Finally, we evaluate dual-flow performance under refined parameter settings informed by this analysis.

\begin{figure*}[ht]
\centering

\begin{minipage}[c]{0.95\textwidth}
    \makebox[0.24\linewidth]{Cumulative Throughput} \hfill
    \makebox[0.24\linewidth]{Queue Occupancy} \hfill
    \makebox[0.24\linewidth]{ECN Marks} \hfill
    \makebox[0.24\linewidth]{Packet Drops}
\end{minipage}
\begin{minipage}[c]{0.04\textwidth}
\end{minipage}

\vspace{0.3em}

\begin{minipage}[c]{0.95\textwidth}
    \includegraphics[width=0.25\linewidth]{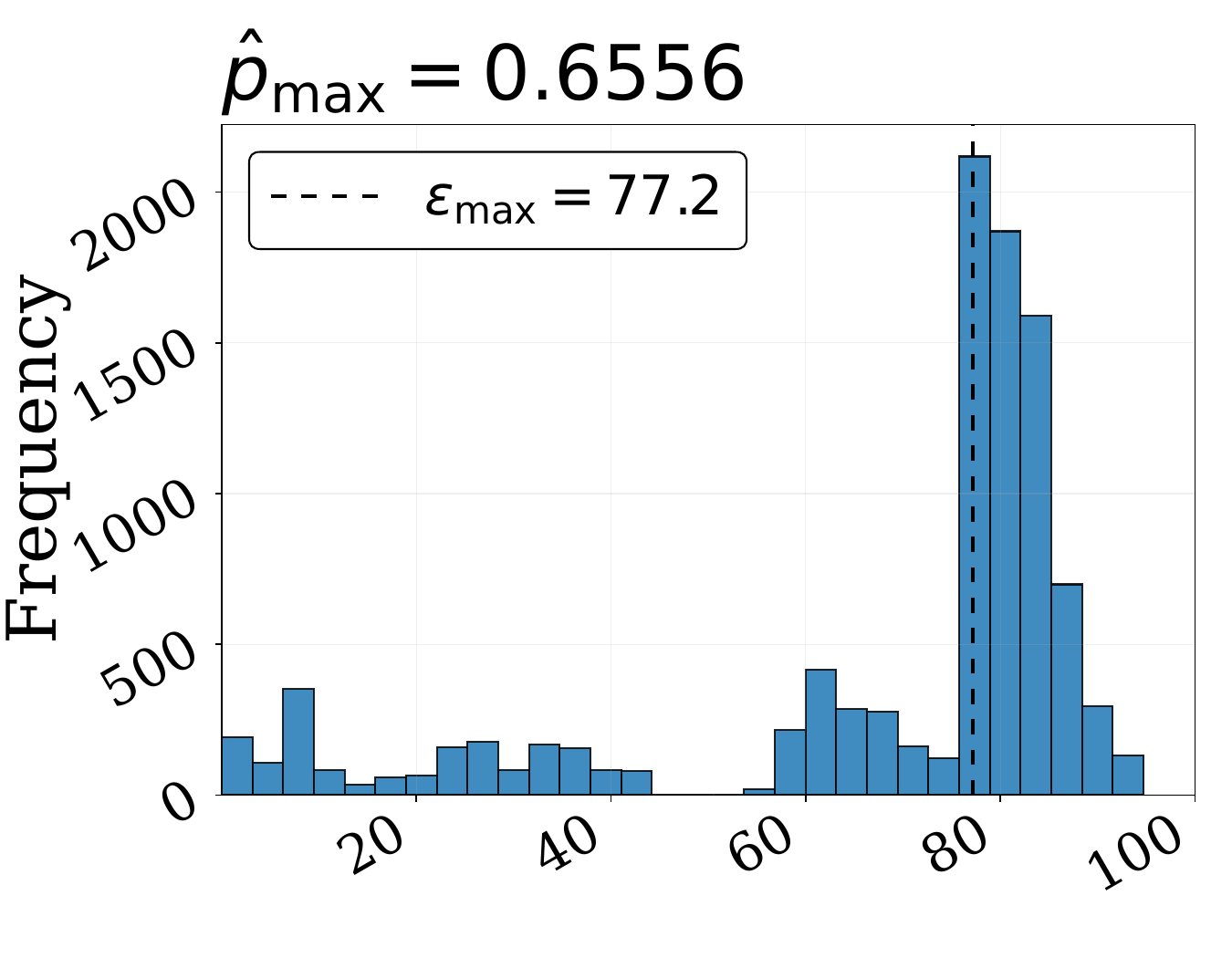}\hfill
    \includegraphics[width=0.25\linewidth]{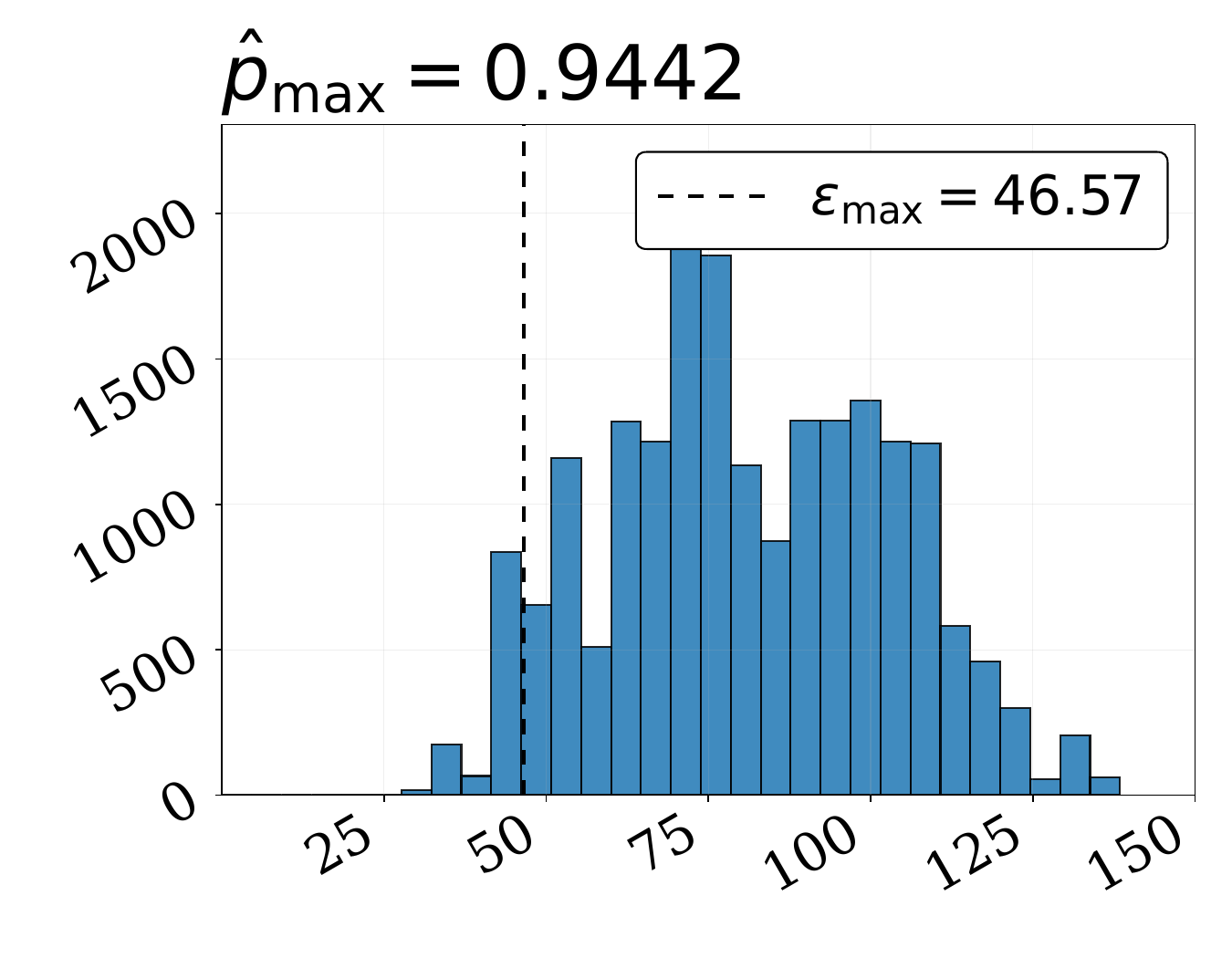}\hfill
    \includegraphics[width=0.25\linewidth]{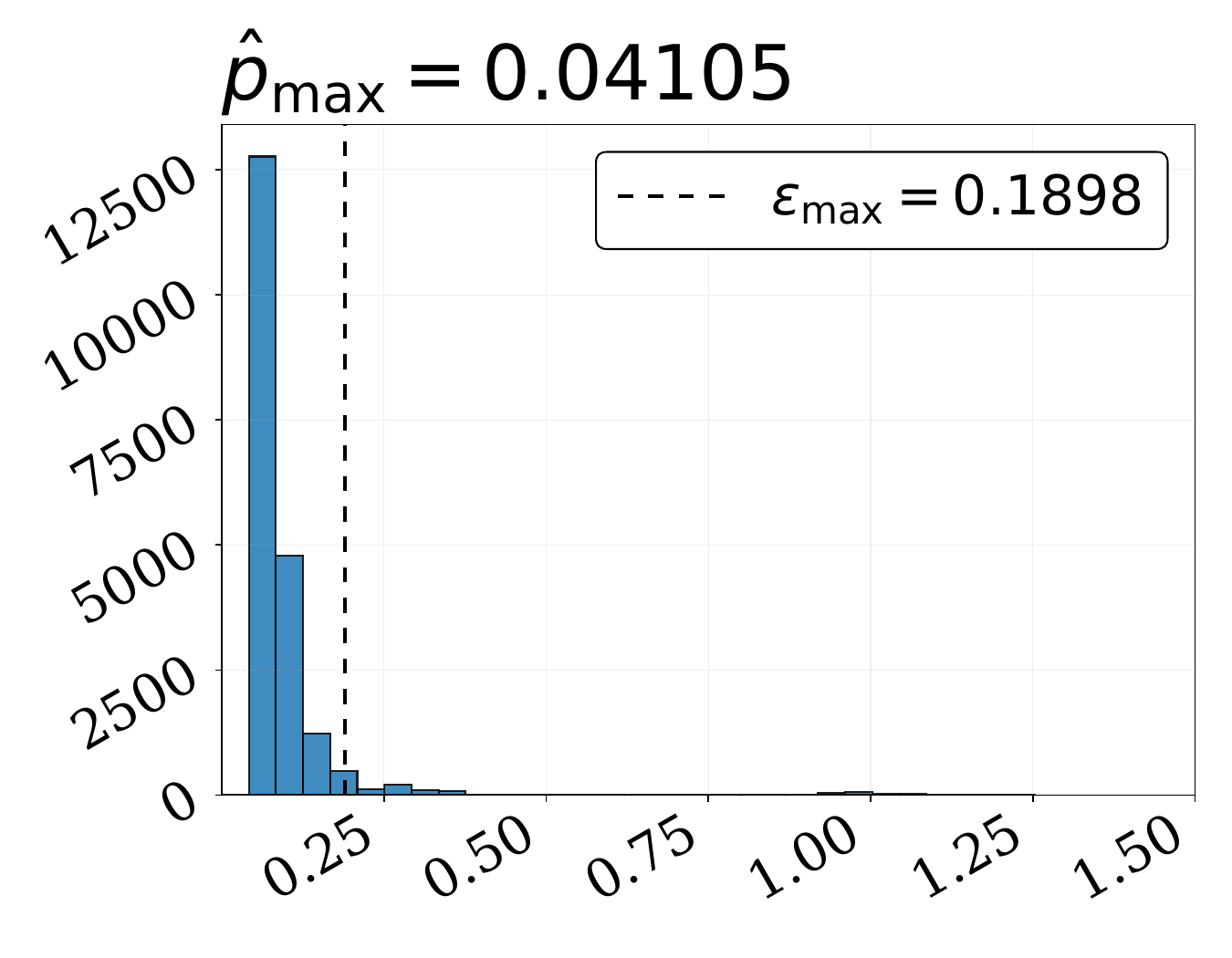}\hfill
    \includegraphics[width=0.25\linewidth]{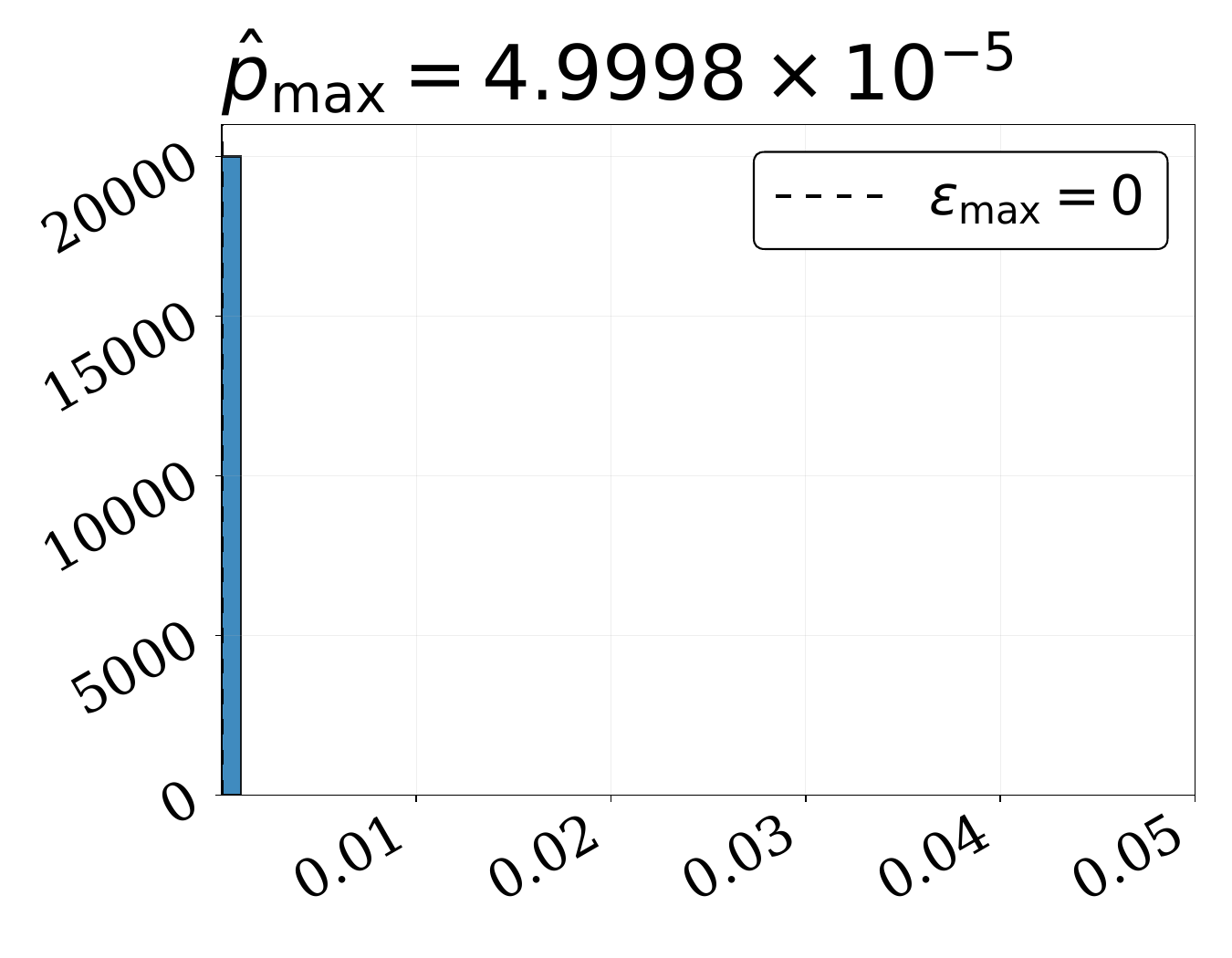}
\end{minipage}
\begin{minipage}[c]{0.04\textwidth}
    \centering\rotatebox{90}{High BDP}
\end{minipage}


\begin{minipage}[c]{0.95\textwidth}
    \includegraphics[width=0.25\linewidth]{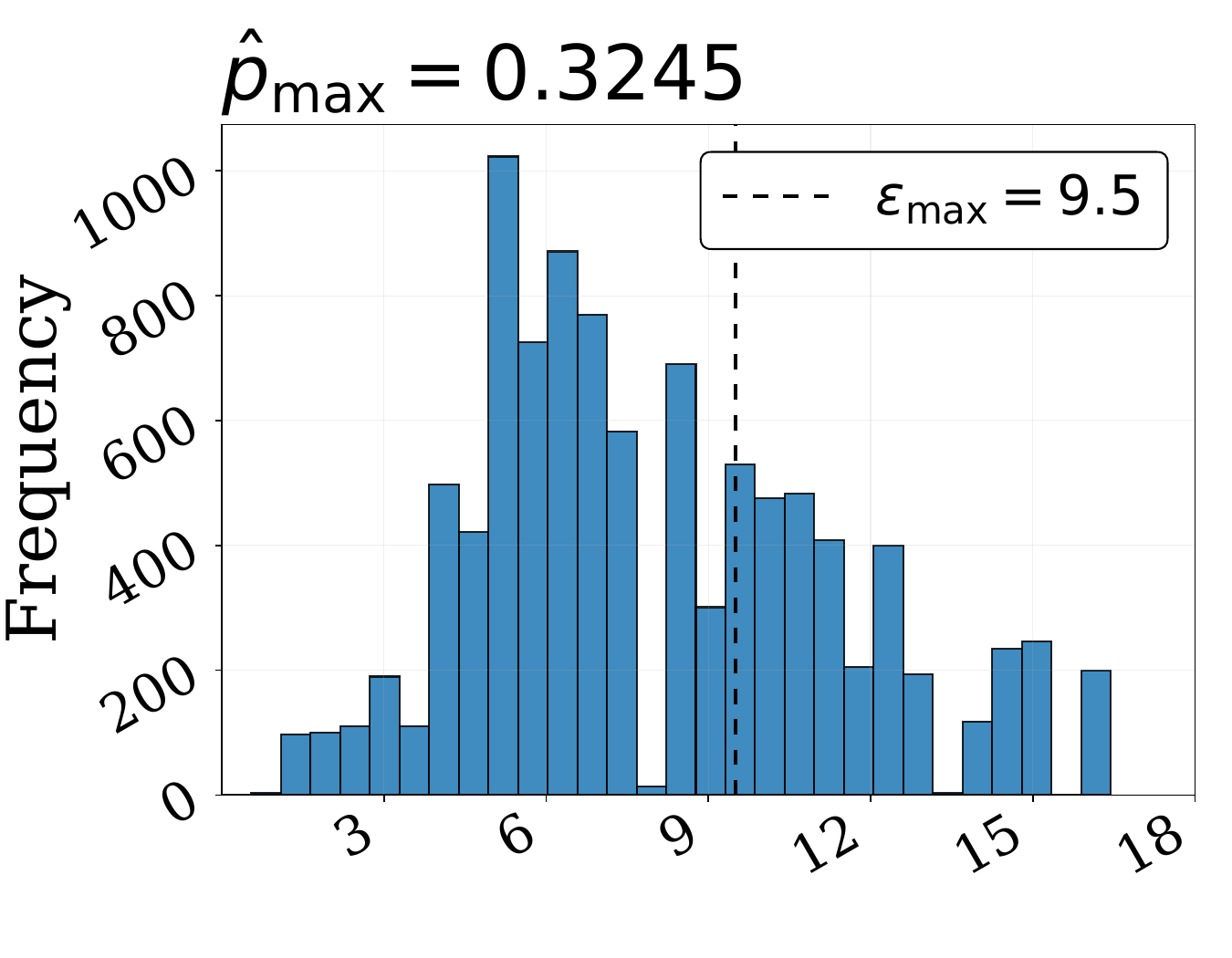}\hfill
    \includegraphics[width=0.25\linewidth]{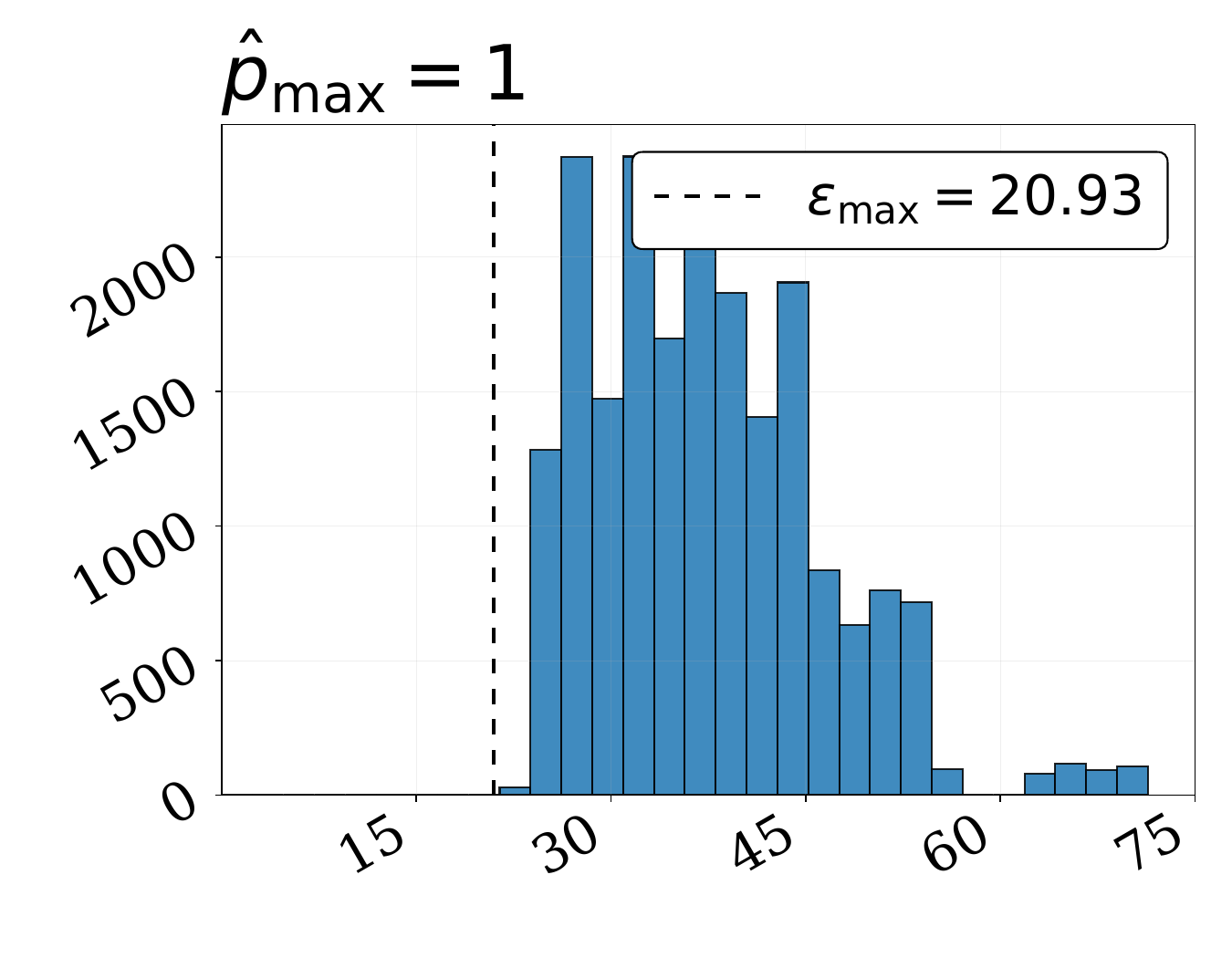}\hfill
    \includegraphics[width=0.25\linewidth]{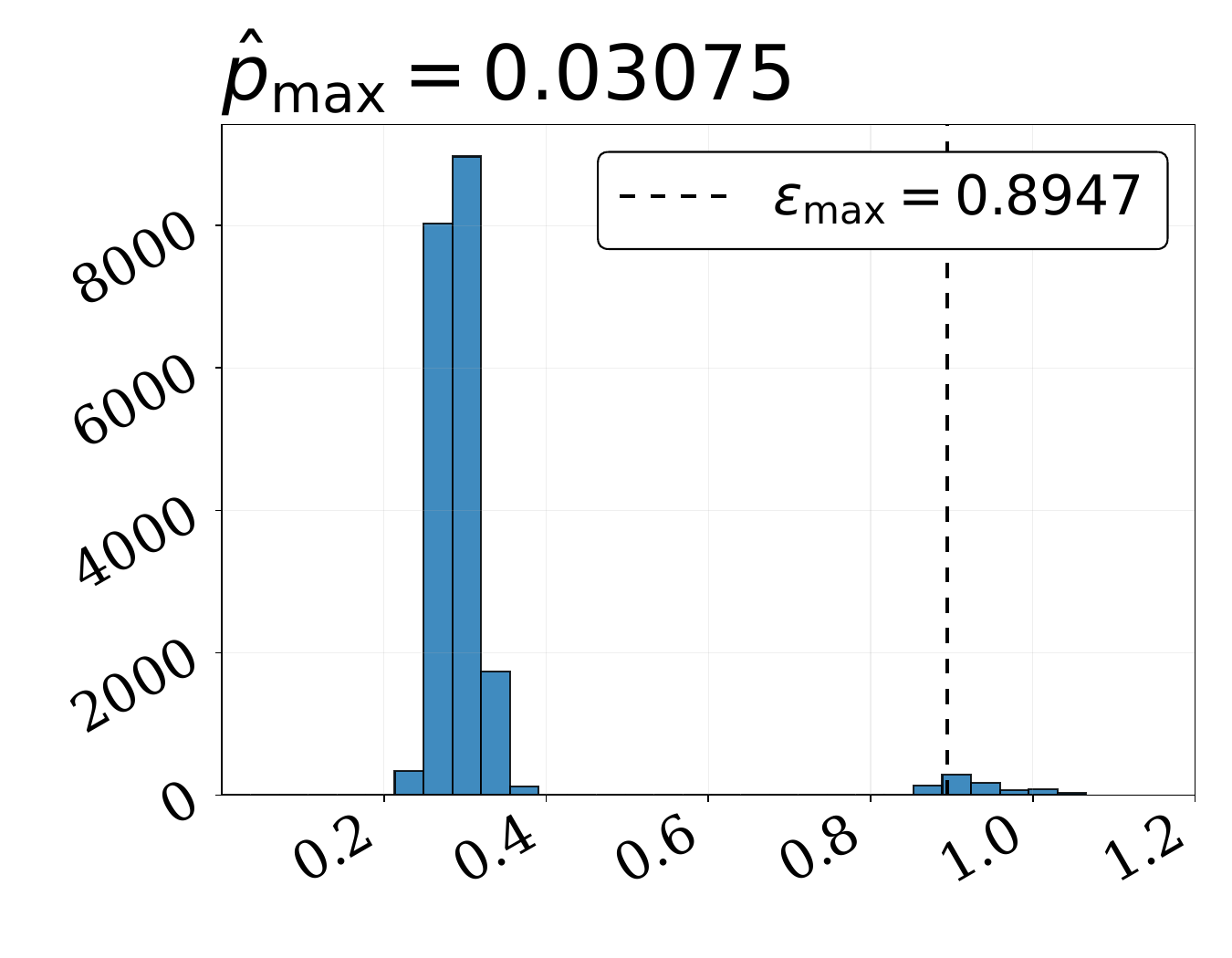}\hfill
    \includegraphics[width=0.25\linewidth]{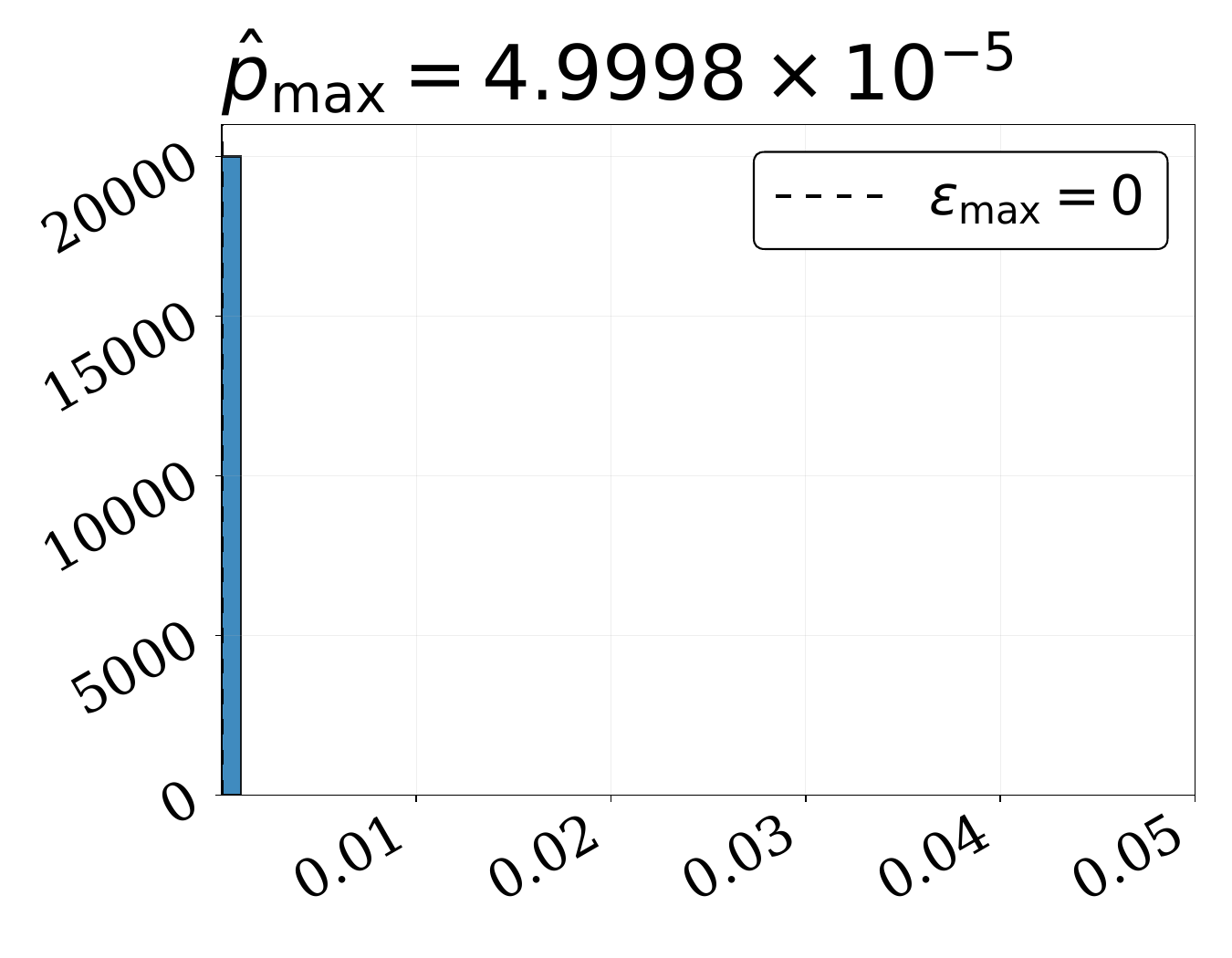}
\end{minipage}
\begin{minipage}[c]{0.04\textwidth}
    \centering\rotatebox{90}{Medium BDP}
\end{minipage}


\begin{minipage}[c]{0.95\textwidth}
    \includegraphics[width=0.25\linewidth]{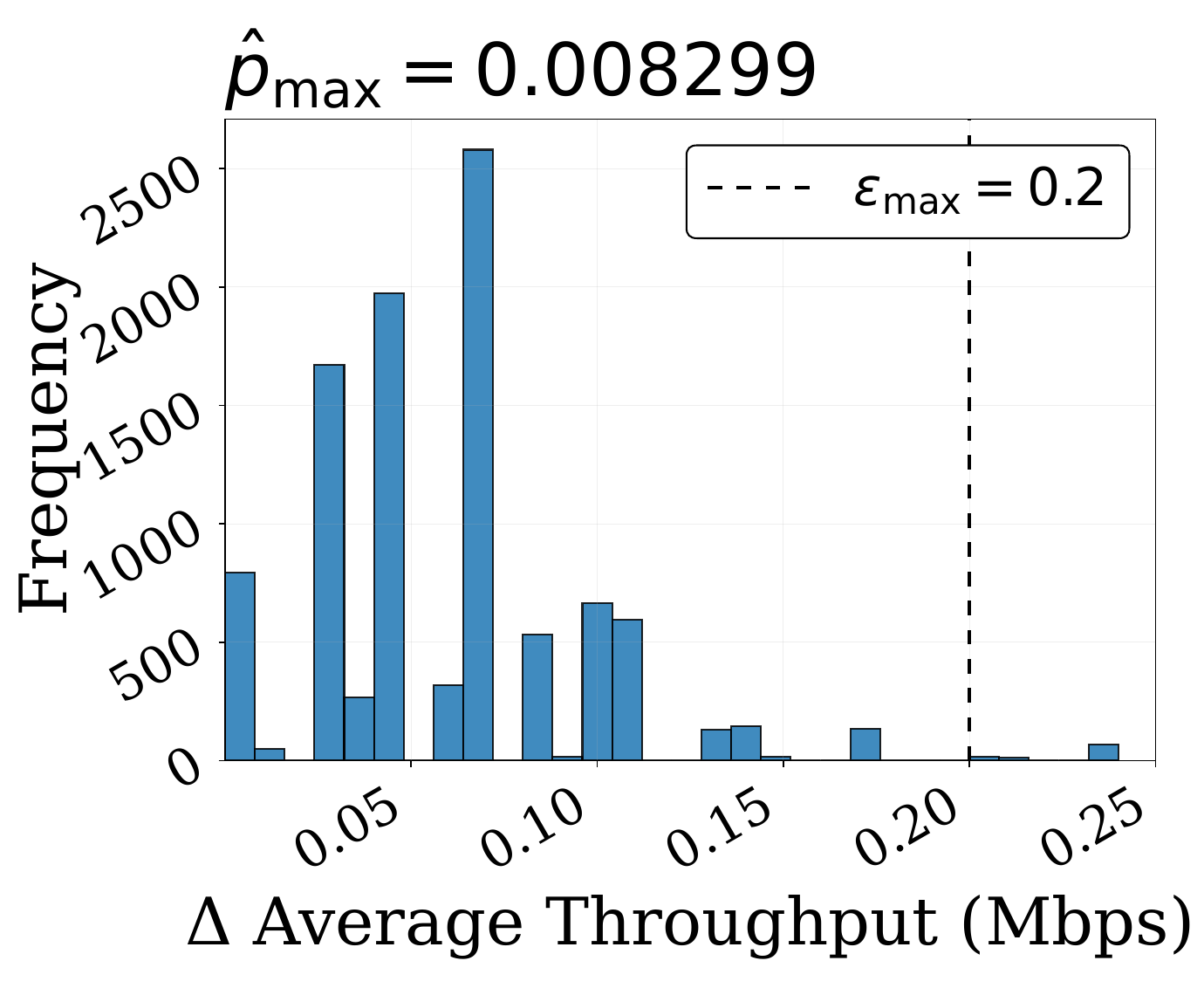}\hfill
    \includegraphics[width=0.25\linewidth]{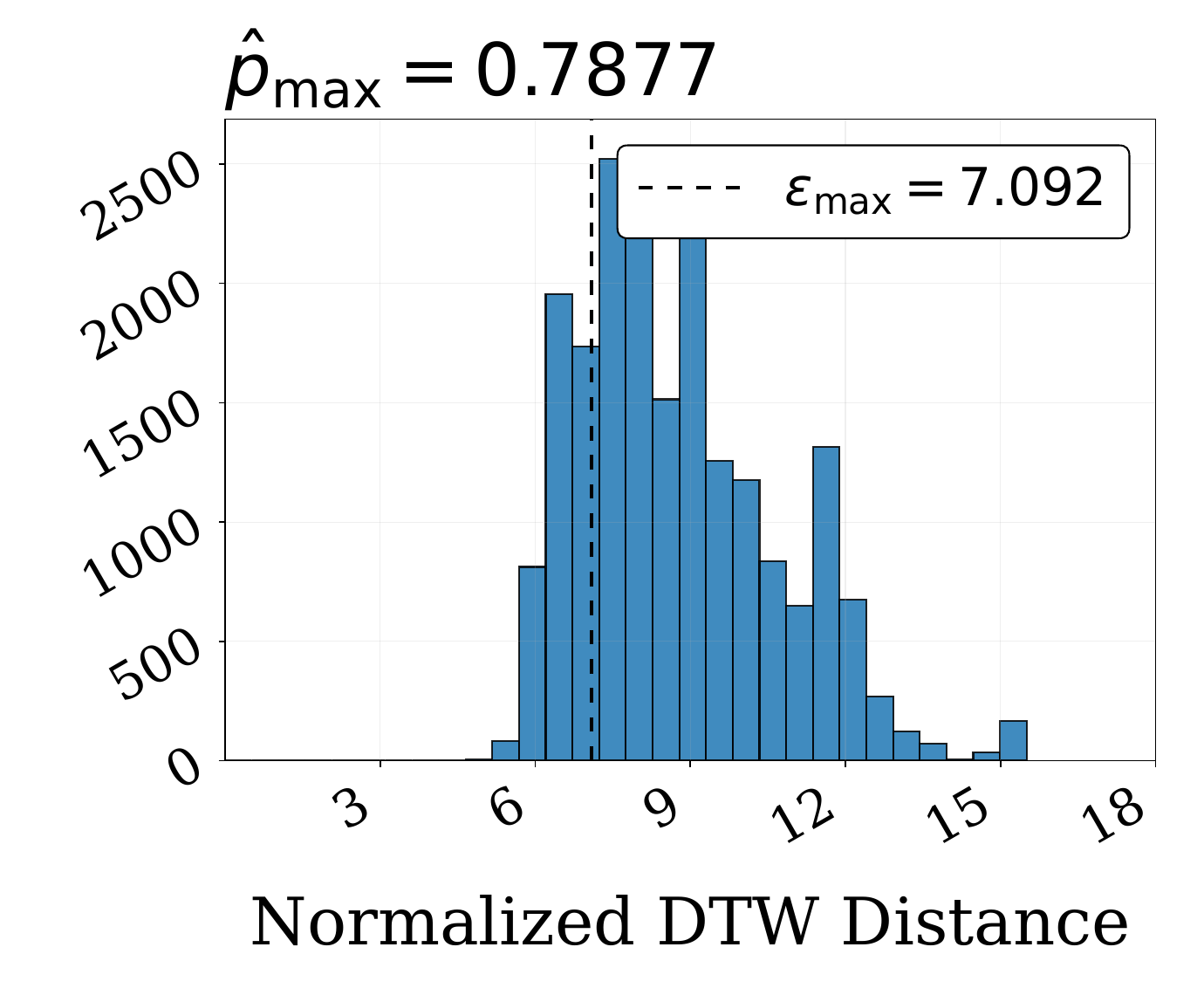}\hfill
    \includegraphics[width=0.25\linewidth]{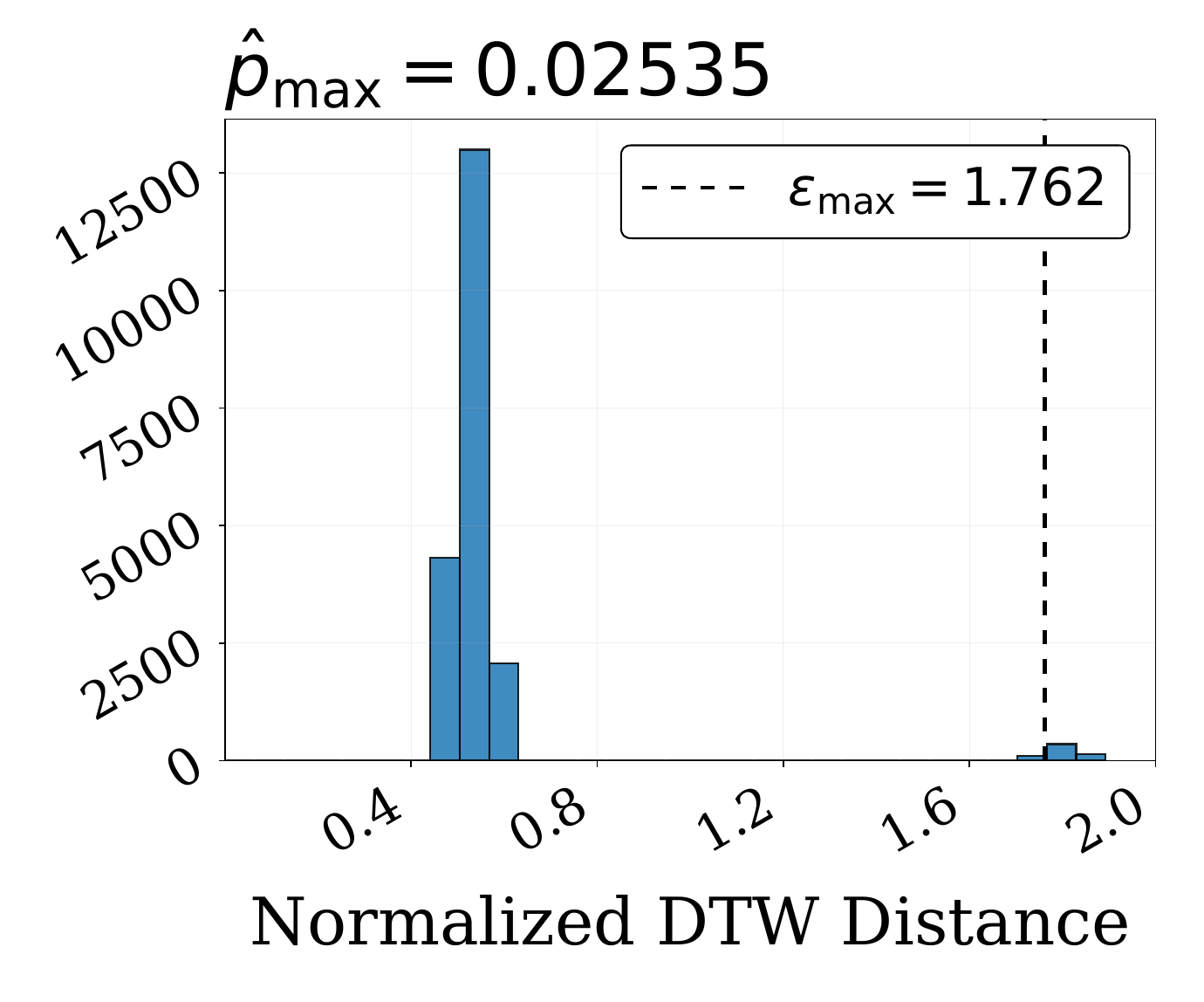}\hfill
    \includegraphics[width=0.25\linewidth]{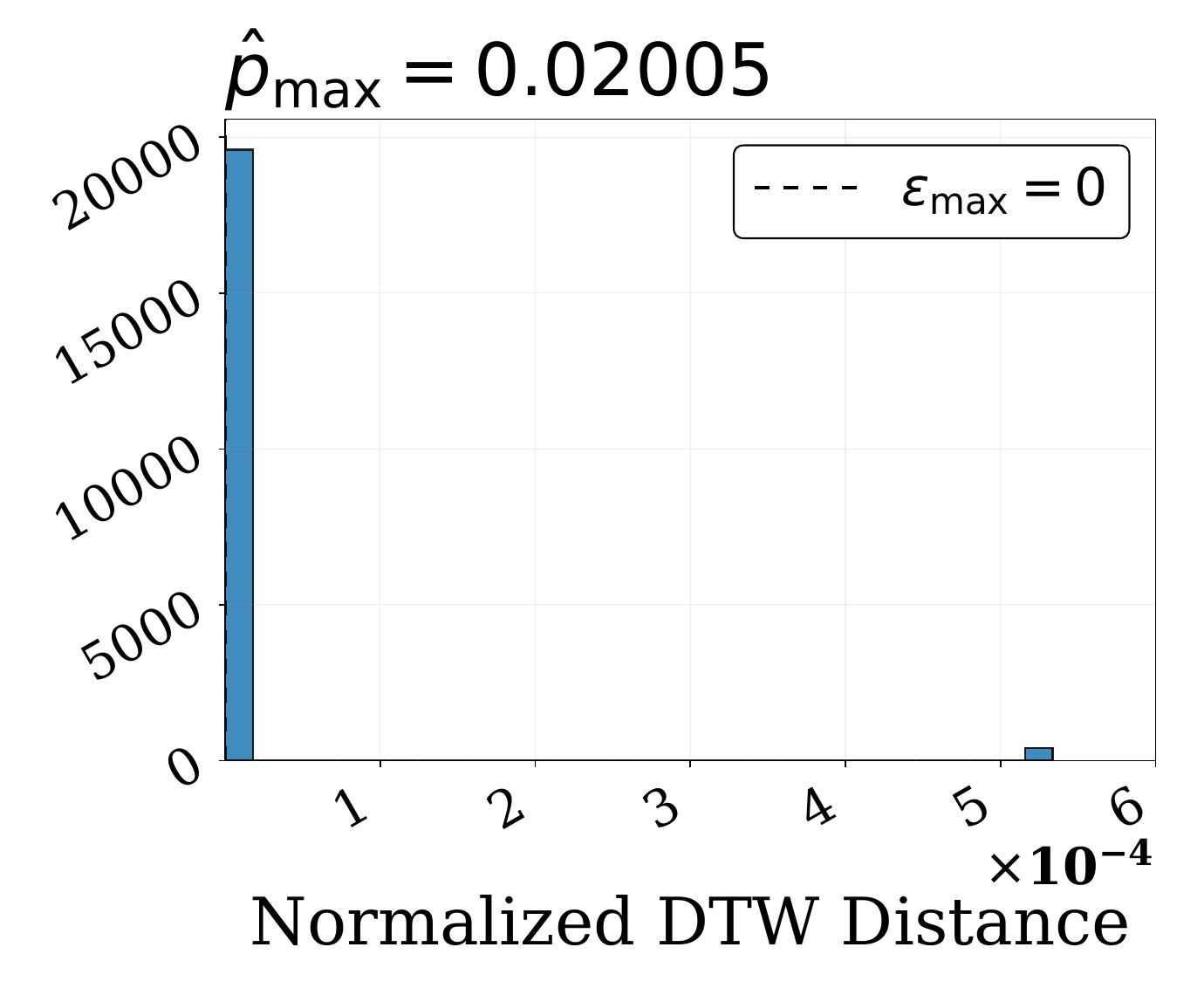}
\end{minipage}
\begin{minipage}[c]{0.04\textwidth}
    \centering\rotatebox{90}{Low BDP}
\end{minipage}

\caption{Validation of the TCP Prague $\times$ TCP Cubic runs in High, Medium, and Low BDP scenarios - Default DualPI2 parameters.}
\label{fig:multi_bdp_grid_original}
\end{figure*}

\vspace*{0.1in}
\noindent \textbf{Dual-flow experiments with default DualPI2 parameters.} In the first part of our evaluation, we collect traces from 100 runs for the dual flows traffic pattern (TCP Prague $\times$ TCP Cubic) in all three BDP regimes on each of the Mahimahi module and the Linux kernel qdisc. In this set of experiments, we use the default DualPI2 parameters described in \S\ref{sec:validation-setup}. The historgrams in Fig.~\ref{fig:multi_bdp_grid_original}  show the empirical distribution of each of our test statistics in all three BDP regimes. Each distribution is accompanied by the corresponding $\varepsilon_{max}$ and $\widehat{p}_{\max}$. The figure shows that the queue occupancy test statistic fails to reject the null hypothesis $H_0$ across all BDPs, indicating a considerable mismatch in the instantaneous state of DualPI2 between the Mahimahi and the kernel setting. This result is due to system differences between the kernel environment and the emulated Mahimahi environment. In particular, Mahimahi's network traces encode the bandwidth as a succession of delivery opportunities at every millisecond, effectively implementing a token bucket that refills at 1 ms intervals, creating small traffic bursts, while packet delivery in the kernel execution is more smooth and evenly-spaced. We explore this observation in more detail in Appendix~\ref{app:cadence}. Meanwhile, two test statistics reject the null hypothesis $H_0$ in all BDPs within the acceptable threshold of 5\%: packet drops and ECN marks. This indicates practical 
behavioral similarity with regard to these signal metrics. Looking at the traces more closely, we can see that the packet drops are close to zero across all experiments. This is expected even for classic traffic, since TCP Cubic supports classic ECN, which means the congestion signal is reinterpreted from packet drops to ECN marks. Considering the throughput test statistic, the test only rejects the null hypothesis for the low BDP regime (12 Mbps, RTT = 20 ms), indicating the achieved throughput in Mahimahi is very different from that of the kernel setting in both the medium and high BDP regimes.

\begin{figure}[b!]
\centering
    \begin{subfigure}[b]{0.32\columnwidth}
        \centering
        \includegraphics[width=\linewidth]{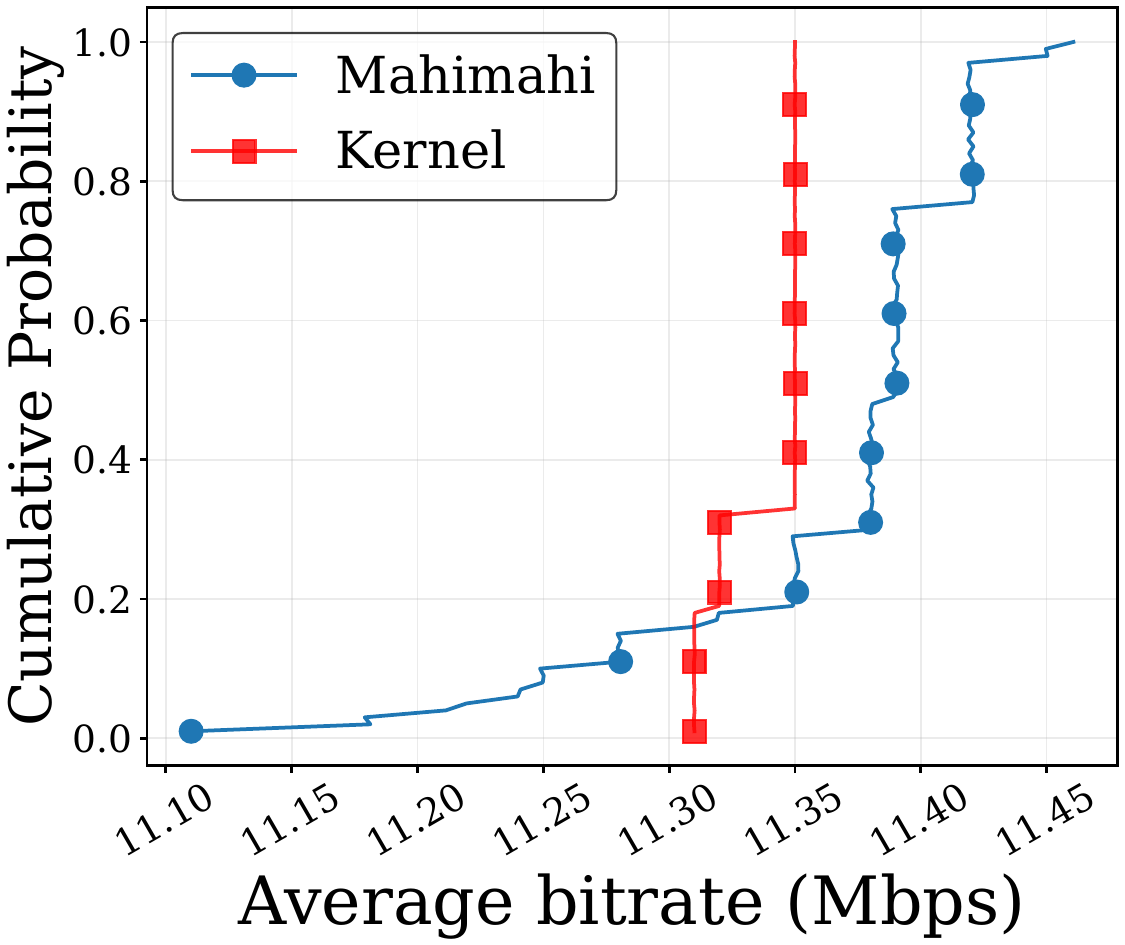}
        \caption{Low BDP}
    \end{subfigure}
    \hfill
    \begin{subfigure}[b]{0.32\columnwidth}
        \centering
        \includegraphics[width=\linewidth]{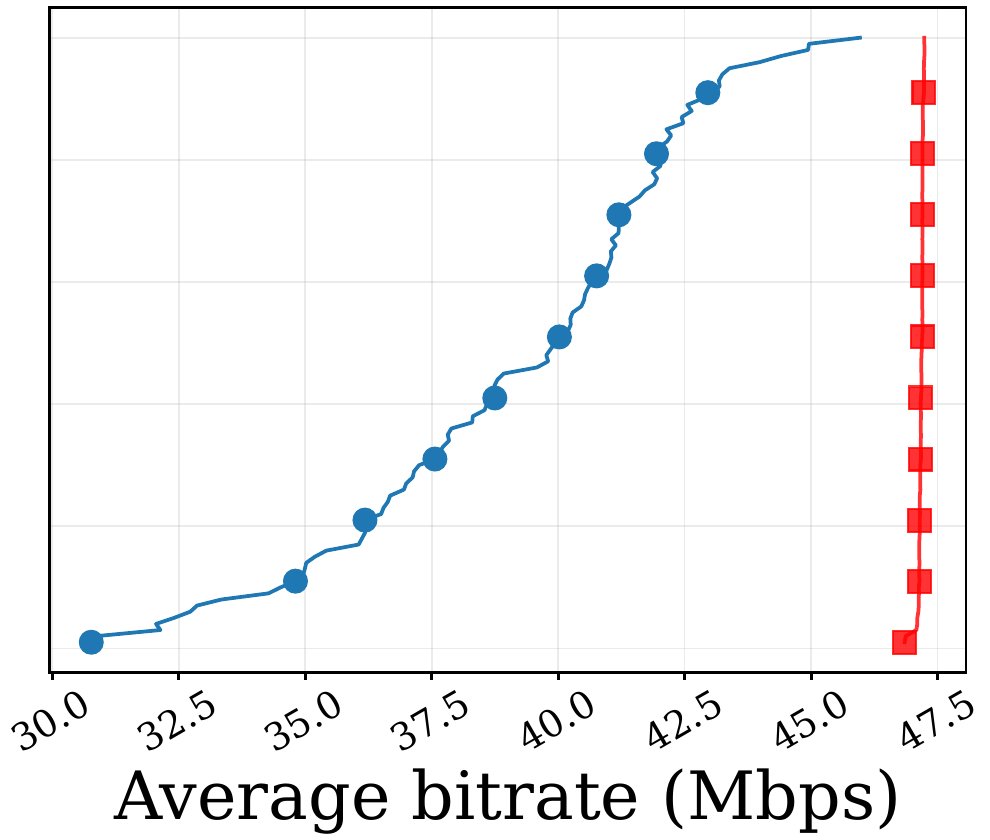}
        \caption{Medium BDP}
    \end{subfigure}
    \hfill
    \begin{subfigure}[b]{0.32\columnwidth}
        \centering
        \includegraphics[width=\linewidth]{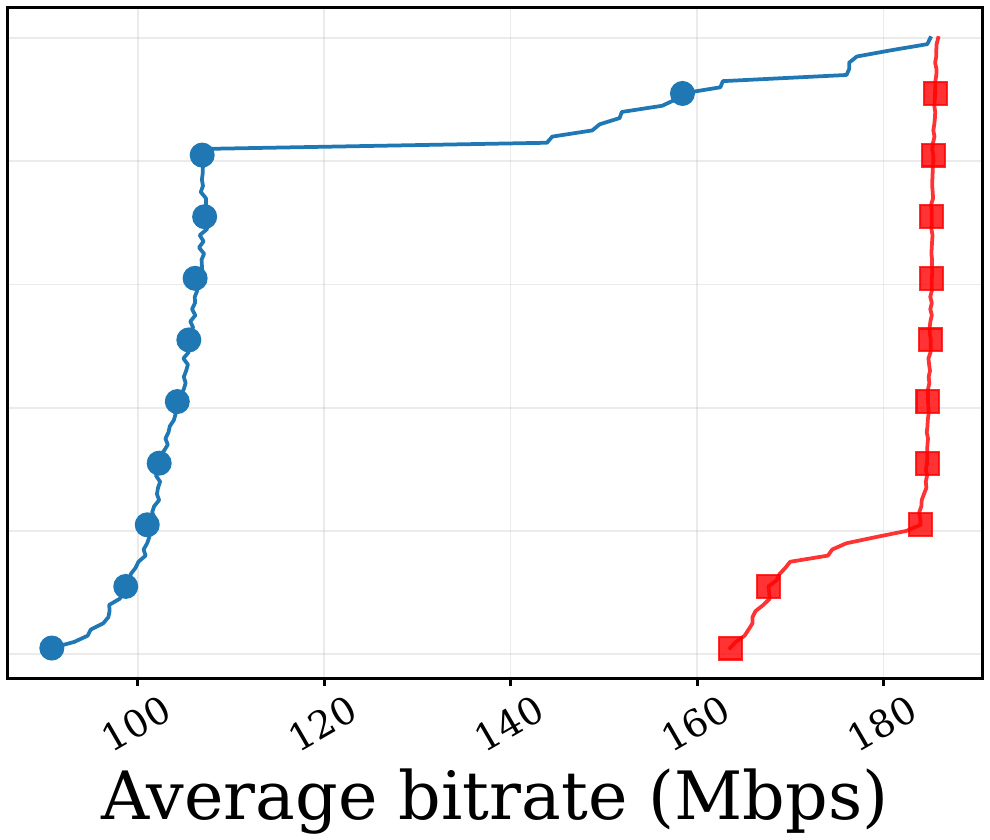}
        \caption{High BDP}
    \end{subfigure}
\caption{Dual-flow cumulative throughput with default DualPI2 parameters in all BDP regimes}
\label{fig:all-throughput-cdf}
\end{figure}

To more deeply analyze the throughput discrepancy, we plot the CDF of achieved throughput in both systems across all BDPs, shown in Fig.~\ref{fig:all-throughput-cdf}. Mahimahi clearly fails to saturate the bottleneck capacity in the medium and high BDP scenarios, under-utilizing the link compared to the kernel qdisc.

The achieved throughput through DualPI2 is the result of queue dynamics, particularly the ECN marking behavior. Although the result of the hypothesis testing with ECN marks does not a priori indicate a mismatch in ECN marking behavior, we know that excessive ECN signaling causes the sender to throttle transmission, leading to reduced throughput. This is a direct consequence of the queue delay (sojourn time) exceeding \texttt{step\_thresh} for L4S traffic, and \texttt{target} for classic traffic. 

These observations reveal that DualPI2’s default parameterization, while appropriate for kernel deployment, does not transfer directly to an emulated execution environment. The interaction between Mahimahi’s timing and queueing mechanisms and DualPI2’s control thresholds alters AQM dynamics, thereby influencing throughput. This highlights the environment-dependent sensitivity of DualPI2 parameters and demonstrates that careful parameter selection is essential for achieving comparable behavior across platforms.

\vspace*{0.1in}
\noindent \textbf{Single-flow experiments with varying DualPI2 parameters.} 
To understand the impact of these parameters, we conduct single-flow experiments in Mahimahi DualPI2, isolating L4S and classic traffic to assess how \texttt{step\_thresh} affects L4S throughput and \texttt{target} affects classic throughput. For a fair comparison, we also run a series of single flow experiments on the kernel, using the default parameters of DualPI2. The objective is to determine whether adjusting these parameters enhances behavioral alignment with the Linux kernel implementation.

\begin{figure}[t]
\centering
    \begin{subfigure}[b]{0.32\columnwidth}
        \centering
        \includegraphics[width=\linewidth]{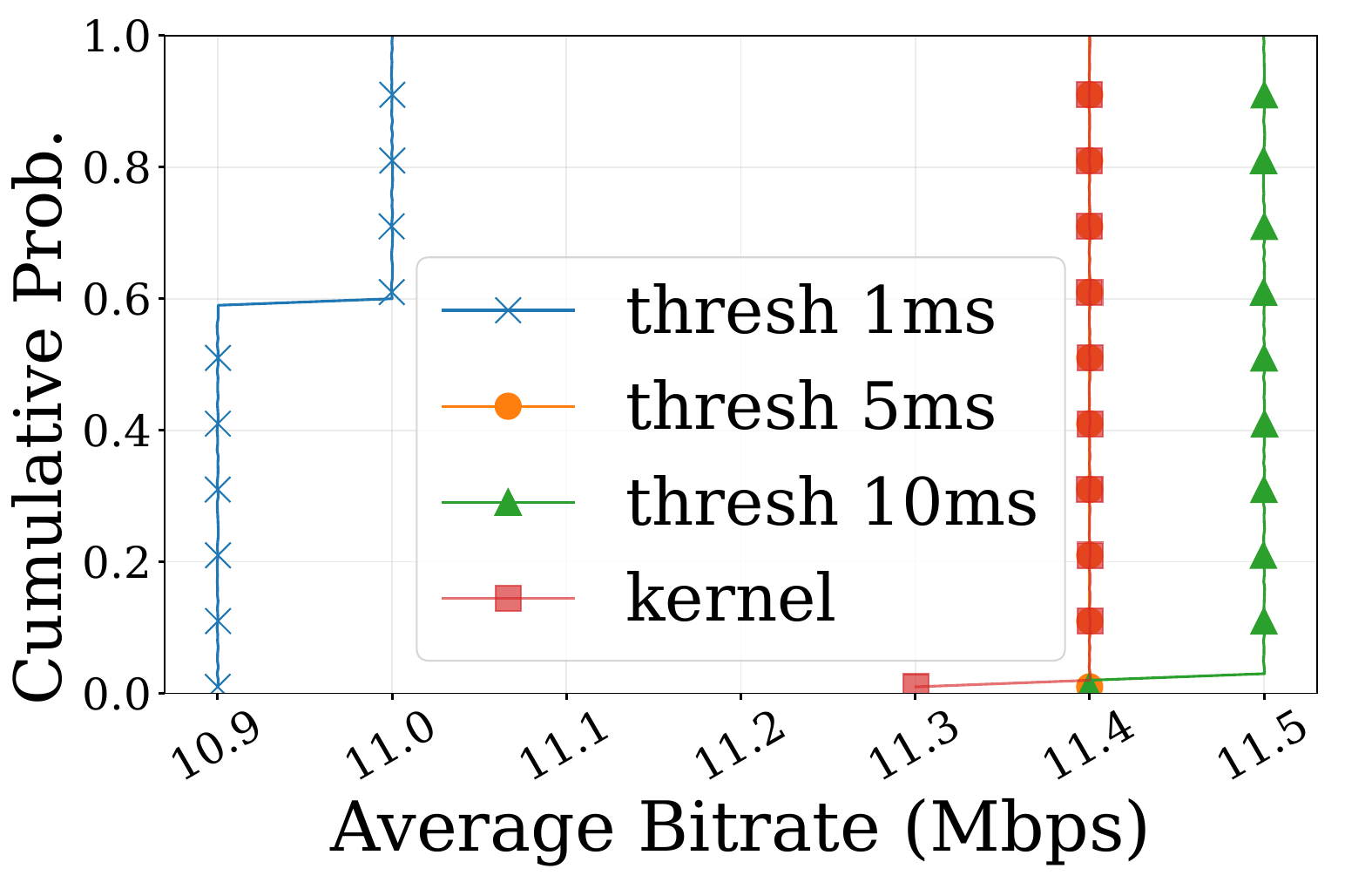}
        \caption{Low BDP}
        \label{fig:thr-l4s-low-bdp}
    \end{subfigure}
    \hfill
    \begin{subfigure}[b]{0.32\columnwidth}
        \centering
        \includegraphics[width=\linewidth]{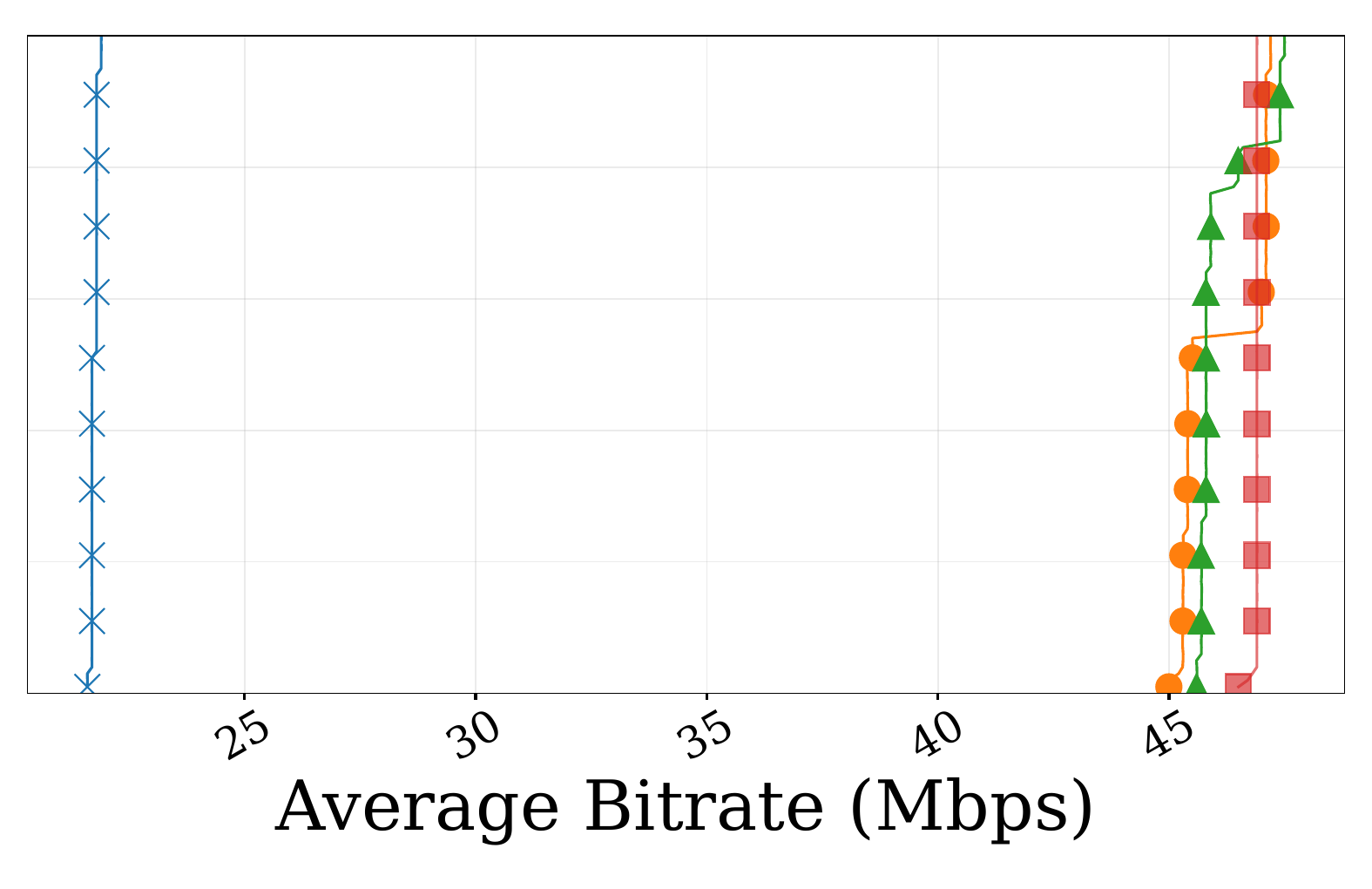}        \caption{Medium BDP}
        \label{fig:thr-l4s-mid-bdp}
    \end{subfigure}
    \hfill
    \begin{subfigure}[b]{0.32\columnwidth}
        \centering
                \includegraphics[width=\linewidth]{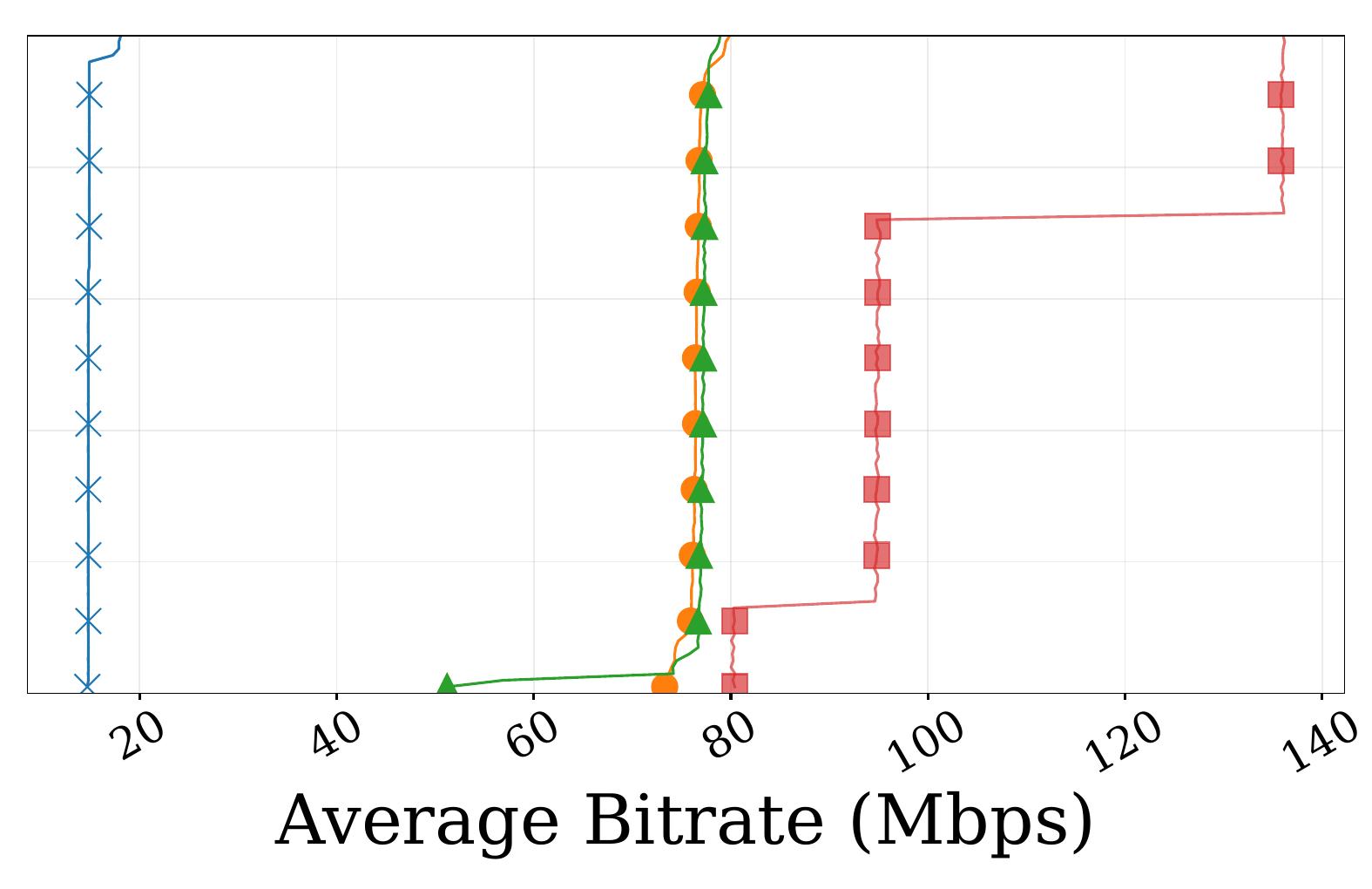}
        \caption{High BDP}
        \label{fig:thr-l4s-high-bdp}
    \end{subfigure}
    
\caption{L4S-only throughput varying \texttt{step\_thresh} in all BDP regimes}
\vspace*{-0.1in}
\label{fig:thr-l4s-thresh}
\end{figure}

\begin{figure}[ht]
\centering
    \begin{subfigure}[b]{0.32\columnwidth}
        \centering
        \includegraphics[width=\linewidth]{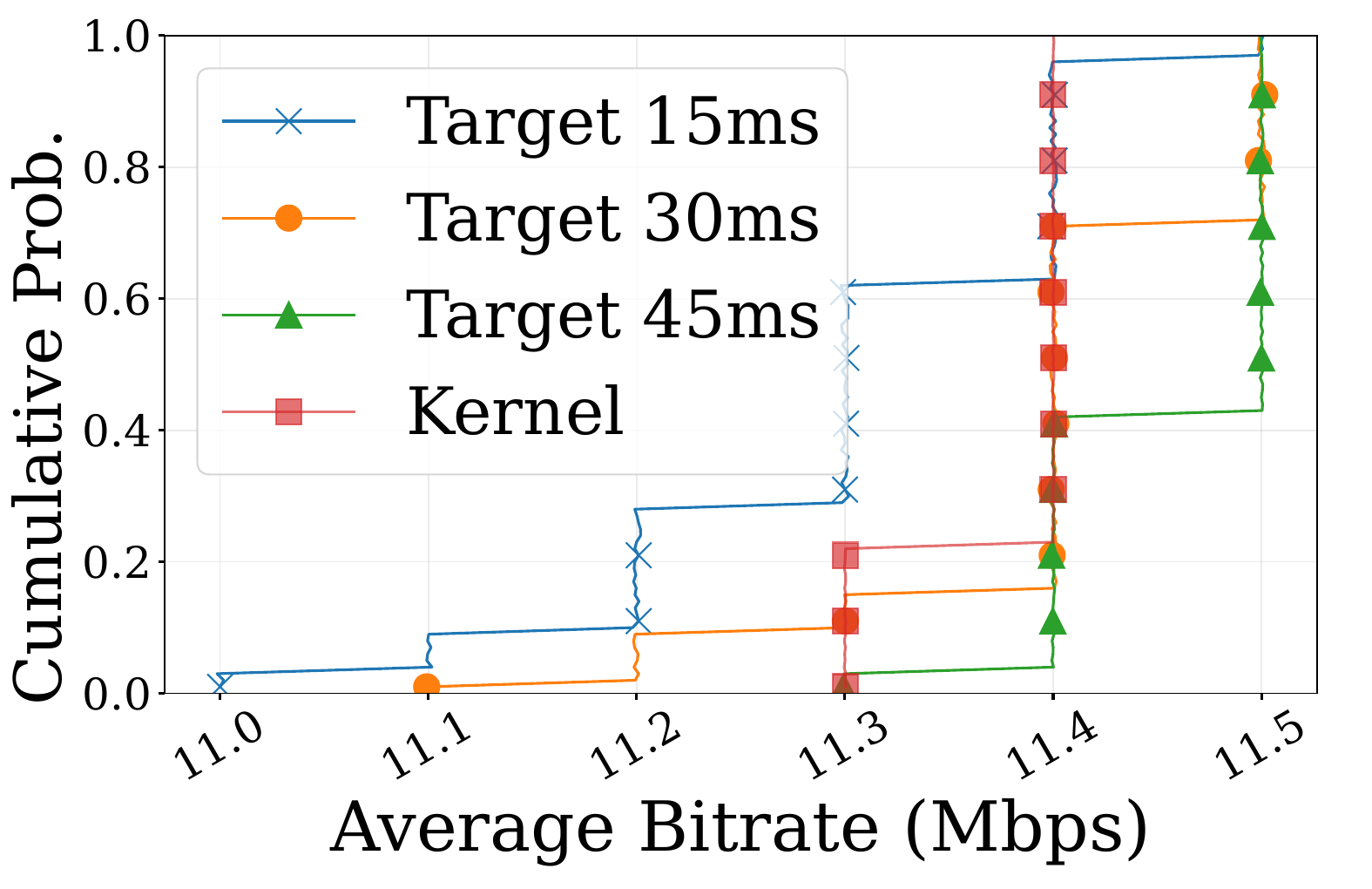}
        \caption{Low BDP}
    \end{subfigure}
    \hfill
    \begin{subfigure}[b]{0.32\columnwidth}
        \centering
        \includegraphics[width=\linewidth]{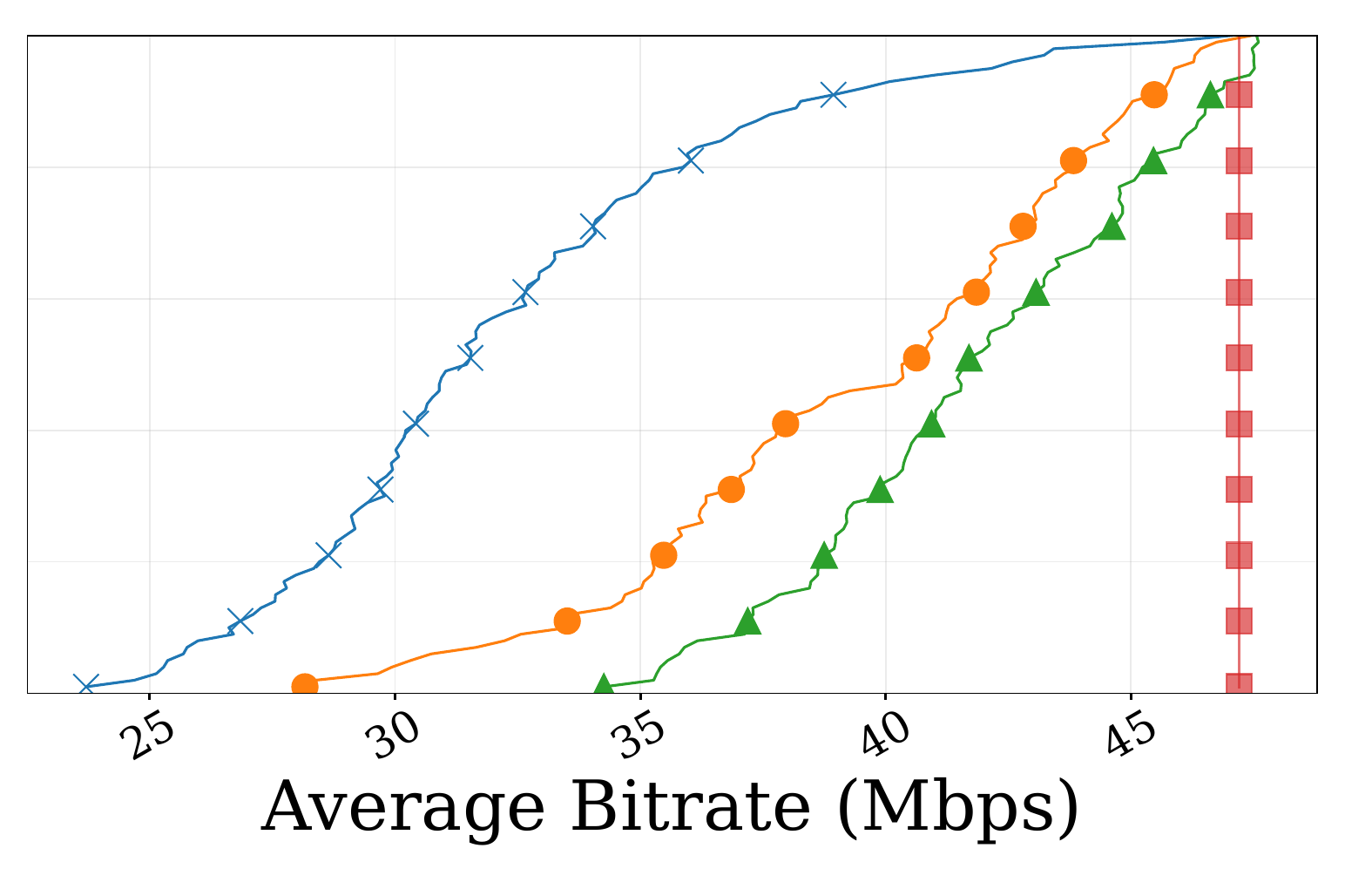}        \caption{Medium BDP}
    \end{subfigure}
    \hfill
    \begin{subfigure}[b]{0.32\columnwidth}
        \centering
        \includegraphics[width=\linewidth]{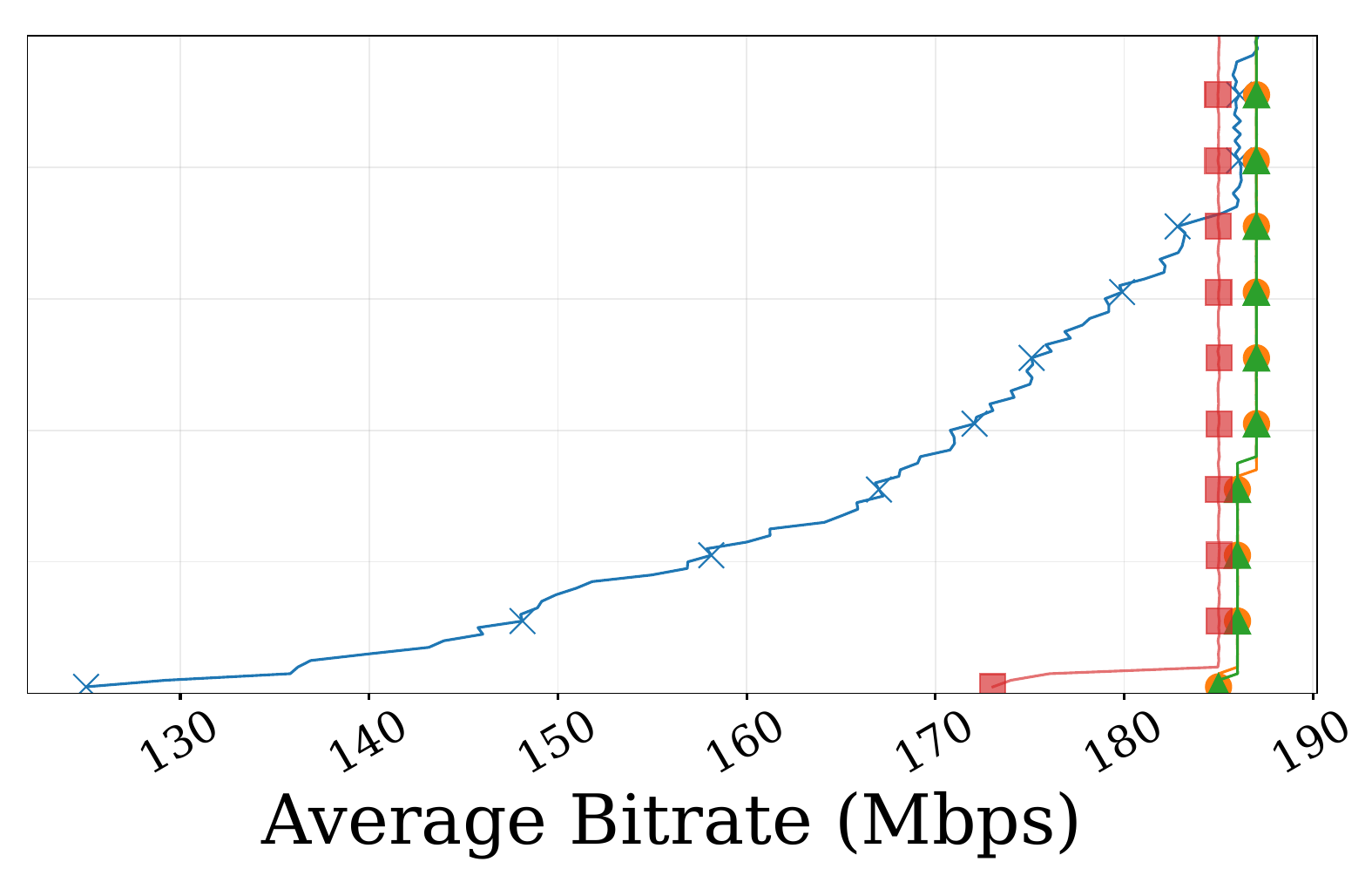}        \caption{High BDP}
    \end{subfigure}
    
\caption{Classic-only throughput varying \texttt{target} in all BDP regimes}
\vspace*{-0.15in}
\label{fig:thr-classic-target}
\end{figure}

In the L4S-only experiment, we sequentially run 100 \texttt{iperf3} flows using TCP Prague for two additional values of \texttt{step\_thresh}: 5 and 10 ms. Similarly, we run 100 classic-only \texttt{iperf3} flows sequentially using TCP Cubic for two additional values of \texttt{target}: 30 and 45 ms. We plot the variation of the achieved throughput in each one of these scenarios as shown in Fig.~\ref{fig:thr-l4s-thresh} for L4S traffic and Fig.~\ref{fig:thr-classic-target} for classic traffic. In each scenario, we also plot the corresponding kernel experiment results that use the default parameters for comparison.

We run the statistical test comparing each modified DualPI2 run on Mahimahi with its kernel counterpart that uses the default parameters. We compile the results of the L4S-only and classic-only experiments in Tables~\ref{tab:l4s-thresh-validation} and~\ref{tab:classic-target-validation}, respectively, where we show for each BDP regime and value of the considered DualPI2 parameter, the statistical test results for each metric in terms of $\widehat{p}_{\max}$, denoted as $p$, and $\varepsilon_{max}$, denoted as $eps$ for brevity. The ``OK'' column refers to the result of the statistical test, which either fails to reject $H_0$ ($\times$),  implying a significant behavioral misalignment in terms of the corresponding metric, or rejects $H_0$ ($\checkmark$), implying similarity under the 5\% exceedance criterion.

\begin{table*}[b]
  \centering
  \caption{Validation results of single L4S flows varying \texttt{step\_thresh} across BDP scenarios}
  \label{tab:l4s-thresh-validation}
  
  \renewcommand{\arraystretch}{1.5}

  \setlength{\tabcolsep}{0pt} 
  \scriptsize
  
  \begin{tabular*}{\textwidth}{|>{\centering\arraybackslash}p{0.095\textwidth}|*{9}{>{\centering\arraybackslash}p{0.035\textwidth}|>{\centering\arraybackslash}p{0.035\textwidth}|>{\centering\arraybackslash}p{0.028\textwidth}|}  }
    \hline
    & \multicolumn{9}{c|}{\textbf{Low BDP (12 Mbps, 20 ms)}} & \multicolumn{9}{c|}{\textbf{Medium BDP (50 Mbps, 40 ms)}} & \multicolumn{9}{c|}{\textbf{High BDP (200 Mbps, 100 ms)}} \\ \hline

    \specialrule{0.5pt}{0pt}{0pt}
    
    \textbf{Step\_thresh} & \multicolumn{3}{c|}{1 ms} & \multicolumn{3}{c|}{5 ms} & \multicolumn{3}{c|}{10 ms} & \multicolumn{3}{c|}{1 ms} & \multicolumn{3}{c|}{5 ms} & \multicolumn{3}{c|}{10 ms} & \multicolumn{3}{c|}{1 ms} & \multicolumn{3}{c|}{5 ms} & \multicolumn{3}{c|}{10 ms} \\ \hline
    
    \textbf{Metric} & \textbf{eps} & \textbf{p} & \textbf{OK} & \textbf{eps} & \textbf{p} & \textbf{OK} & \textbf{eps} & \textbf{p} & \textbf{OK} & \textbf{eps} & \textbf{p} & \textbf{OK} & \textbf{eps} & \textbf{p} & \textbf{OK} & \textbf{eps} & \textbf{p} & \textbf{OK} & \textbf{eps} & \textbf{p} & \textbf{OK} & \textbf{eps} & \textbf{p} & \textbf{OK} & \textbf{eps} & \textbf{p} & \textbf{OK} \\ \hline
    
    Throughput & 0.10 & 1.00 &  $\times$ & 0 & 0.010 & $\checkmark$ & 0 & 0.980 & $\times$ & 0.20 & 1.00 & $\times$ & 1.90 & 0.0001 & $\checkmark$ & 1.70 & 0.0001 & $\checkmark$ & 55.80 & 1.00 & $\times$ & 55.80 & 0.280 & $\times$ & 55.80 & 0.280 & $\times$ \\ \hline

    Queue occup. & 0.232 & 1.00 & $\times$ & 0.368 & 1 & $\times$ & 0.363 & 1 & $\times$ & 0.218 & 0.886 & $\times$ & 0.617 & 1 & $\times$ & 0.710 & 1 & $\times$ & 0.144 & 0.949 & $\times$ & 0.925 & 1 & $\times$ & 0.865 & 1 & $\times$\\ \hline
    ECN Marks & 3.40 & 0.040 & $\checkmark$ & 1.088 & 1 & $\times$ & 2.822 & 0.006 & $\checkmark$ & 0.004 & 0.998 & $\times$ & 0.957 & 0.932 & $\times$ & 1.583 & 0.040 & $\checkmark$ & 0.163 & 1.00 & $\times$ & 0.198 & 0.984 & $\times$ & 1.154 & 0.003 & $\checkmark$ \\ \hline
    Drops & 0 & \tiny{0.00005} & $\checkmark$ & 0 & \tiny{0.00005} & $\checkmark$ & 0 & \tiny{0.00005} & $\checkmark$ & 0 & \tiny{0.00005} & $\checkmark$ & 0 & \tiny{0.00005} & $\checkmark$ & 0 & \tiny{0.00005} & $\checkmark$ & 0 & \tiny{0.00005} & $\checkmark$ & 0 & \tiny{0.00005} & $\checkmark$ & 0 & \tiny{0.00005} & $\checkmark$\\ \hline
  \end{tabular*}

  \vspace{0.4cm}

  \caption{Validation results of single classic flows varying \texttt{target} across BDP scenarios}
  \label{tab:classic-target-validation}

  \begin{tabular*}{\textwidth}{|>{\centering\arraybackslash}p{0.095\textwidth}|*{9}{>{\centering\arraybackslash}p{0.035\textwidth}|>{\centering\arraybackslash}p{0.035\textwidth}|>{\centering\arraybackslash}p{0.028\textwidth}|}  }
    \hline
    & \multicolumn{9}{c|}{\textbf{Low BDP (12 Mbps, 20 ms)}} & \multicolumn{9}{c|}{\textbf{Medium BDP (50 Mbps, 40 ms)}} & \multicolumn{9}{c|}{\textbf{High BDP (200 Mbps, 100 ms)}} \\ \hline

    \specialrule{0.5pt}{0pt}{0pt}
    
    \textbf{Target} & \multicolumn{3}{c|}{15 ms} & \multicolumn{3}{c|}{30 ms} & \multicolumn{3}{c|}{45 ms} & \multicolumn{3}{c|}{15 ms} & \multicolumn{3}{c|}{30 ms} & \multicolumn{3}{c|}{45 ms} & \multicolumn{3}{c|}{15 ms} & \multicolumn{3}{c|}{30 ms} & \multicolumn{3}{c|}{45 ms} \\ \hline
    
    \textbf{Metric} & \textbf{eps} & \textbf{p} & \textbf{OK} & \textbf{eps} & \textbf{p} & \textbf{OK} & \textbf{eps} & \textbf{p} & \textbf{OK} & \textbf{eps} & \textbf{p} & \textbf{OK} & \textbf{eps} & \textbf{p} & \textbf{OK} & \textbf{eps} & \textbf{p} & \textbf{OK} & \textbf{eps} & \textbf{p} & \textbf{OK} & \textbf{eps} & \textbf{p} & \textbf{OK} & \textbf{eps} & \textbf{p} & \textbf{OK} \\ \hline
    
    Throughput & 0.30 & 0.013 & $\checkmark$ & 0.20 & 0.0101 & $\checkmark$ & 0.10 & 0.128 &  $\times$  & 13.90 & 0.670 & $\times$ & 9.00 & 0.030 & $\checkmark$ & 9.50 & 0.140 & $\times$ & 46.00 & 0.049 & $\checkmark$ & 12.80 & 0.130 & $\times$ & 9.00 & 0.030 & $\checkmark$ \\ \hline

    Queue occup. & 7.583 & 1 & $\times$ & 7.522 & 1 & $\times$ & 9.096 & 1 & $\times$ & 22.73 & 1 & $\times$ & 23.79 & 1 & $\times$ & 75.52 & 0.786 & $\times$ & 67.93 & 0.993 & $\times$ & 55.7 & 0.895 & $\times$ & 23.73 & 1 & $\times$ \\ \hline
    ECN Marks & 0.0084 & 0.999 & $\times$ & 0.007 & 0.999 & $\times$ & 0.007 & 0.984 & $\times$ & 0.004 & 0.998 & $\times$ & 0.0005 & 0.018 & $\checkmark$ & 0.003 & 0.997 & $\times$ & 0.001 & 0.617 & $\times$ & 0.004 & 0.976 & $\times$ & 0.0005 & 0.002 & $\checkmark$\\ \hline
    Drops & 0.009 & 0.066 &$\checkmark$ & 0.009 & 0.062 & $\checkmark$ & 0.013 & 0.092 & $\times$ & 0.051 & 0.024 & $\checkmark$ & 0 & \tiny{0.00005} & $\checkmark$ & 0.042 & 0.040 & $\checkmark$ & 0 & \tiny{0.00005} & $\checkmark$ & 0.034 & 0.020 & $\checkmark$ & 0 & \tiny{0.00005} & $\checkmark$ \\ \hline
  \end{tabular*}
  
\end{table*}

The results in Table~\ref{tab:l4s-thresh-validation} show that, in the L4S-only case, setting \texttt{step\_thresh} to 5 ms improves the statistical test outcome of the throughput for both the low and medium BDP scenarios compared to the default value of 1 ms. As for the high BDP scenario of 200~Mbps, increasing \texttt{step\_thresh} to 5 ms does not yield any improvement. At 10 ms, while  the outcome for the ECN mark metric improves, there is no change with regard to achieved throughput. Concerning classic-only traffic scenarios,  Table~\ref{tab:classic-target-validation} shows that setting \texttt{target} to 30 ms improves the statistical test outcome for the medium BDP scenario, while it conserves the same outcome for the low BDP scenario. In the high BDP case, a \texttt{target} value of 45 ms improves the outcome for all considered metrics with the exception of queue occupancy.

\vspace*{0.1in}
\noindent \textbf{Dual-flow experiments with refined DualPI2 parameters.} Bringing together the results of these experiments, we rerun the dual-flow scenarios, combining the values of \texttt{step\_thresh} and \texttt{target} that optimize the statistical results for each BDP regime prioritizing throughput, namely: 

\vspace*{0.06in}
 \textbf{Low BDP}: \texttt{step\_thresh} = 5 ms, \texttt{target} = 30 ms;

 \textbf{Medium BDP}: \texttt{step\_thresh} = 5 ms, \texttt{target} = 30 ms; and

  \textbf{High BDP}: \texttt{step\_thresh} = 10 ms, \texttt{target} = 45 ms. 

\begin{figure*}[ht]
\centering

\begin{minipage}[c]{0.95\textwidth}
    \makebox[0.24\linewidth]{Cumulative Throughput} \hfill
    \makebox[0.24\linewidth]{Queue Occupancy} \hfill
    \makebox[0.24\linewidth]{ECN Marks} \hfill
    \makebox[0.24\linewidth]{Packet Drops}
\end{minipage}
\begin{minipage}[c]{0.04\textwidth}
\end{minipage}

\vspace{0.3em}

\begin{minipage}[c]{0.95\textwidth}
    \includegraphics[width=0.25\linewidth]{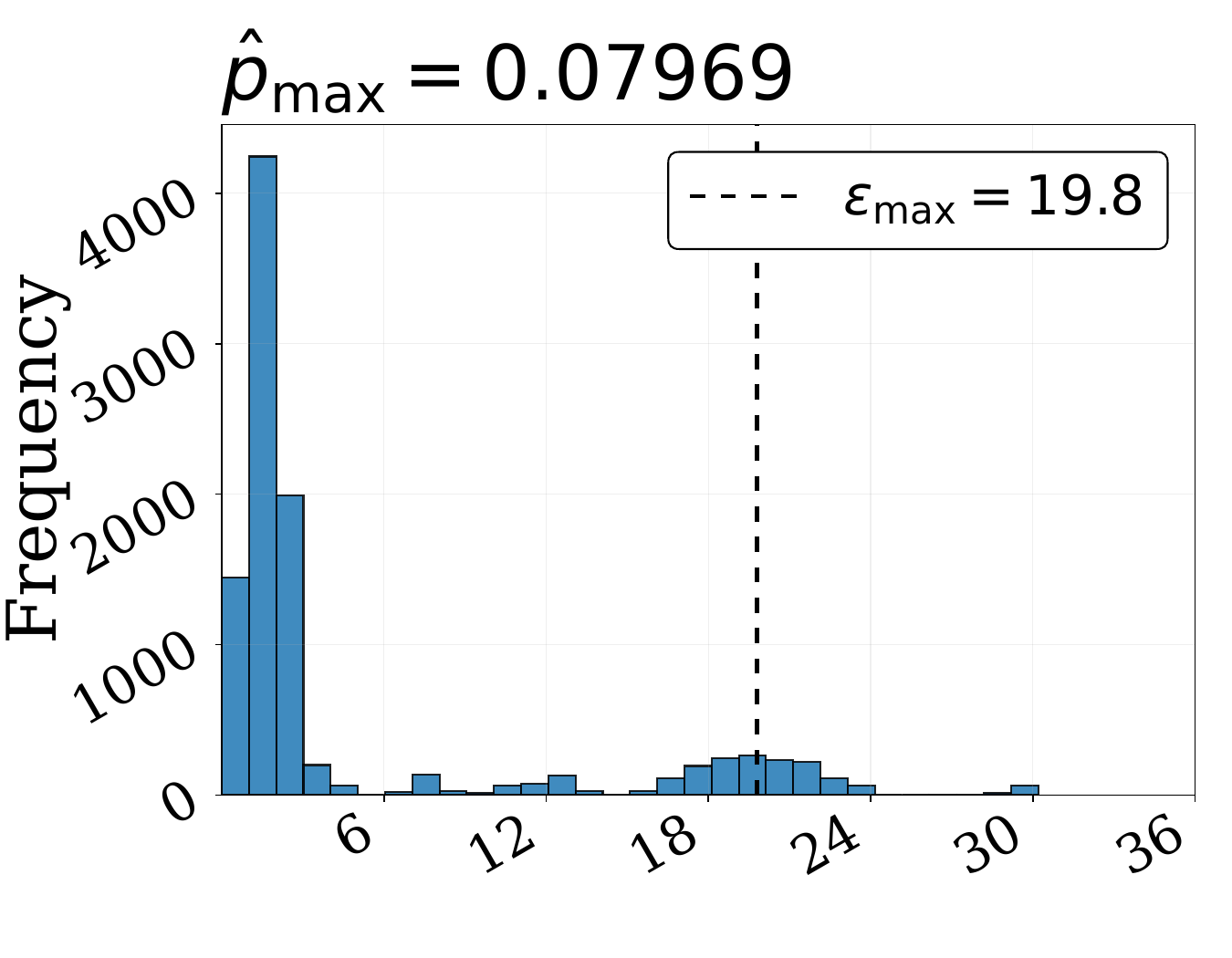}\hfill
    \includegraphics[width=0.25\linewidth]{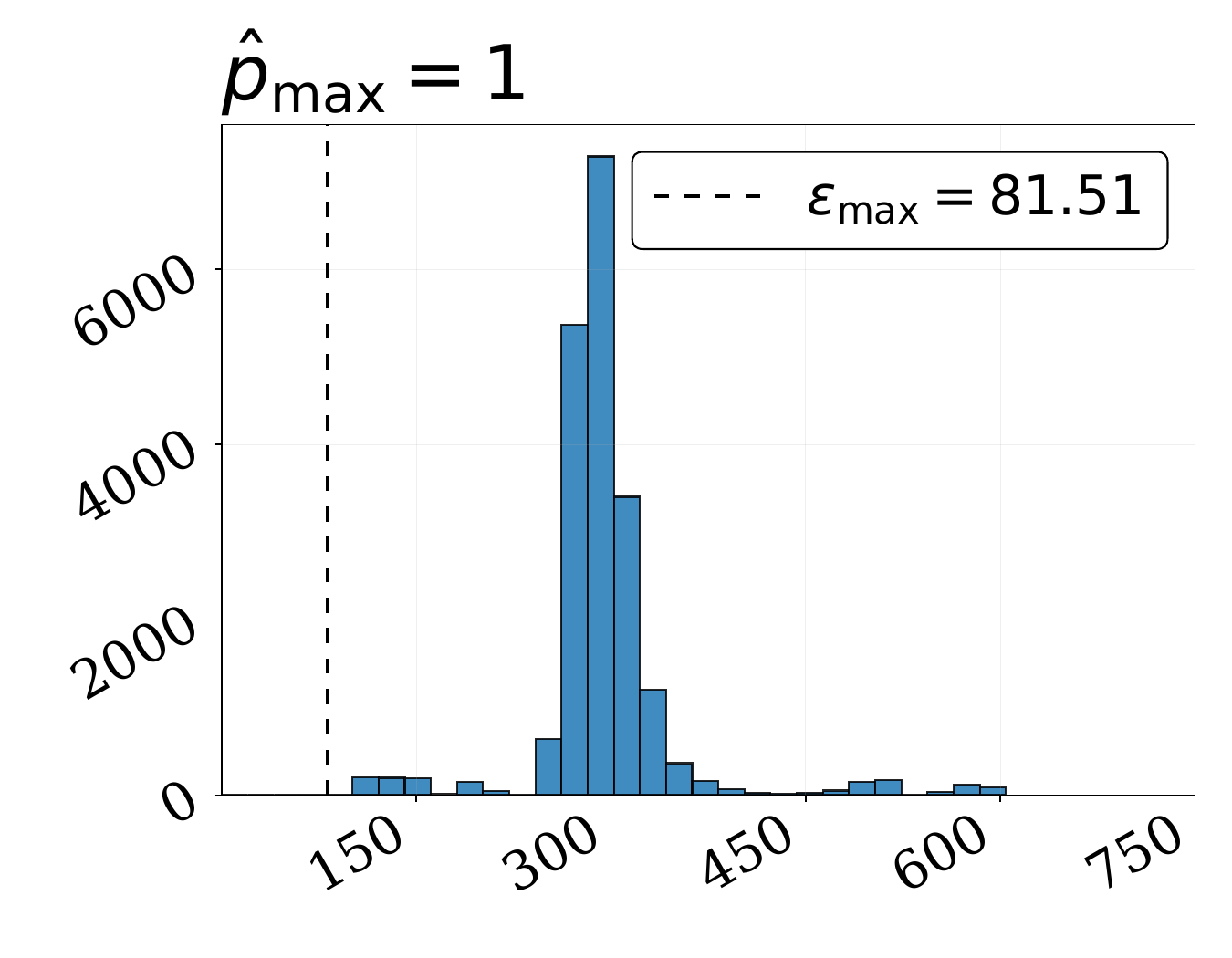}\hfill
    \includegraphics[width=0.25\linewidth]{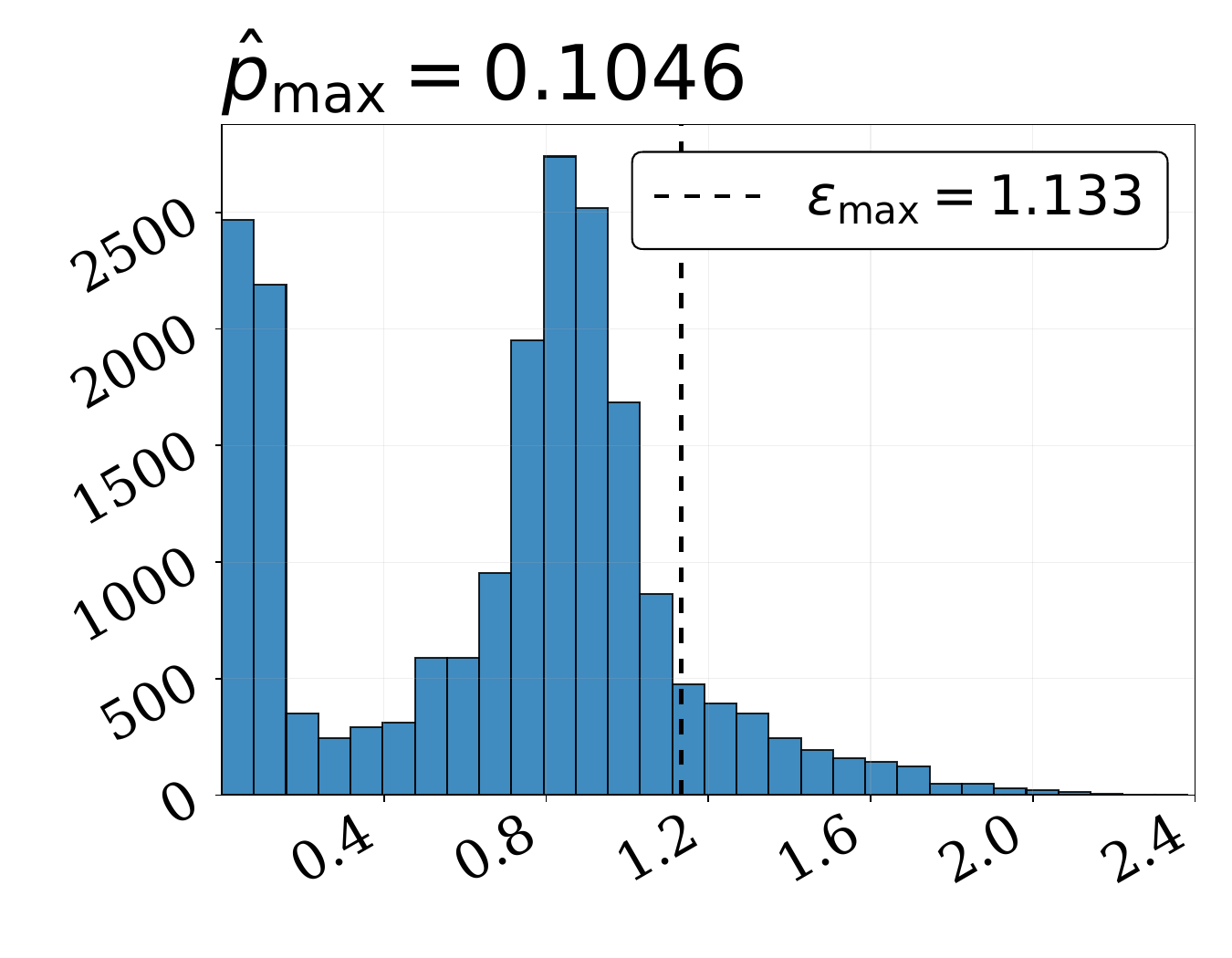}\hfill
    \includegraphics[width=0.25\linewidth]{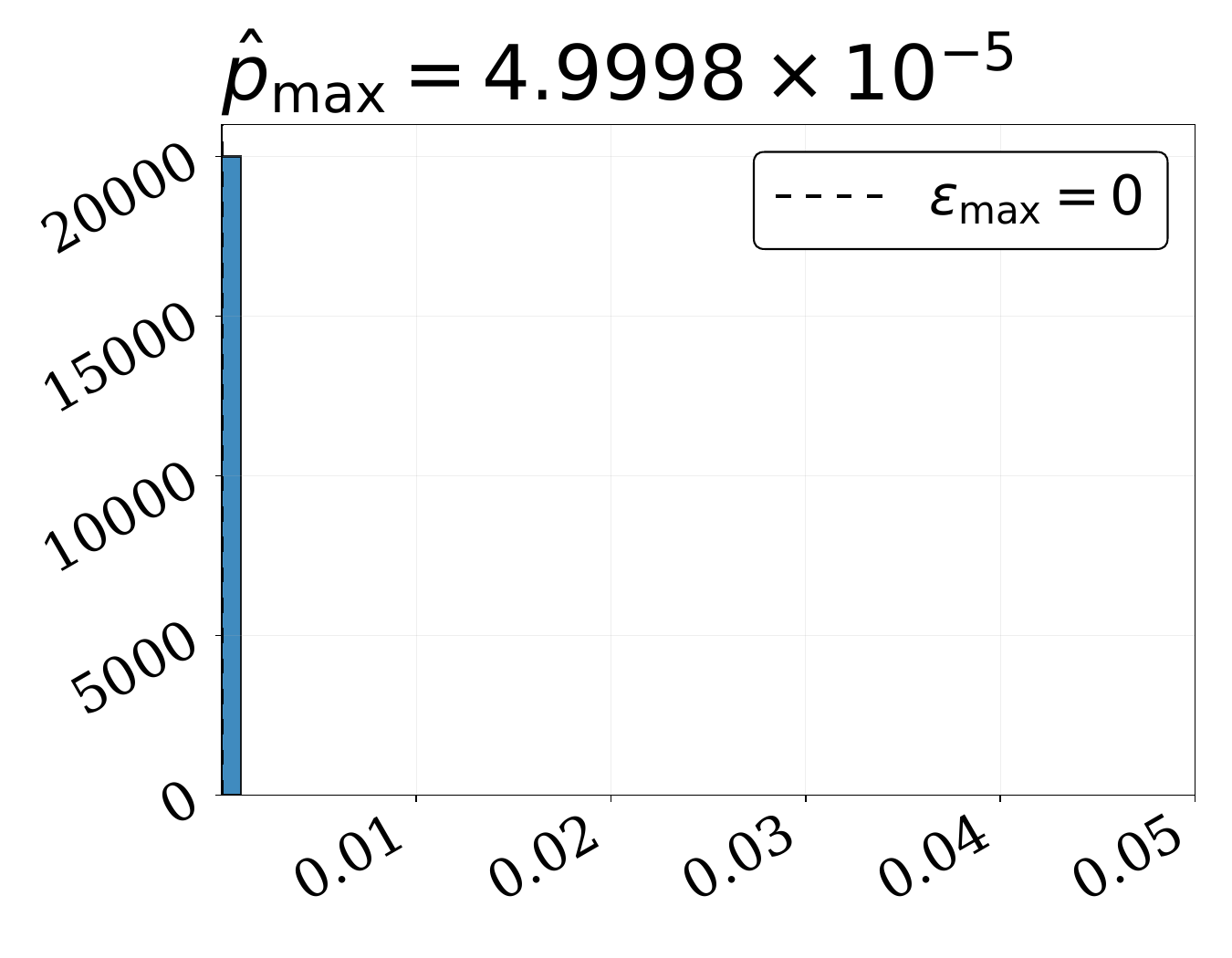}
\end{minipage}
\begin{minipage}[c]{0.04\textwidth}
    \centering\rotatebox{90}{High BDP}
\end{minipage}


\begin{minipage}[c]{0.95\textwidth}
    \includegraphics[width=0.25\linewidth]{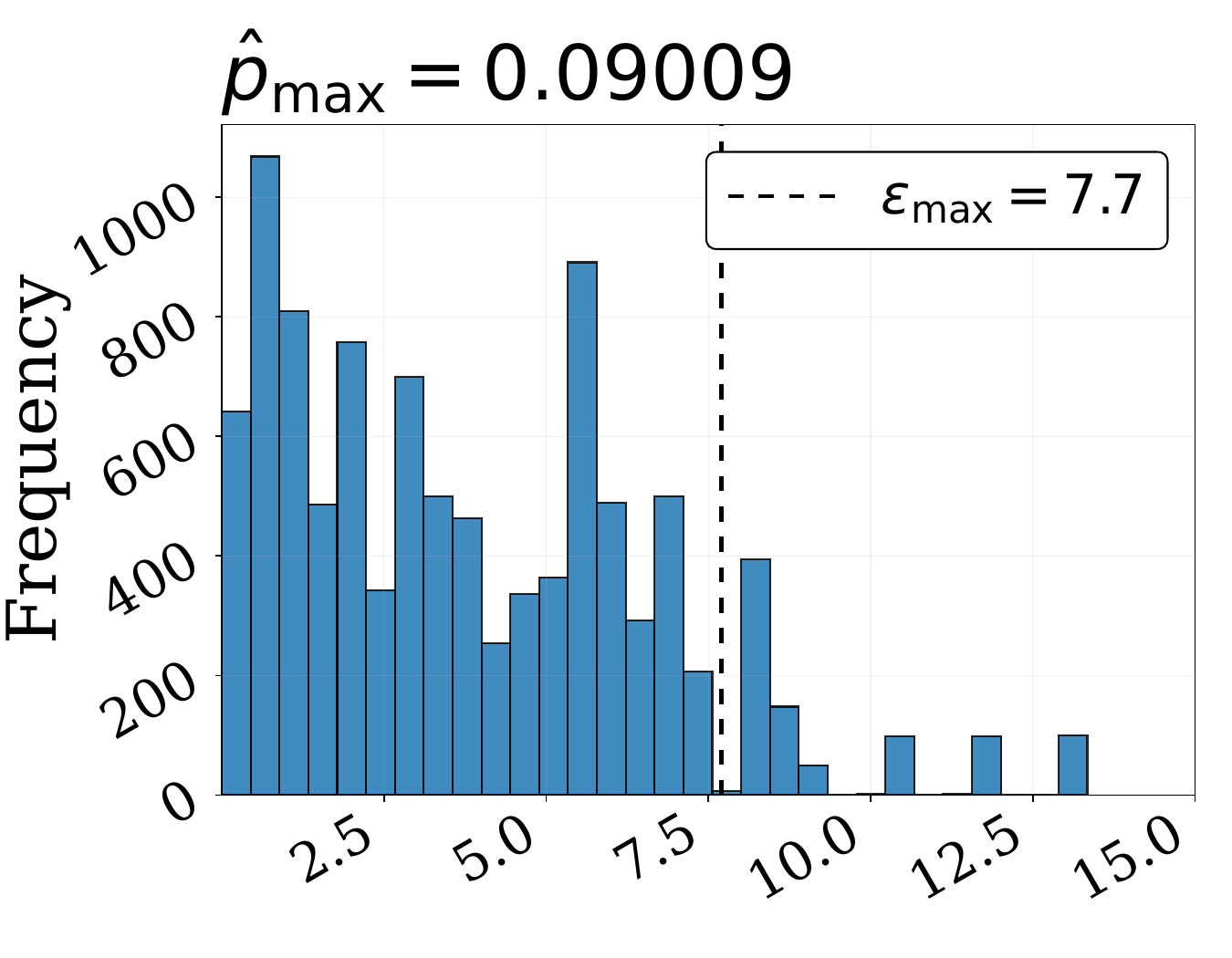}\hfill
    \includegraphics[width=0.25\linewidth]{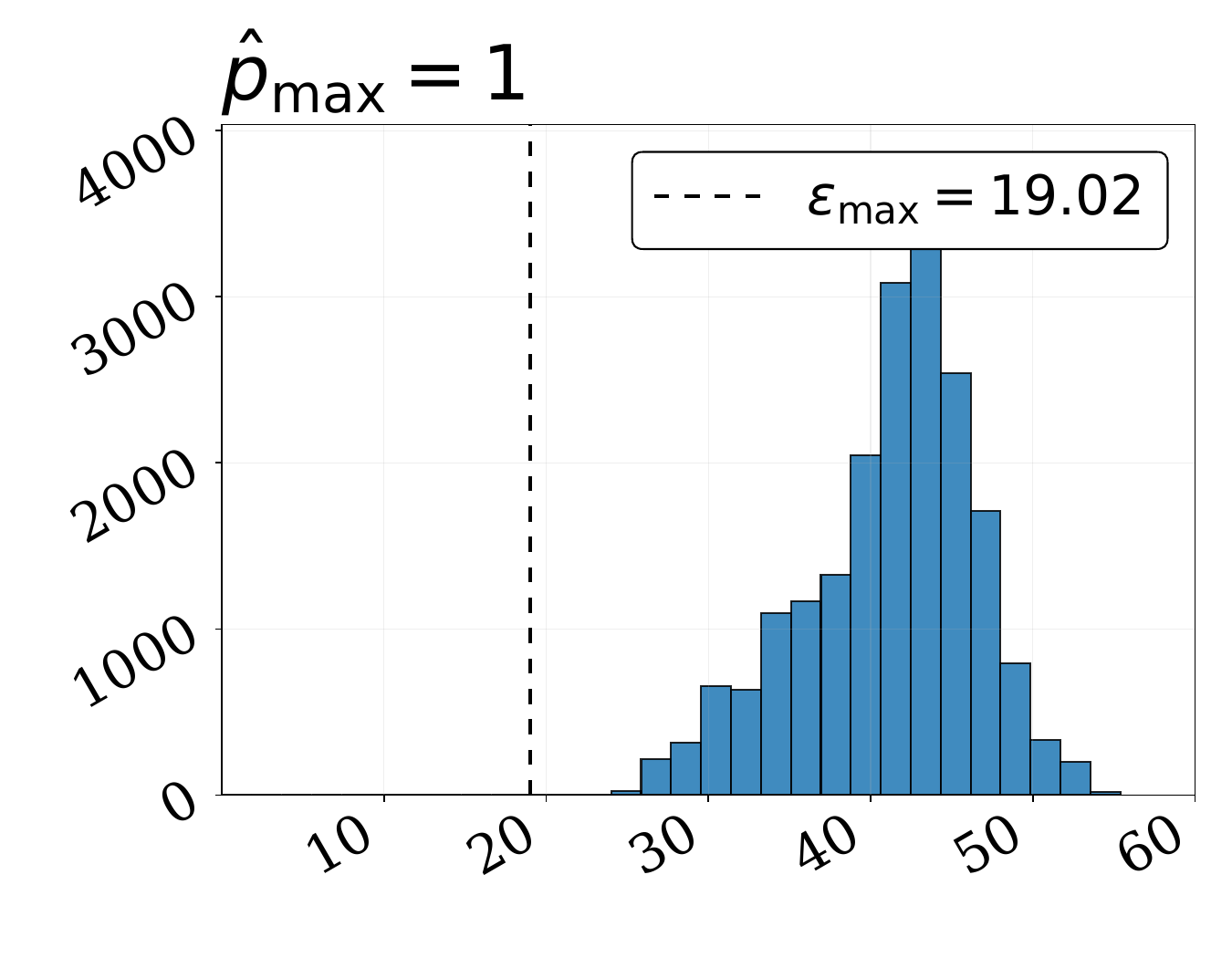}\hfill
    \includegraphics[width=0.25\linewidth]{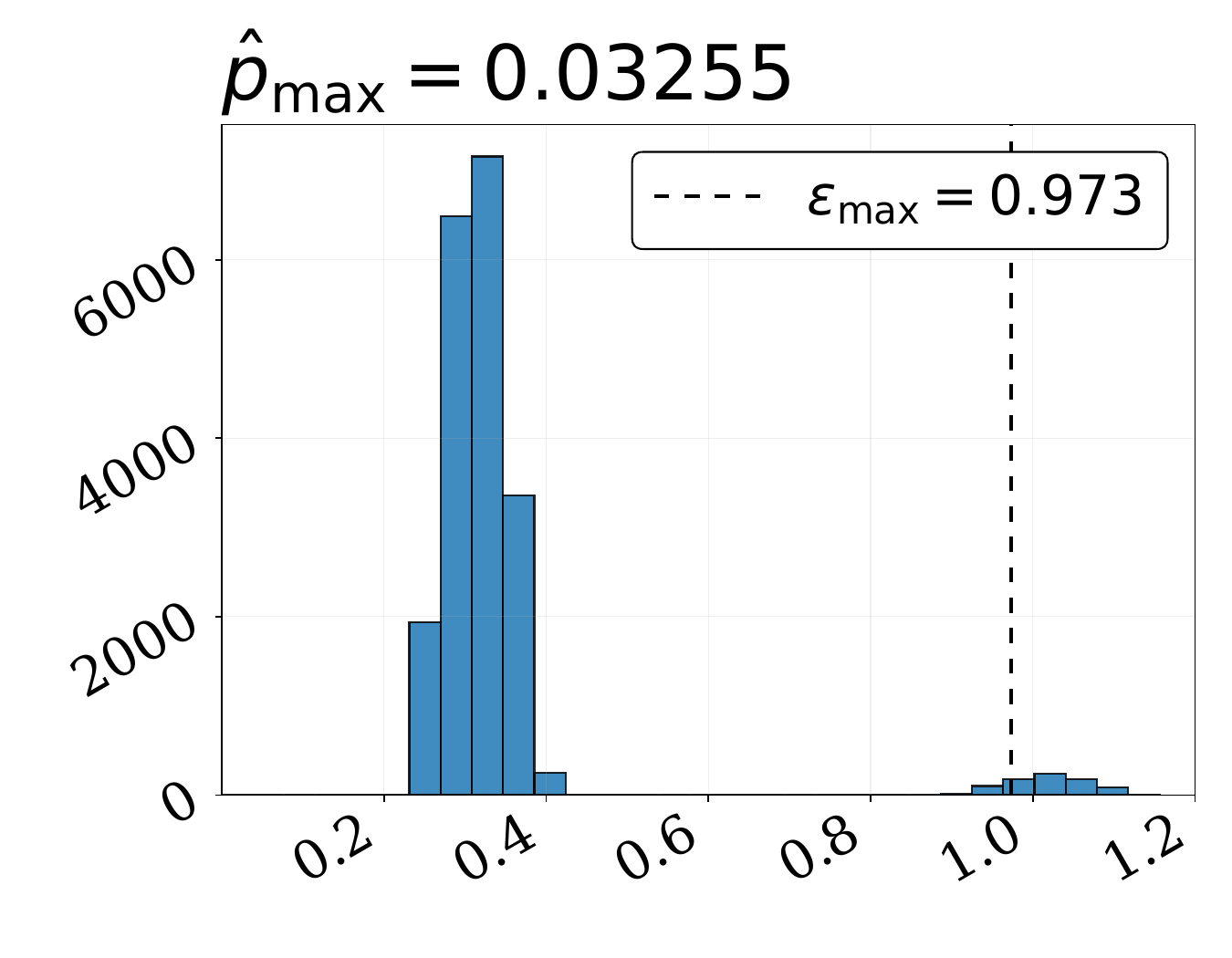}\hfill
    \includegraphics[width=0.25\linewidth]{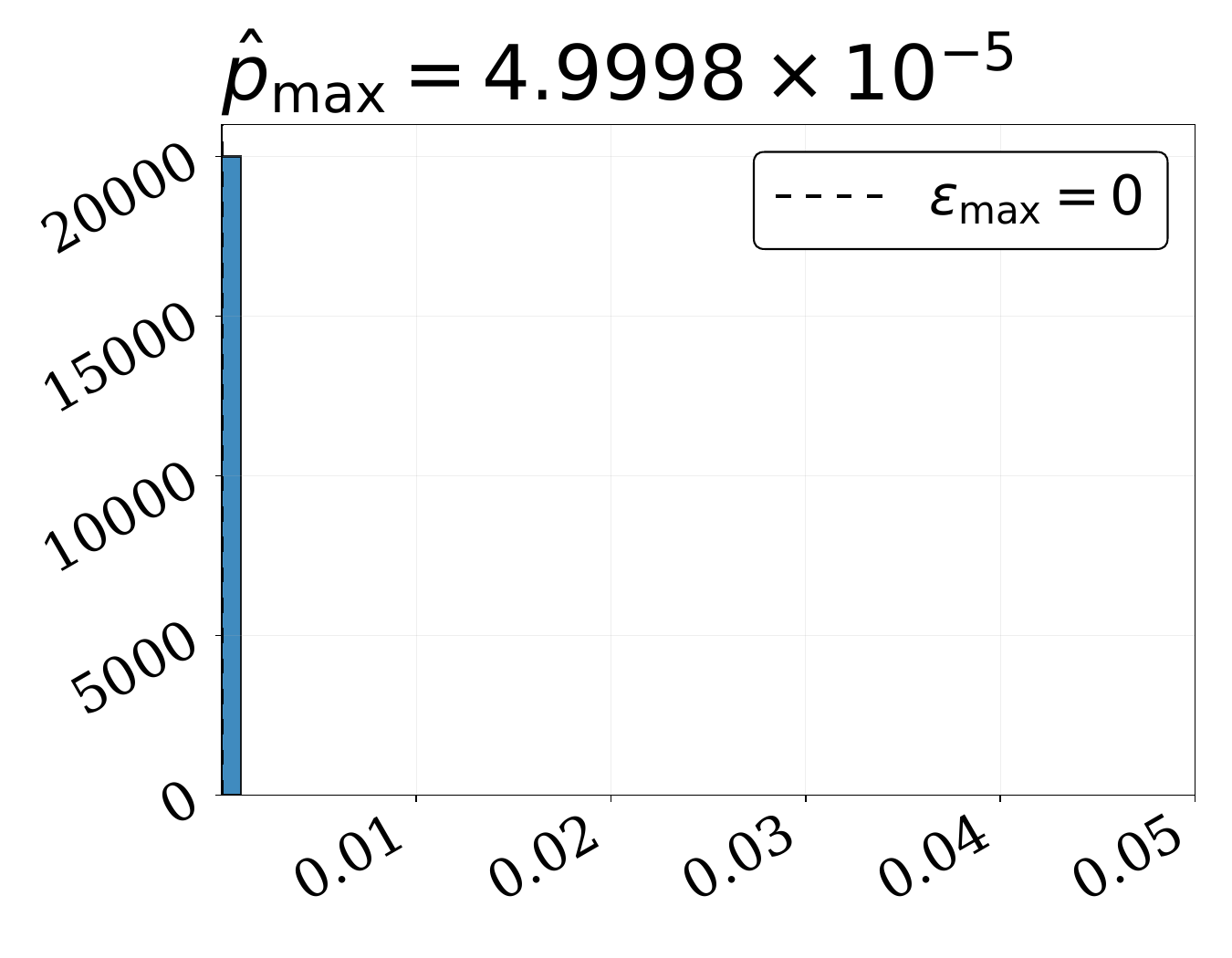}
\end{minipage}
\begin{minipage}[c]{0.04\textwidth}
    \centering\rotatebox{90}{Medium BDP}
\end{minipage}


\begin{minipage}[c]{0.95\textwidth}
    \includegraphics[width=0.25\linewidth]{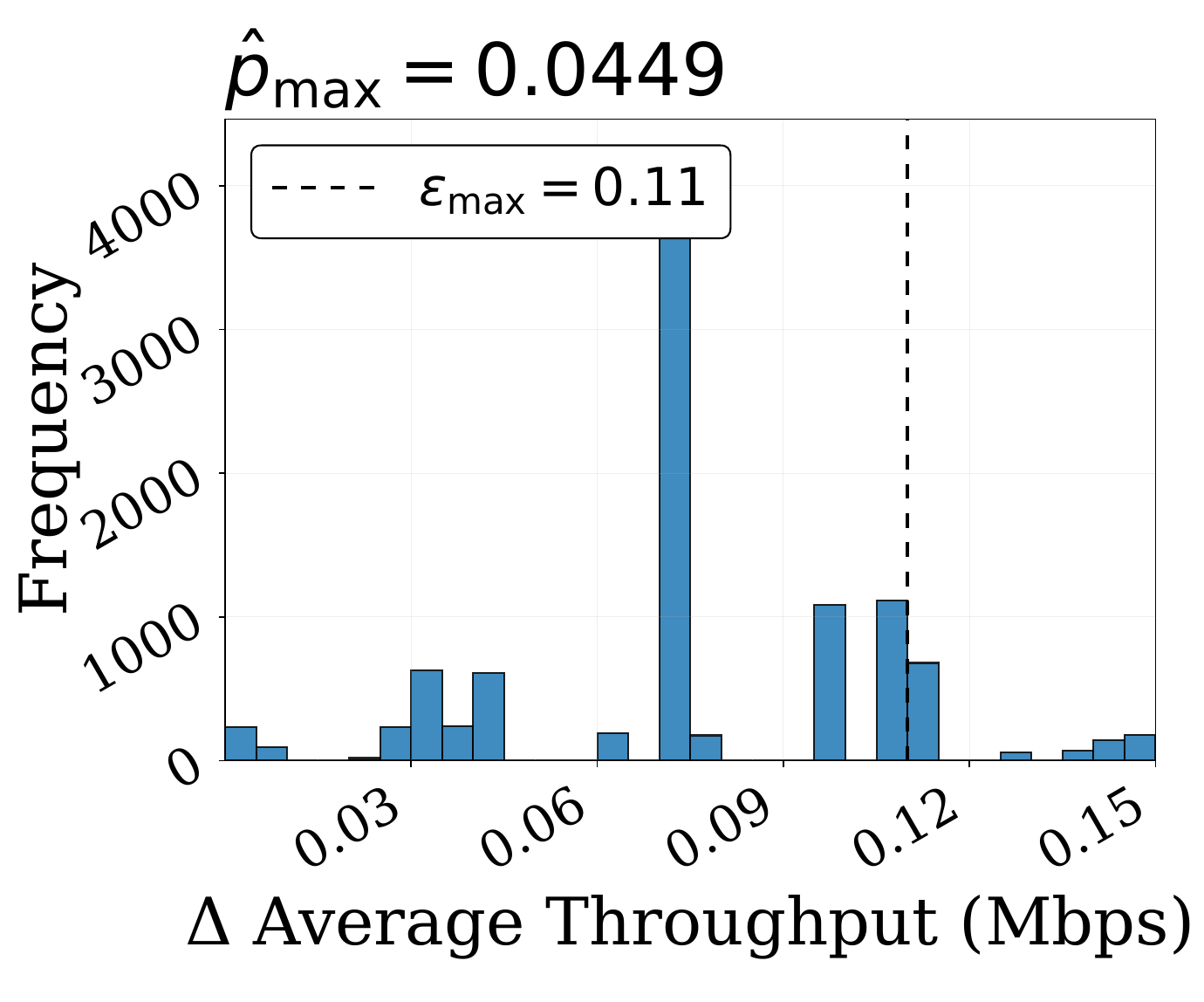}\hfill
    \includegraphics[width=0.25\linewidth]{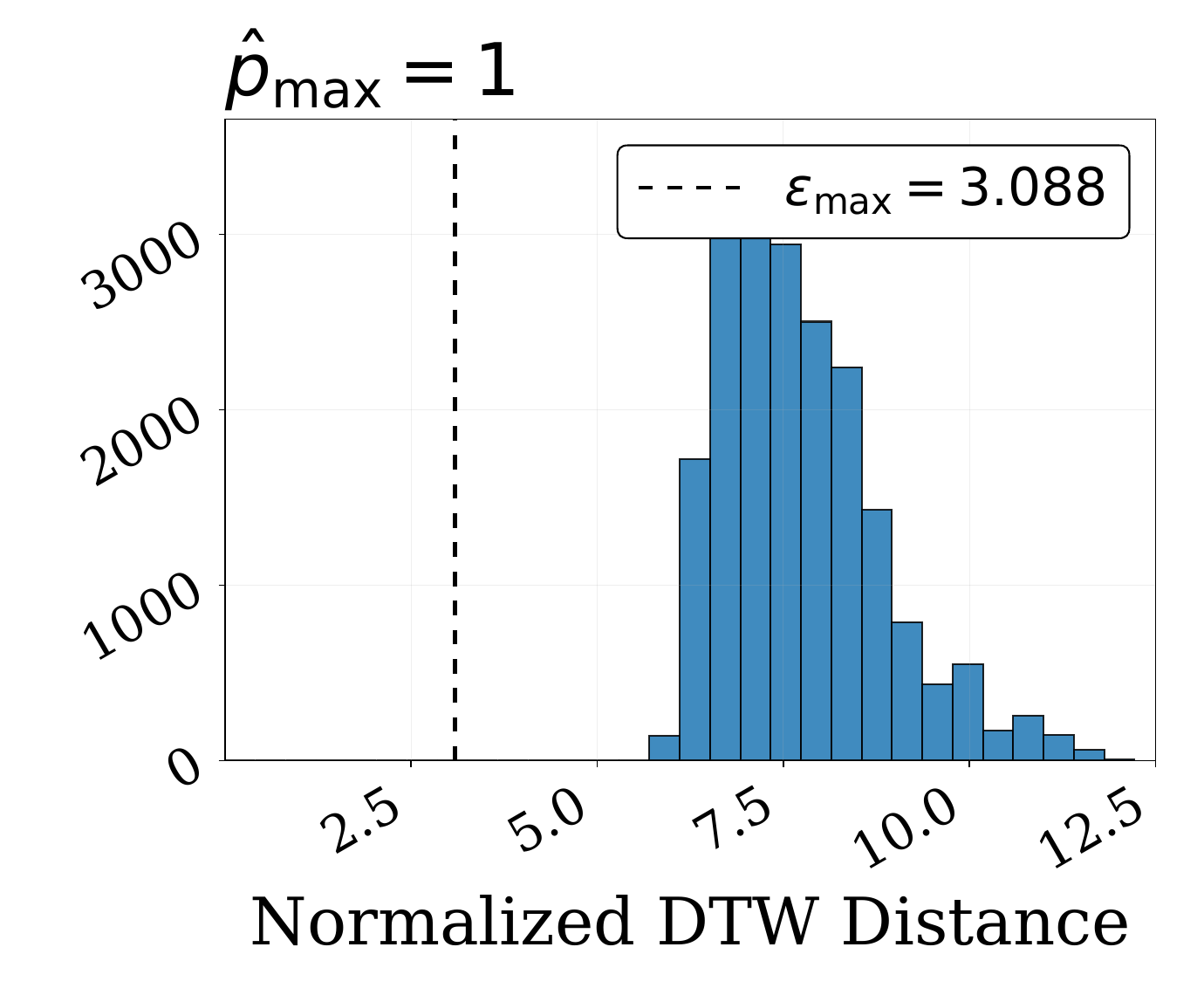}\hfill
    \includegraphics[width=0.25\linewidth]{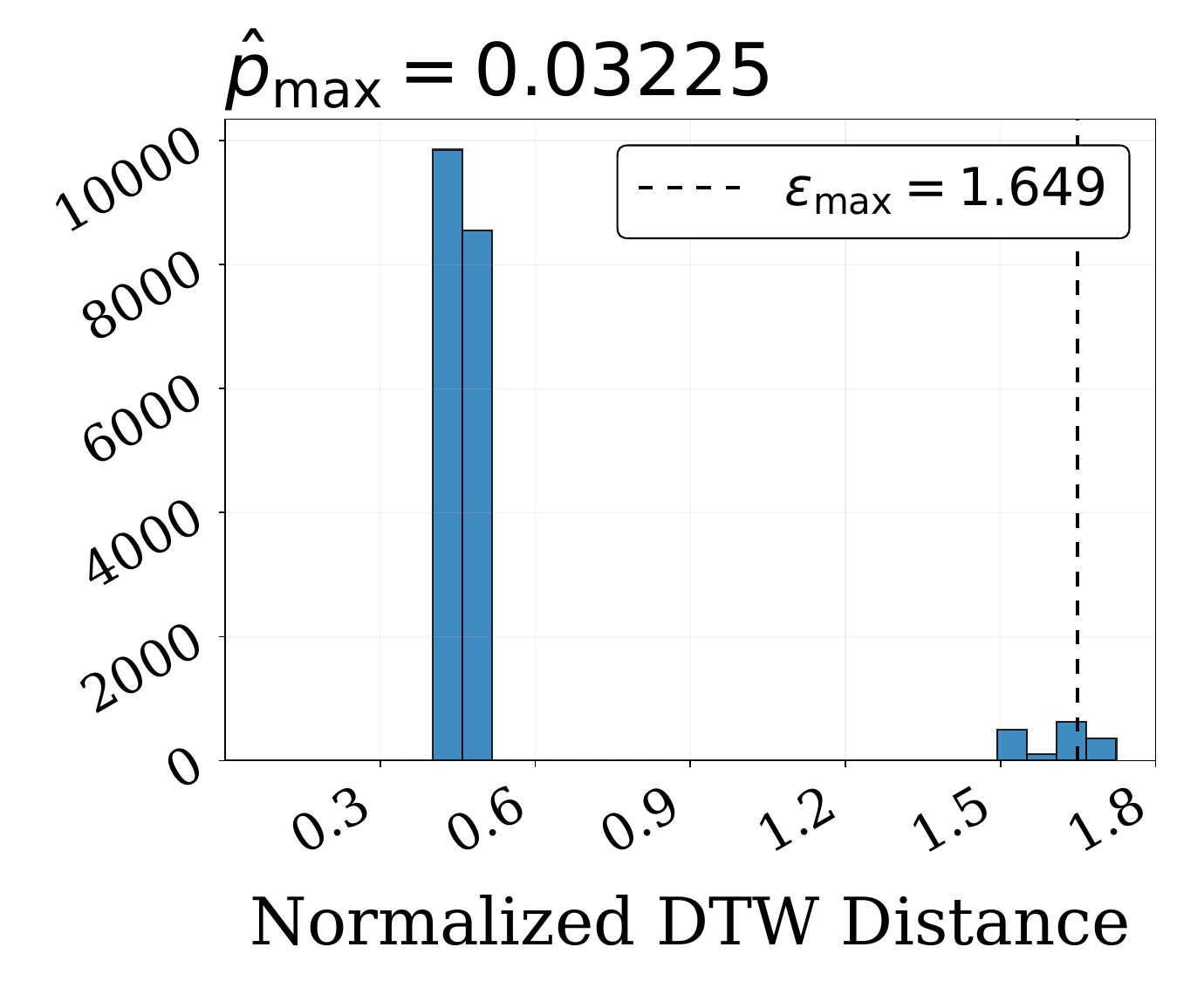}\hfill
    \includegraphics[width=0.25\linewidth]{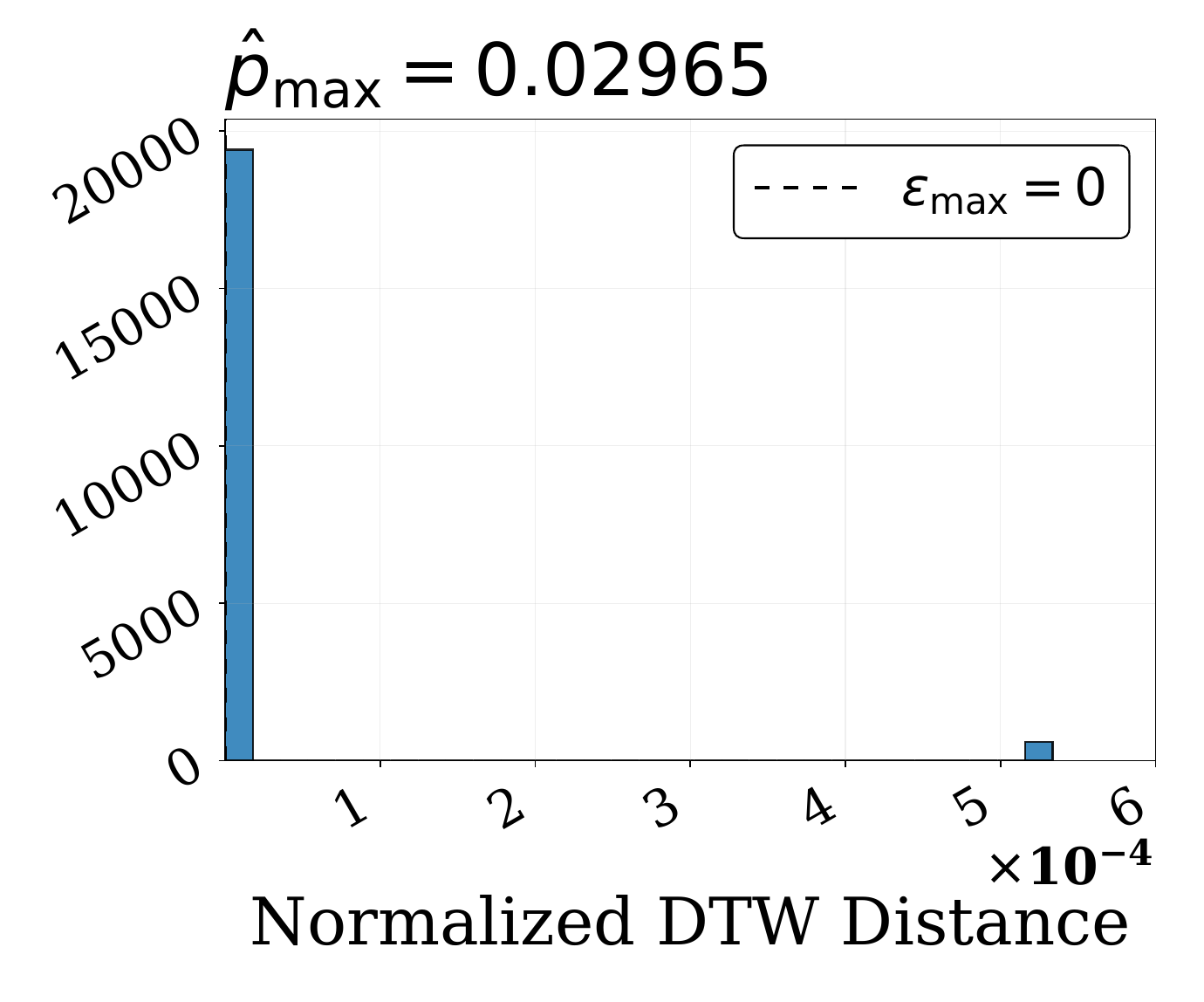}
\end{minipage}
\begin{minipage}[c]{0.04\textwidth}
    \centering\rotatebox{90}{Low BDP}
\end{minipage}

\caption{Validation of the TCP Prague $\times$ TCP Cubic runs across BDP scenarios - refined Mahimahi DualPI2 parameters.}
\label{fig:multi_bdp_grid_finale}
\end{figure*}

\vspace*{0.06in}
\noindent For the medium BDP scenario, although the value of \texttt{step\_thresh} that achieves the greatest  metric alignment in the L4S-only experiments is 10~ms, we empirically find that the 5 ms value yields the best results in terms of throughput for the dual-flow traffic pattern. 
The empirical distribution of each of our test statistics using the above optimal parameters for each BDP regime is summarized in Fig.~\ref{fig:multi_bdp_grid_finale}. We observe that, except for queue occupancy, the tuned parameters improve statistical alignment across most metrics in the medium and high BDP regimes. However, while throughput alignment improves at 50 and 200 Mbps, the resulting $\widehat{p}_{\max}$ remains above the rejection threshold, indicating persistent behavioral differences. Furthermore, the parameter adjustments introduce a tradeoff in the high BDP scenario: ECN marking behavior, previously aligned with the kernel, exhibits increased divergence under the tuned configuration. These results motivate further investigation, not only into the interaction with emulation timing granularity, but also in its coupling mechanism to better understand the sources of persistent divergence under high load.

To begin to understand the systemic differences between the Mahimahi and the kernel namespace-based experiments, we investigate the packet dequeue cadence in both systems and report results in Appendix~\ref{app:cadence}. Briefly, the discrepancies we observed are partly due to the relatively coarse network channel emulation in Mahimahi compared to the kernel. 

In Appendix~\ref{app:boostrap}, we introduce a bootstrap-based confidence interval framework to more precisely characterize this divergence and to formally assess, with statistical significance, whether one configuration demonstrates improvement over another. Our results show that, at moderate and high bandwidths, the refined configuration delivers clear, statistically significant improvements in throughput consistency over the default parameter setting. Nonetheless, the persistence of discrepancies at high BDP suggests that structural differences between Mahimahi’s emulation model and the kernel’s packet scheduling and timing mechanisms become more pronounced under higher traffic intensity.

\vspace*{0.05in}
Overall, our findings demonstrate that while cross-platform alignment can be improved through targeted parameter tuning, structural differences between Mahimahi and kernel execution environments introduce regime-dependent effects that must be carefully considered in reproducible L4S experimentation.

\section{Limitations}\label{sec:limitations}
In this section, we discuss the known limitations of our behavioral comparison of the Mahimahi DualPI2 module and the Linux implementation.

In terms of the traffic types considered, our experiments only test a single congestion control protocol per traffic type, namely TCP Cubic for classic flows and TCP Prague for L4S flows. This selection may not be representative of the congestion control distribution on the Internet, and does not provide insight into the behavioral equivalence of the two systems under other congestion controllers. In particular, for L4S traffic, we do not test Google's BBRv2 or BBRv3, which offer ECN/L4S support and are growing in prevalence~\cite{bbr-boon-bane, ietf-ccwg-bbr-05}. Furthermore, prior work has described fairness issues using the DualPI2 reference implementation under certain conditions, especially when a BBR flow is involved~\cite{adopt-not-adopt, bbr-boon-bane, l4s_partial_maprg}. In our experiments, the Cubic and Prague dual-flow scenarios exhibit relatively fair link sharing, although we do not formally quantify fairness or statistically validate it against the Linux implementation.

Regarding the network configurations tested, we only evaluate static scenarios across the different BDPs, with constant throughput and RTT values, because these settings are easier to configure and reproduce in both Mahimahi and the Linux namespace-based experiments than dynamic scenarios with time-varying throughput or RTT. Moreover, the Mahimahi emulator only allows for a constant RTT value per execution, and does not define dynamic RTT trace files that can be replayed in the same way as throughput traces.

\section{Related work}
\label{sec:related}

\subsection{Network Emulation \& Simulation Tools}
\label{sec:mahimahi-related-work}
Our work builds on Mahimahi, a widely used user-space network emulator that supports the record-and-replay of packet-level  network traces, originally introduced by Netravali et al.~\cite{mahimahi-atc, mahimahi-ccr}. While it was initially developed to emulate cellular conditions using real-world mobile bandwidth traces, Mahimahi is not limited to mobile scenarios. Users can define custom network traces, including constant-rate links or synthetic profiles, enabling its use across a broad range of network conditions. Its extensibility, support for time-varying network behavior and ease of use have made it the go-to emulator for reproducible protocol experimentation, and has enabled a plethora of important work in networked systems and protocol design and evaluation~\cite{tcp-cellular, abc, alohamora, pensieve, vesper, cellfusion, floo, vantage, robust-dash, salient-vr, ng-scope, in-situ, neural-video, prior, low-latency-http, compact, salsify}. Other emulation tools include NetEm~\cite{netem} and dummynet~\cite{dummynet}, which allow bandwidth and delay control but rely on static configurations that do not capture the time-varying nature of mobile networks. Mininet~\cite{mininet} emulates complete network topologies by creating virtual hosts and links using Linux namespaces. These systems are valuable for repeatable testing, but none currently support L4S-compatible AQMs.

Aside from emulators, it is worth mentioning  that the ns-3 simulator~\cite{ns3-paper} has started to integrate L4S-related support~\cite{ns3-L4S-support}; The DualPI2 AQM, although available as an external module~\cite{ns3_dualq_pi2}, has not yet been included in the mainline release. Recent work has implemented the L4S end-host stack in ns-3~\cite{ns3-l4s-implem} including TCP Prague, and validated it against a Linux L4S testbed. However, testing real-world applications through ns-3 poses additional challenges not found in Mahimahi-based experimentation. First, although Direct Code Execution (DCE) enables existing userspace applications to run within ns-3, applications must be rebuilt in a DCE-compatible form~\cite{ns3-dce-intro}, and compatibility depends on whether the required POSIX/library/system calls are supported by DCE. In particular, missing function symbols or system calls may require adding the corresponding support manually~\cite{ns3-dce-syscalls}. This adds uncertainty when integrating complex applications such as real multimedia systems, which are especially relevant for evaluating L4S. Moreover, DCE only supports C/C++ applications~\cite{ns3-dce-intro}, limiting the range of applications that can be integrated into the ns-3 ecosystem. By contrast, Mahimahi enables applications to run without modifying or rebuilding their binaries, and our integrated DualPI2 module allows low-friction end-to-end testing of complex multimedia applications under L4S. Second, trace-driven experiments are simpler in Mahimahi: each trace line represents a packet-delivery opportunity, making cellular traces straightforward to replay~\cite{mahimahi-atc}. In ns-3, variable-link behavior can be emulated by dynamically changing link attributes such as the point-to-point device DataRate, or by using more detailed wireless/channel models, adding to the complexity of the experiment preparation task~\cite{ns3-p2p-datarate}.

Lastly, proprietary tools have also been used for L4S experimentation. For instance, Ericsson employed an internal simulator in early evaluations of SCReAM over L4S networks~\cite{brunello-master, brunello-ieee}. While such setups enable advanced testing, they are not publicly available, limiting their usefulness for broader validation or follow-up work.

\vspace*{0.1in}
To our knowledge, our work is the first to introduce modular L4S capabilities into an open-source, record-and-replay emulator like Mahimahi, allowing the community to both evaluate existing protocols and experiment with new algorithmic designs under controlled, trace-driven network conditions.

\subsection{Prior Work on L4S}
\label{sec:l4s-prior}
One of the earlier research in the L4S applications sphere is work by Brunello et al~\cite{brunello-master, brunello-ieee} on L4S in 5G networks. In particular, this work explores the adoption of L4S for latency-critical Augmented Reality (AR) video gaming traffic using the Self-Clocked Rate Adaptation for Multimedia (SCReAM) congestion controller~\cite{rfc8298, scream-paper}. The study shows that enabling L4S significantly reduces end-to-end delay compared to non-L4S operation, while maintaining application-layer throughput above the minimum requirements of high-rate, latency-sensitive applications, even under high system load. Although a fundamental throughput–latency trade-off remains, L4S consistently achieves lower packet loss. This study also investigates practical deployment challenges in 5G networks, such as encryption at the Radio Link Control (RLC) layer, which prevents packet marking within certain network nodes. To address this issue, the authors implement the L4S marking strategy at the Packet Data Convergence Protocol (PDCP) layer within the gNB, treating it as the effective network marking point. Still in the 5G context, the L4Span Radio Access Network (RAN) architecture~\cite{l4span} was proposed to enable end-to-end low-latency signaling on 5G's last-mile link. It exposes RAN queue state to predict millisecond-scale queue occupancy and applies ECN marking for both low-latency and classic flows. The design aims to minimize required RAN modifications while remaining 3GPP and O-RAN compliant. A prototype is implemented in C++ on the open-source srsRAN platform. In the authors’ reported evaluation, their design reduces one‑way delay of both low‑latency and classic flows by up to 98\% while maintaining near line‑rate throughput. A number of other work have shown similar performance gains for various applications: Extended Reality (XR)~\cite{l4sxr-steininger, split-render-xr}, cloud gaming~\cite{l4s-cloud-gaming-graff}, volumetric video conferencing~\cite{l4s-volumetric}, real-time streaming~\cite{l4s-private-5g-video, rtc-role-l4s, real-time-stream}. 

Some other research explored the effect of a partial deployment of the L4S ecosystem, pointing to L4S benefits not being guaranteed under partial deployment. Prior work~\cite{adopt-not-adopt} looks into the deployment of L4S-compatible congestion controllers amid partial L4S deployment on bottleneck nodes. The paper concludes that an L4S sender (using TCP Prague or BBRv2‑style scalable congestion controller) cannot be assured it will not harm or be harmed by other flows when the sender does not know the bottleneck type and other cross‑traffic conditions. In a preceding study~\cite{switch-not-switch}, findings highlight cases where TCP Prague’s throughput or fairness can degrade in single‑queue or incompatible coexistence settings (e.g., with non‑ECN‑responsive cross traffic).

Besides, there have been several attempts at devising novel congestion control protocols that leverage L4S' signaling. Among these works, we can cite aL4S-CC~\cite{real-time-stream}. The evaluation against GCC and SCReAMv2 reveals an improved average link utilization by 4\% compared to GCC and 17.9\% compared to SCReAMv2. Another such work by Pan et al. proposes L4S-GCC~\cite{rtc-role-l4s} for RTC/WebRTC: it combines ECN-derived queue information with delay-gradient logic to improve responsiveness and recovery; the authors report a reduction in stalling rate by 1.51–2.80\% vs. GCC and an improvement in bandwidth utilization by 11.4–31.4\% under delay jitter.

While these efforts have shown promising results, they remain hard to reproduce and extend. Our goal is to lower the barrier to entry for L4S research by embedding DualPI2 into a flexible user-space emulator.

\section{Conclusion}
Although a Linux implementation of DualPI2 is available, controlled and reproducible experimentation on L4S mechanisms benefits from a modular user-space implementation. By adding the DualPI2 AQM to the Mahimahi emulator, we provide an extensible module that enables rapid iteration on control parameters, facilitates component-level experimentation, and supports application-focused studies without kernel modification and across diverse network conditions.

Through statistical behavioral characterization and cross platform comparison, we show that alignment between emulated and kernel executions is neither automatic nor invariant across BDP regimes. Instead, DualPI2’s effective dynamics depend on the interaction between control parameters and the underlying execution environment, leading to regime-dependent behavior. Our analysis demonstrates that targeted parameter tuning can improve alignment in specific BDP conditions, while also revealing structural differences that persist under higher load. Table~\ref{tab:param-summary} summarizes the DualPI2 parameters that showed, with a 5\% statistical significance threshold, the best alignment with the Linux kernel reference implementation for each BDP regime.

\begin{table}[ht]
\centering
\caption{Mahimahi DualPI2 parameters for each BDP regime}
\renewcommand{\arraystretch}{1.15}
\setlength{\tabcolsep}{4pt}
\begin{tabular}{|l|c|c|c|c|}

\hline
\textbf{Regime} & \textbf{RTT} & \textbf{Bandwidth} & \textbf{\texttt{step\_thresh}} &
\textbf{\texttt{target}} \\
\hline\hline
Low-BDP  & 20 ms       & 12 Mbps  & 5 ms & 30 ms \\
\hline
Medium-BDP  & 40 ms   & 50 Mbps  & 5 ms & 30 ms \\
\hline
High-BDP & 100 ms & 200 Mbps & 10 ms & 45 ms \\
\hline
\end{tabular}

\label{tab:param-summary}
\end{table}

Collectively, this work lowers the barrier to experimental L4S research while providing an empirical understanding of cross-platform behavioral discrepancies. We hope it serves both as a practical tool for the community and as a foundation for deeper investigation into the parameter sensitivity of DualPI2 and the interaction between AQM control logic and emulation environments.

\begin{acks}
 We are grateful to Dave Täht for insightful discussions in the early stages of this work. We especially thank Dr. Sadjad Fouladi for his technical guidance and encouragement as we navigated the Mahimahi emulator project. We also acknowledge Dr. Prateesh Goyal for his advice on AQM development in Mahimahi. Generative AI tools were used to support scripting for experiments and statistical validation, plot generation, and language polishing. This work was partly funded by the Comcast Innovation Fund, grant 2024-1030.

\end{acks}

\balance
\bibliographystyle{ACM-Reference-Format}
\bibliography{reference}

\clearpage
\section*{Appendix}
\appendix
\section{Mahimahi Vs. Kernel Dequeue Cadence}
\label{app:cadence}

\begin{figure}[ht]
  \centering
  \includegraphics[width=0.47\textwidth]{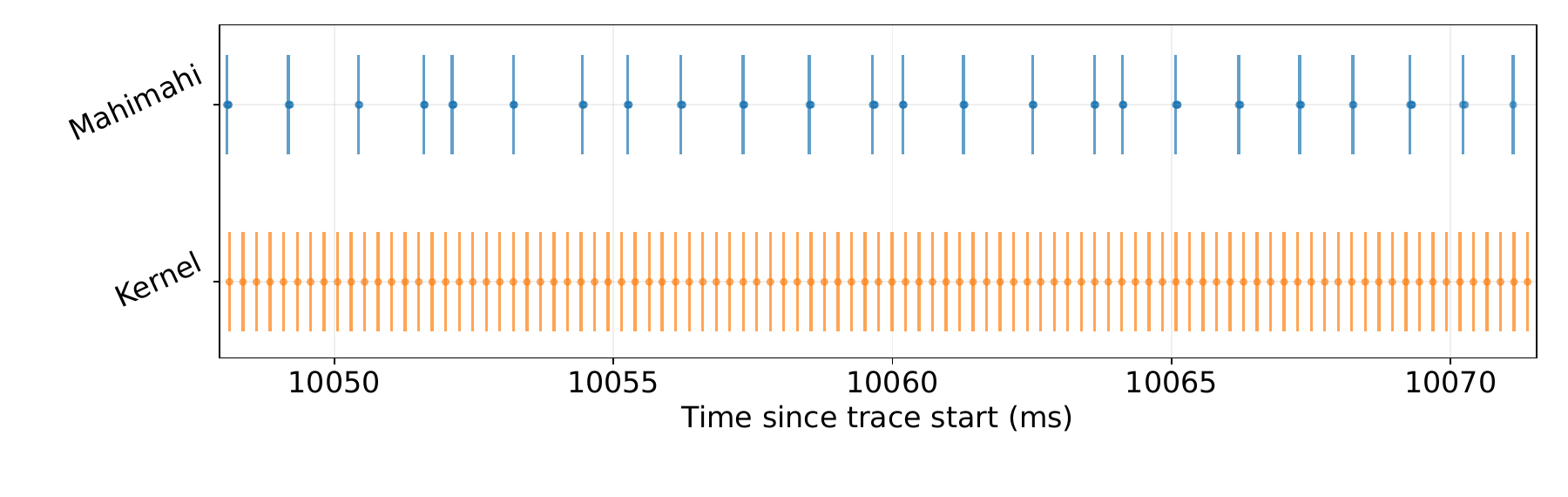}
  \caption{Dequeue events from Mahimahi and kernel. Each vertical line represents the beginning of a burst, and each dot represents a packet. For Mahimahi, each visible dot is in fact multiple dots stacked on top of each other.}
  \label{fig:dequeue_raster}
\end{figure}

To investigate the statistical misalignment described in \S\ref{sec:eval} between the DualPI2 Mahimahi module and the kernel reference implementation, we collect dequeue traces, which consist of a series of timestamps corresponding to packet dequeue events from both systems. Fig.~\ref{fig:dequeue_raster} shows a snippet of a dequeue trace obtained by running an \texttt{iperf3} flow using TCP Prague through the DualPI2 AQM with the default parameters outlined in \S\ref{sec:validation-setup}, in the medium BDP scenario (50 Mbps with an RTT of 40 ms).

The figure shows that the packet dequeue in Mahimahi is visibly more sparse than the smooth and regular dequeue in the kernel setting. Mahimahi has a granularity of 1 ms, resulting from how the network traces are encoded. Indeed, the trace is a series of millisecond timestamps, where if a given millisecond appears in the trace \texttt{n} times, then a maximum of \texttt{n} MTU-size packets (1500 bytes by default) can be released during that one millisecond. This naturally creates a burst of packets every millisecond, and the higher the throughput, the higher the amplitude of the burst. 

To quantify the dequeue period in both Mahimahi and the kernel settings, we run a period extraction heuristic on the dequeue traces and obtain a period of 998 $\mu$s for Mahimahi, confirming a 1 ms granularity, and 242 $\mu$s for the kernel, corresponding to dequeue events that are about four times more frequent than in Mahimahi. 

\begin{figure}[H]
  \centering
  \includegraphics[width=0.47\textwidth]{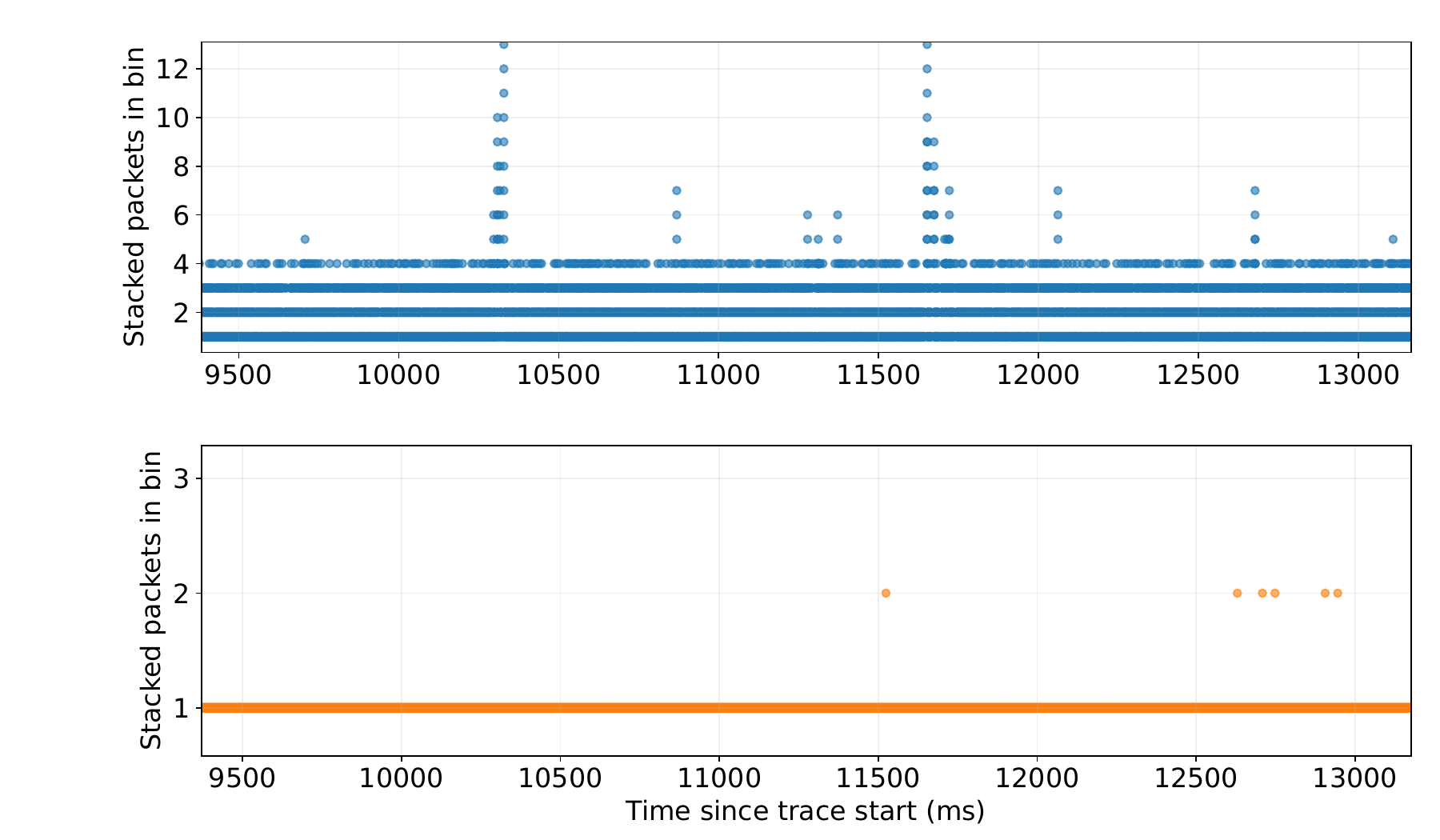}
  \caption{Packets dequeued in an interval of 120 $\mu$s. The top figure represents Mahimahi, while the bottom represents the kernel.}
  \label{fig:packet_dot_plot}
\end{figure}

Fig.~\ref{fig:packet_dot_plot} shows a sample dot plot of dequeued packets over an arbitrarily small interval (we choose 120 $\mu$s), where each dot is a single packet. In the top graph, representing Mahimahi, frequent bursts of up to 12 packets can be observed. By contrast, bursts in the kernel setting are both less frequent and smaller, with a maximum of two packets per interval.

Together, these observations highlight the inherent difficulty of emulating a low-latency AQM such as DualPI2. One major consequence of the 1 ms default granularity of Mahimahi is that the L4S queue's default \texttt{step\_thresh} of 1 ms cannot be precisely enforced, as packets build up in the queue while waiting for the next millisecond mark to be released. This helps explain why the throughput results improved when we set \texttt{step\_thresh} and \texttt{target} to higher values for higher BDP regimes. Adjusting the DualPI2 parameters, as we do in this work, therefore provides a practical path toward using an emulated DualPI2 module and opens up opportunities for rapid end-to-end L4S research and experimentation.

 Alternatively, efforts aimed at calibrating emulation tools such as Mahimahi for low-latency experiments could be considered, as low-latency research is only going to receive increasing attention.

\section{Bootstrap Confidence Intervals for Further Statistical Validation}
\label{app:boostrap}

While $\widehat{p}_{\max}$ provides a point estimate of the 
exceedance rate, it is computed from a finite number of 
experimental runs and is therefore subject to sampling 
variability. To ensure that our conclusions are not driven 
by a particular realization of the data, we quantify the 
uncertainty in $\widehat{p}_{\max}$ using nonparametric 
bootstrap resampling with replacement~\cite{efron1979bootstrap}. 
This procedure approximates the sampling distribution of 
$\widehat{p}_{\max}$ under repeated experiments, allowing us 
to assess how the exceedance probability would vary under 
repeated sampling of runs.

\subsection{Bootstrap Resampling Implementation}

In our implementation, resampling is performed at the run level within each system. For each replicate, we sample runs with replacement from the kernel set and independently from the Mahimahi set, preserving the original sample size in each group. Using these resampled run sets, we recompute all pairwise within-system and cross-system differences, re-estimate the tolerance threshold $\varepsilon_{\max}$, and recompute the exceedance statistic $\widehat{p}_{\max}$. Repeating this procedure across $B$ replicates yields an empirical bootstrap distribution of $\widehat{p}_{\max}$ from which confidence intervals are constructed.

We use $B=2000$ bootstrap replicates, which provides stable percentile
confidence intervals: the Monte Carlo error of the bootstrap quantiles
decreases as $O(B^{-1/2})$, and $B=2000$ yields sufficiently fine
resolution for the 2.5th and 97.5th percentiles (indices 50 and 1950),
ensuring that our conclusions are not sensitive to the number of
replicates.

\subsection{Bootstrap Confidence Interval}

Let $\widehat{p}_{\max}^{(1)}, \dots, \widehat{p}_{\max}^{(B)}$ 
denote the bootstrap replicates obtained from $B$ resamples. 
We construct a 95\% percentile bootstrap confidence interval as:

\[
\left[
\mathrm{Quantile}_{0.025}
\big( \widehat{p}_{\max}^{(1:B)} \big),
\;
\mathrm{Quantile}_{0.975}
\big( \widehat{p}_{\max}^{(1:B)} \big)
\right]
\]

This interval captures the central 95\% of the empirical bootstrap 
distribution. The lower bound represents a conservative estimate of 
the smallest plausible exceedance probability under repeated 
sampling, while the upper bound represents the largest plausible 
exceedance probability consistent with the observed variability.

\subsection{Decision Rule}

We formalize the validation criterion as a one-sided test of the 
exceedance probability:

\vspace*{-0.1in}
\[
H_0: p_{\max} \ge 0.05
\quad\text{versus}\quad
H_1: p_{\max} < 0.05.
\]

We reject $H_0$ if the upper endpoint of the 95\% bootstrap 
confidence interval for $\widehat{p}_{\max}$ lies below 0.05. 
In this case, the data provides statistically significant evidence 
that the cross-system exceedance rate is below the prescribed 5\% 
tolerance. Conversely, if the confidence interval overlaps or 
exceeds 0.05, we do not reject $H_0$. This means there is not 
sufficient statistical evidence to conclude that the exceedance 
probability is below 5\%.

To assess whether the Optimized configuration improves upon the 
Default configuration, we compare their bootstrap 95\% confidence 
intervals for $p_{\max}$. We declare a statistically significant 
improvement only if the entire Optimized confidence interval lies 
strictly below the entire Default confidence interval. In this case, 
even accounting for sampling variability captured by the bootstrap, 
all plausible values of $p_{\max}$ under the Optimized configuration 
are smaller than those under Default, providing statistical evidence 
that the exceedance probability has been reduced.

\subsection{Results}

We summarize the validation results in Table~\ref{tab:combined}, which reports, for each bandwidth and variable, the bootstrap 95\% confidence intervals for $p_{\max}$ under both the Default and Optimized configurations. The table additionally indicates whether the Optimized configuration demonstrates a strict improvement over Default (non-overlapping confidence intervals) and whether statistical significance at the 5\% level is achieved according to the decision rule.

\begin{table*}[t]
\centering
\caption{Bootstrap 95\% CIs for $\widehat{p}_{\max}$: Default vs.\ Optimized across bandwidths and variables. \checkmark\,=\,yes; $\times$\,=\,no.}
\label{tab:combined}
\setlength{\tabcolsep}{4pt}
\small
\begin{tabular}{llcccccc}
\toprule
& & \multicolumn{2}{c}{Bootstrap 95\% CI for $\widehat{p}_{\max}$} & & \multicolumn{2}{c}{Sig.\ ($\alpha = 0.05$)} \\
\cmidrule(lr){3-4} \cmidrule(lr){6-7}
Bandwidth & Variable & Default & Optimized & Improved & Default & Optimized \\
\midrule
\multirow{4}{*}{12\,Mbps}
 & packets                & [0.613, 0.954] & [1.000, 1.000] & $\times$ & $\times$ & $\times$ \\
 & throughput             & [0.000, 0.018] & [0.000, 0.280] & $\times$ & $\checkmark$ & $\times$ \\
 & ecn\_mark              & [0.019, 0.418] & [0.023, 0.046] & $\times$ & $\times$ & $\checkmark$ \\
 & packet\_dropped\_total & [0.000, 0.020] & [0.000, 0.039] & $\times$ & $\checkmark$ & $\checkmark$ \\
\midrule
\multirow{4}{*}{50\,Mbps}
 & packets                & [0.975, 1.000] & [1.000, 1.000] & $\times$ & $\times$ & $\times$ \\
 & throughput             & [0.222, 0.470] & [0.050, 0.169] & $\checkmark$ & $\times$ & $\times$ \\
 & ecn\_mark              & [0.022, 0.051] & [0.025, 0.324] & $\times$ & $\times$ & $\times$ \\
 & packet\_dropped\_total & [0.000, 0.000] & [0.000, 0.000] & $\times$ & $\checkmark$ & $\checkmark$ \\
\midrule
\multirow{4}{*}{200\,Mbps}
 & packets                & [0.814, 1.000] & [1.000, 1.000] & $\times$ & $\times$ & $\times$ \\
 & throughput             & [0.338, 0.779] & [0.000, 0.115] & $\checkmark$ & $\times$ & $\times$ \\
 & ecn\_mark              & [0.015, 0.812] & [0.076, 0.158] & $\times$ & $\times$ & $\times$ \\
 & packet\_dropped\_total & [0.000, 0.000] & [0.000, 0.000] & $\times$ & $\checkmark$ & $\checkmark$ \\
\bottomrule
\end{tabular}
\end{table*}

Table~\ref{tab:combined} reports bootstrap 95\% confidence intervals for $\widehat{p}_{\max}$ under both configurations. Statistical significance, defined by the upper bound of the confidence interval lying strictly below the 5\% tolerance threshold, is achieved only in isolated cases. Specifically, significance is observed for throughput and packet\_dropped\_total at 12\,Mbps under the Default configuration; for ecn\_mark and packet\_dropped\_total at 12\,Mbps under the Optimized configuration; and for packet\_dropped\_total at 50 and 200\,Mbps under both configurations. In all remaining cases, the confidence intervals overlap or exceed the 0.05 threshold, and we therefore fail to reject $H_0$, indicating insufficient statistical evidence that the exceedance probability remains below the prescribed tolerance.

Although the Optimized configuration does not demonstrate a universal improvement across all variables, it achieves clear and strict gains in throughput at higher bandwidths. At 50\,Mbps, the Optimized confidence interval $[0.050, 0.169]$ lies entirely below the Default interval $[0.222, 0.470]$, indicating a non-overlapping and strictly lower exceedance probability. An even stronger separation appears at 200\,Mbps, where the Optimized interval $[0.000, 0.115]$ is fully contained below the Default interval $[0.338, 0.779]$. In both cases, the entire Optimized confidence band is strictly smaller than the corresponding Default band, providing statistical evidence of improvement under our non-overlapping CI criterion. These results indicate that at moderate and high bandwidths, the Optimized configuration substantially reduces the probability of cross-system deviations exceeding within-system variability, yielding improved statistical consistency in throughput behavior.

\subsection{Sample Size Justification for 100 Flows per Configuration}
\label{app:sample-size}

To justify the adequacy of 100 flows per configuration, we examine the bootstrap 95\% confidence intervals computed from the collected data. In several cases, the intervals are already extremely narrow and lie entirely below the 0.05 tolerance threshold, including throughput at 12\,Mbps under Default ([0.000, 0.018]) and packet\_dropped\_total at 12\,Mbps under both configurations ([0.000, 0.020] and [0.000, 0.039]), as well as packet\_dropped\_total at 50 and 200\,Mbps where the intervals collapse to $[0.000, 0.000]$. 

\begin{figure}[H]
    \centering
    \begin{subfigure}[b]{0.155\textwidth}
        \includegraphics[width=\textwidth]{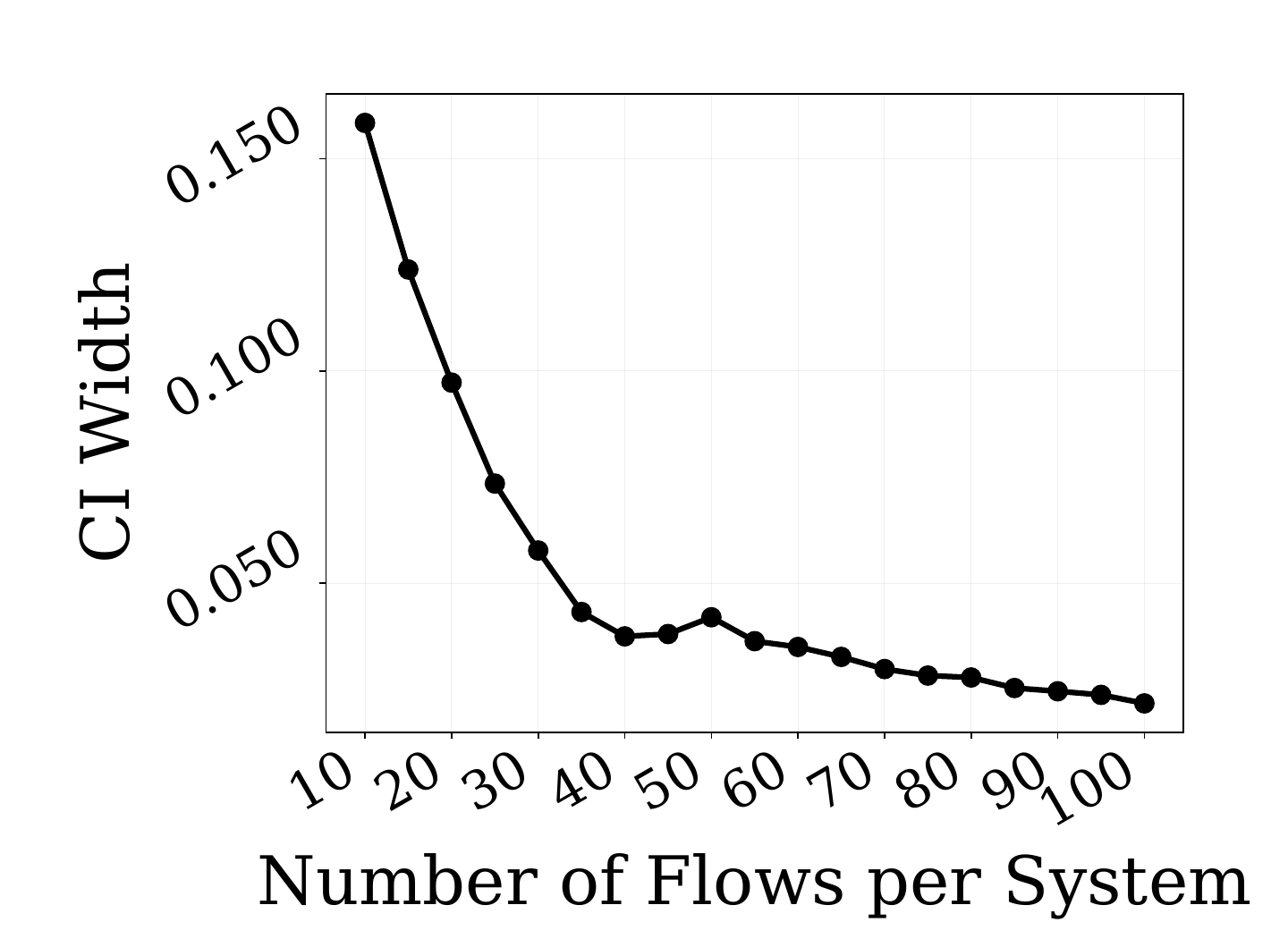}
        \caption{12\,Mbps finale}
    \end{subfigure}
    \begin{subfigure}[b]{0.155\textwidth}
        \includegraphics[width=\textwidth]{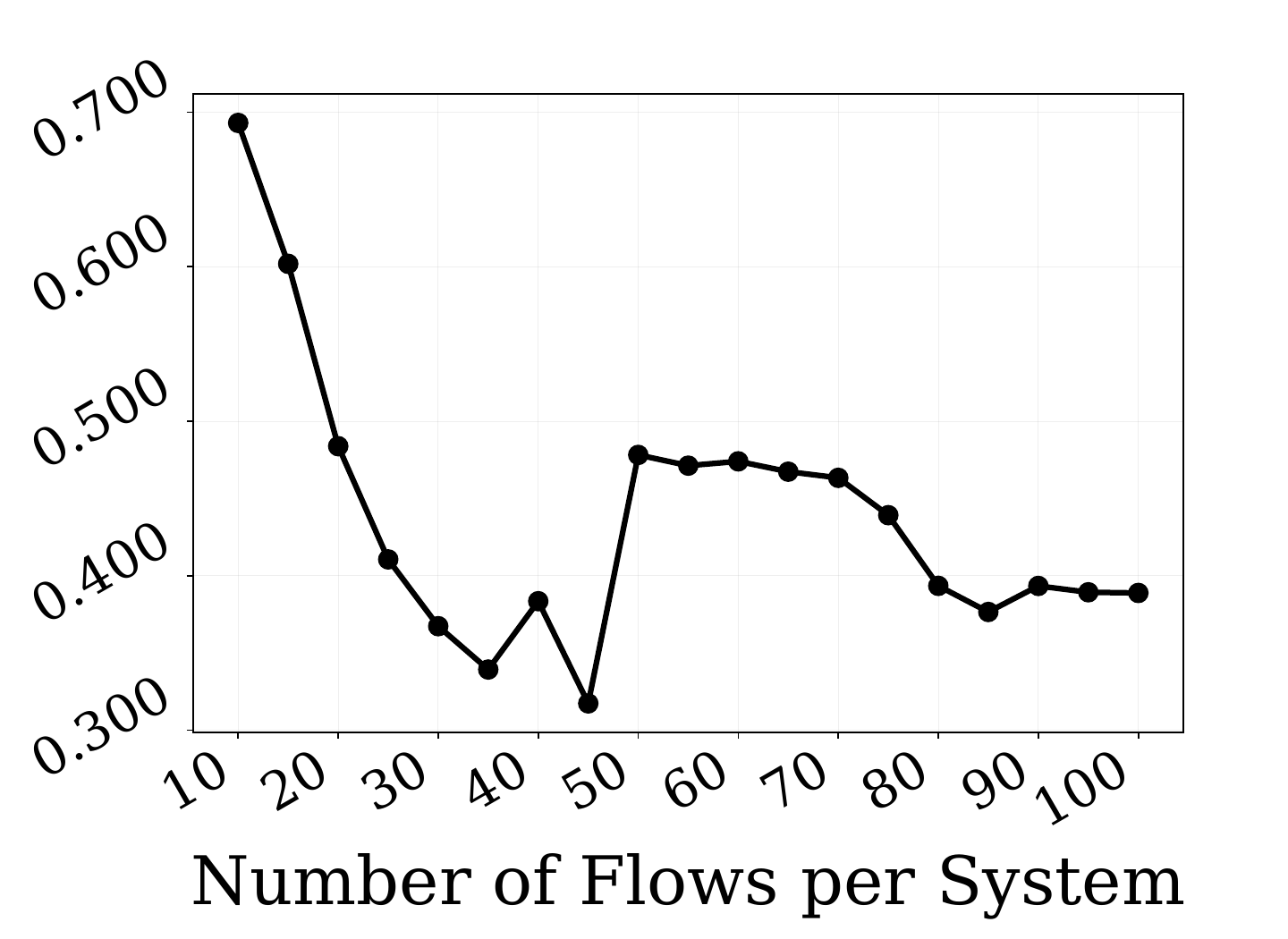}
        \caption{12\,Mbps}
    \end{subfigure}
    \begin{subfigure}[b]{0.155\textwidth}
        \includegraphics[width=\textwidth]{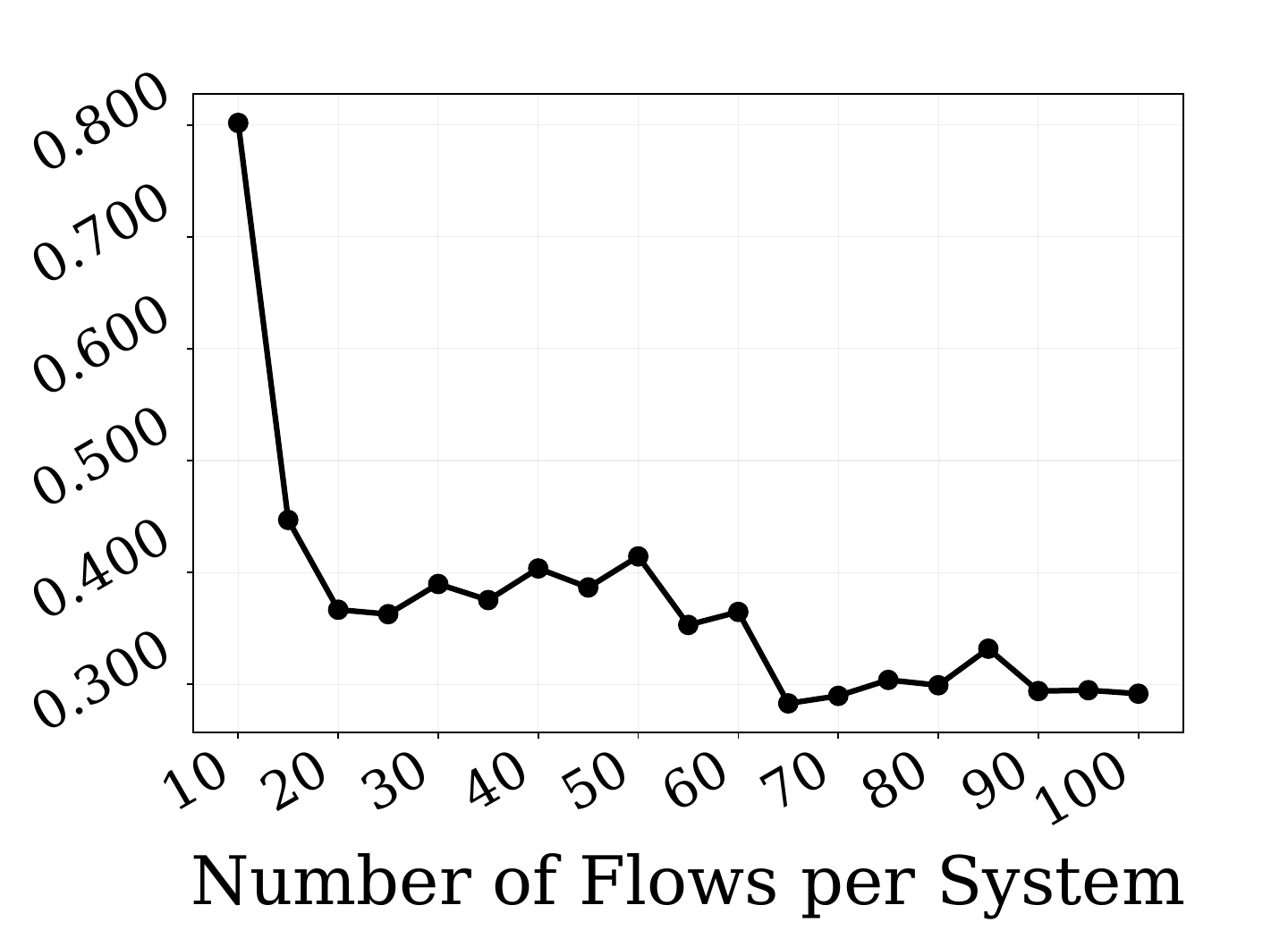}
        \caption{50\,Mbps finale}
    \end{subfigure}
    \begin{subfigure}[b]{0.155\textwidth}
        \includegraphics[width=\textwidth]{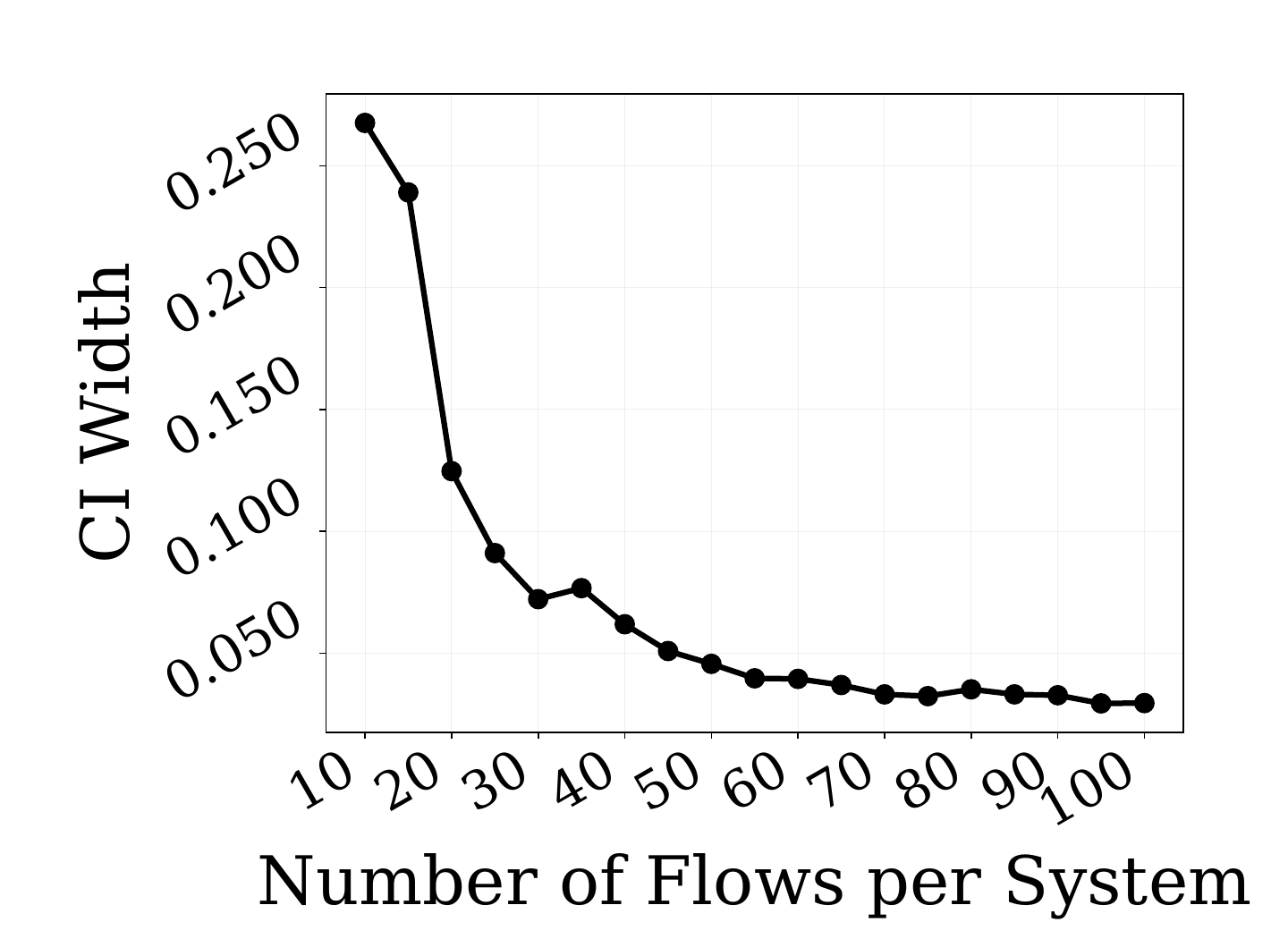}
        \caption{50\,Mbps}
    \end{subfigure}
    \begin{subfigure}[b]{0.155\textwidth}
        \includegraphics[width=\textwidth]{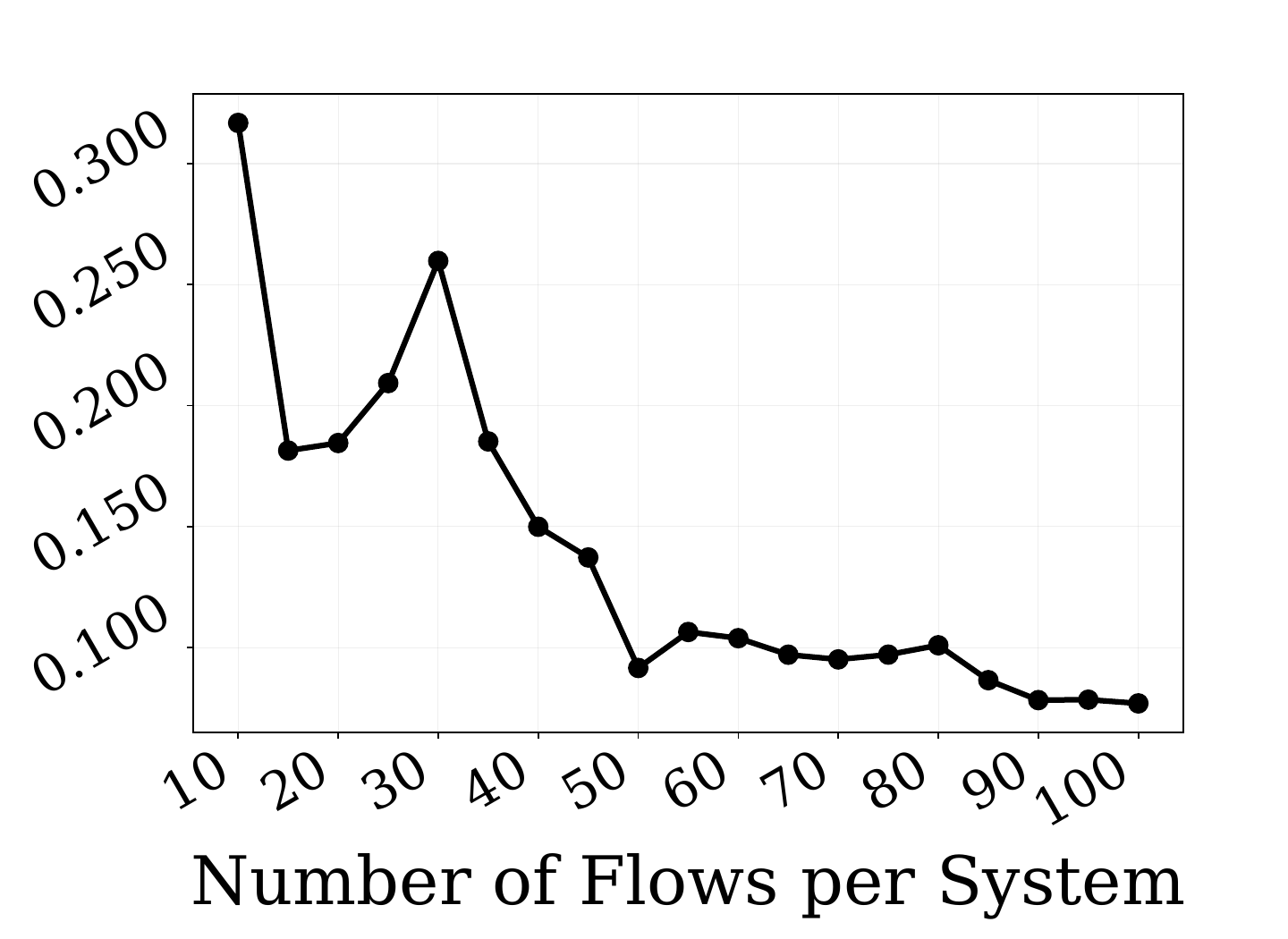}
        \caption{200\,Mbps finale}
    \end{subfigure}
    \begin{subfigure}[b]{0.155\textwidth}
        \includegraphics[width=\textwidth]{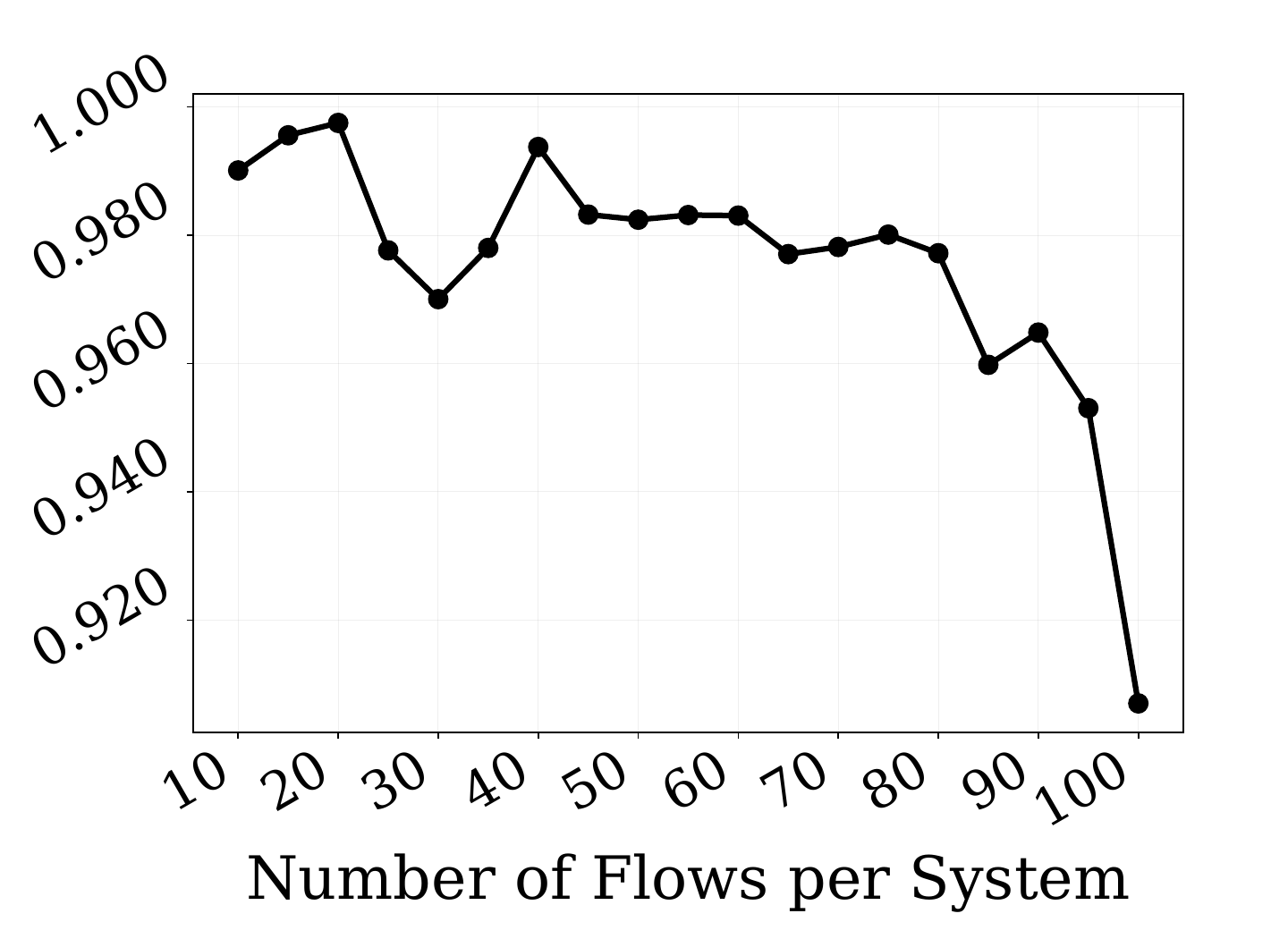}
        \caption{200\,Mbps}
    \end{subfigure}
    \caption{Bootstrap 95\% CI width vs.\ number of flows $n$ for ECN mark rate across all bandwidth configurations.}
    \label{fig:ci_width_all}
\end{figure}

Because these intervals are both tight and fully contained below the tolerance level, additional flows would be unlikely to alter the decision outcome. This indicates that, for these scenarios, the current sample size already provides stable and statistically decisive estimates of $p_{\max}$. 

However, the relatively wide ECN-mark confidence intervals could be interpreted as indicating insufficient sampling and the potential need for additional runs. However, when plotted against the number of flows (see Fig.~\ref{fig:ci_width_all}), the confidence intervals clearly shrink and stabilize as $n$ approaches 100, demonstrating convergence rather than instability.

 \noindent The slight dip observed in panel~(f) at $n = 100$ is most likely due to bootstrap resampling variability, since it is not part of a consistent downward trend and neighboring sample sizes exhibit similar confidence interval widths. 

Overall, this behavior justifies the choice of $n = 100$ flows for these configurations, as the interval widths have largely plateaued by this point. At the same time, it highlights an important characteristic of ECN marks: they exhibit inherently higher variability compared to other metrics, suggesting that their dispersion reflects natural stochastic behavior rather than insufficient sampling.

\end{document}